\documentclass[aps,prx,reprint,showpacks,twocolumn,superscriptaddress]{revtex4}
\usepackage[intlimits]{amsmath}
\usepackage{mciteplus} 
\usepackage{feynmp}
\usepackage{slashed}
\usepackage{graphicx}
\usepackage{physics}
\usepackage{epstopdf}
\usepackage{amssymb}
\usepackage{subfigure}
\usepackage{amsmath}
\usepackage{amsfonts}
\usepackage{amssymb}
\usepackage{color}
\usepackage{comment}
\usepackage{tikz}
\usetikzlibrary{shapes}
\usepackage{amsmath}
\usepackage{empheq}
\usepackage{mathrsfs}
\usepackage{ulem}
\usepackage[colorlinks=false, citecolor=black, urlcolor=black]{hyperref}

\usepackage[utf8]{inputenc}
\usepackage{lipsum}

\usepackage{graphicx}
\usepackage{amsmath}
\usepackage{amssymb, mathtools, mathrsfs}
\usepackage{hyperref}

\newcommand\be            {\begin{equation}}
\newcommand\ee            {\end{equation}}

\def\bea{\begin{eqnarray}}
\def\eea{\end{eqnarray}}
\def \Tr {{\textrm{Tr}}}

\newcommand\EN           {\end{equation}}
\newcommand\bes           {\begin{subequations}}
\newcommand\esu           {\end{subequations}}

\def\3pt#1#2#3{{\langle{#1}\vert{#2}\vert{#3}\rangle}}

\DeclareFontFamily{U}{mathb}{\hyphenchar\font45}
\DeclareFontShape{U}{mathb}{m}{n}{
<-6> mathb5 <6-7> mathb6 <7-8> mathb7
<8-9> mathb8 <9-10> mathb9
<10-12> mathb10 <12-> mathb12
}{}
\DeclareSymbolFont{mathb}{U}{mathb}{m}{n}
\DeclareMathSymbol{\llcurly}{\mathrel}{mathb}{"CE}
\DeclareMathSymbol{\ggcurly}{\mathrel}{mathb}{"CF}

\begin{document}

\title{
Continuous majorization in quantum phase space 
for Wigner-positive
 states
\\
and proposals 
for Wigner-negative states}

	 \author{Jan de Boer}
	 \affiliation{Institute for Theoretical Physics and Delta Institute for Theoretical Physics, University of Amsterdam, PO Box 94485, 1090 GL Amsterdam, The Netherlands}
	 
	 \author{Giuseppe Di Giulio}
	 \affiliation{The Oscar Klein Centre and Department of Physics,
	 \\
     Stockholm University, AlbaNova,
10691 Stockholm, Sweden}
	 
	 \author{Esko Keski-Vakkuri}
	 \affiliation{Department of Physics, University of Helsinki, PO Box 64, FIN-00014 University of Helsinki, Finland}
	 \affiliation{Helsinki Institute of Physics, PO Box 64, FIN-00014 University of Helsinki, Finland}
\affiliation{InstituteQ - the Finnish Quantum Institute, University of Helsinki, Finland}	 
	 
	 \author{Erik Tonni}
	 \affiliation{SISSA and INFN Sezione di Trieste, via Bonomea 265, 34136, Trieste, Italy }

\begin{abstract} 
In quantum resource theory, one is often interested in identifying which states serve as the best resources for particular quantum tasks. If a relative comparison between quantum states can be made, this gives rise to a partial order, where states are ordered according to their suitability to act as a resource. In the literature, various different partial orders for a variety of quantum resources have been proposed.
In discrete variable systems, vector majorization of Wigner functions in discrete phase space provides a natural partial order between quantum states. In the continuous variable case, a natural counterpart would be continuous majorization of Wigner functions in quantum phase space. Indeed, this concept was recently proposed and explored (mostly restricting to the single-mode case) in Van Herstraeten, Jabbour, Cerf, Quantum 7, 1021 (2023).  
In this work, we develop the theory of continuous majorization in the general $N$-mode case. In addition, we propose extensions to include states with finite Wigner negativity. 
For the special case of the convex hull of $N$-mode Gaussian states, we prove a conjecture made by Van Herstraeten, Jabbour and Cerf.  We also prove a phase space counterpart of Uhlmann's theorem of majorization. 
\end{abstract}

\maketitle

\section{Introduction}

Continuous variable (CV) systems are described by operators with eigenvalues forming a continuous set. Most of the examples in this class are built out of harmonic oscillators, appearing in many different guises in different areas of physics.
CV methods are therefore central in many applications of quantum information science; well-known reviews on this subject include {\it e.g.} \cite{Serafini17book, Weedbrook12b,Adesso14, Brau-vanLoock}.
In particular we want to mention continuous variable approaches to quantum computing \cite{Lloyd-Braunstein, Menicucci-etal}.

Resource theories provide a framework for describing key quantum information-theoretic resources and how they degrade under information processing constrained by 'free operations' \cite{Chitambar_2019}. In this context, 'free operations' refer to transformations that are allowed without consuming the resource in question—typically those that are easy to implement or considered operationally inexpensive within the theory.
An example is the resource theory of entanglement for bipartite quantum systems where the free operations are local operations  combined with classical communication (LOCC), and the loss of entanglement is tracked quantitatively by the entanglement entropy, the relevant "resource monotone" in this context \cite{Plenio:2007zz}. Furthermore, by Nielsen's theorem \cite{nielsen} LOCC produces and requires a preorder among quantum states, called majorization. Conversely, a majorization relation between a pair of states implies an inequality for their entanglement entropies.

In DV quantum computing, relevant resource theories are resource theories of magic, introduced in \cite{Veitch_2014}. One natural choice of free operations is the class of stabilizer protocols, which includes {\it e.g.} acting with Clifford unitaries; the free states are stabilizer states, while the resource states are magic aka non-stabilizer states \footnote{In quantum computing, an important resource is magic, aka non-stabilizerness, which is needed for advantage over classical computing. For qubits, this result follows from the Gottesman-Knill theorem \cite{Gottesman:1998hu}, which was later extended for qudits and continuous variable systems \cite{Veitch_2012,Mari:2012ypq,Veitch_2013}. Both in the discrete variable (DV) case of qudits and in the CV case, quantum advantage requires computation involving states whose Wigner function takes negative values in some parts of the quantum phase space. Thus the resource magic can also be called Wigner negativity, and indeed this is the convention for continuous variables.}.
In DV resource theories of magic, several magic monotones have been introduced \cite{Veitch_2014, BSS, How-Campb, Wang_2019, seddon2019, LOH22, garcia2024}:
for this work we single out the mana \cite{Veitch_2014}. Recently, a preorder based on vector majorization adapted to discrete Wigner functions in the discrete quantum phase space was investigated by Koukoulekidis and Jennings in  \cite{Koukoulekidis:2021ppu}.
Inspired by these well-established results for DV states, we aim to address similar questions for CV states.
 In this context, a natural choice of free operations would be those that always map Wigner-positive states to Wigner-positive states, $\mathcal{W}_+ \rightarrow \mathcal{W}_+$, but they are difficult to classify. 
A more convenient choice is Gaussian protocols, which map within a subset of $\mathcal{W}_+$ the convex hull of Gaussian states. This class includes {\it e.g.} Gaussian operations and homodyne measurements combined with conditioning based on measurement outcomes.
Gaussian protocols are also a very important and easily implementable class of protocols for many other applications beyond quantum computing. For resource states there are then two possible natural choices: Wigner-negative states, leading to the resource theory of Wigner negativity \cite{parisferraro18}, or non-Gaussian states, leading to the resource theory of non-Gaussianity \cite{Takagi:2018rqp}. 
In both cases, the Wigner logarithmic negativity \cite{Kenfack:2004ges}—the CV analogue of mana—serves as a key monotone, and it is known to be non-increasing under Gaussian protocols \cite{parisferraro18,Takagi:2018rqp}.
However, what seems to be missing in order to have a complete resource theory, is the concept of a preorder.

In this work we address the question what should be an appropriate preorder for either resource theory.
A natural strategy for establishing this preorder is to find a CV generalization of the discrete Wigner function majorization introduced in the DV case.
In the CV case, Wigner functions are real-valued functions on $\mathbb{R}^{2N}$. 
The mathematical theory of continuous majorization, applicable to functions, has been developed over time \cite{HLP29,Chong74,Hickey84,Joe87}. Van Herstraeten, Jabbour, and Cerf later adapted this framework to quantum phase space \cite{vanherstraeten2021continuous}. They proposed it as a natural tool to quantify the 'randomness' of Wigner-positive states, proved several properties, and conjectured that coherent states majorize all other Wigner-positive states under this ordering. Since continuous majorization of Wigner functions is the natural counterpart of vector majorization of discrete Wigner functions, it then appears as a natural candidate for a preorder for the resource theories of non-Gaussianity or Wigner negativity. This would already establish continuous majorization in quantum phase space as a fundamental concept.

In this work we develop the theory of continuous majorization of Wigner functions in quantum phase space, which we shorten to {\it Wigner majorization}, proving significant extensions and results first in the $N$-mode case of Wigner-positive states, in particular proving the conjecture of \cite{vanherstraeten2021continuous}
for the convex hull of $N$-mode Gaussian states. Motivated by resource theories, we investigate continuous majorization between Wigner functions of input and output states under Gaussian protocols, limiting the focus here on Gaussian channels.
We then propose two possible ways to extend the definition to states with finite non-vanishing Wigner negativity and we introduce a new form of preorder among generic CV states.
In this most general case we are able to prove some interesting results such as a quantum phase space counterpart of Uhlmann's theorem of majorization.   

This work is structured as follows. This introduction is followed by Sec.\,\ref{sec:summary} summarizing the main results. Secs.\,\ref{sec:MajorizationPositive},\,\ref{sec:Wignermajo_Channels},\,\ref{sec:MajorizationNegative} give a more technical presentation, followed by conclusions in Sec.\,\ref{sec:Discussion} and Appendices with technical details.

{\it Note added}: After the publication of this work, \cite{Upadhyaya:2025wdt} appeared providing a general definition of majorization for quasi-probability distributions. In Appendix \ref{app:Comparison}, we compare this definition with our Proposal 2, showing that the former generalizes the latter.

\section{Summary of the main results} 
\label{sec:summary}

In this section we provide an overview of the main results reported in this manuscript, postponing their proofs and explicit computations to the forthcoming sections and the appendices.
Before the summary, we review continuous majorization between positive functions and its application to defining the {\it Wigner majorization}, a preorder relation in the space of quantum CV states. 

\subsection{Review on Wigner majorization}
\label{subsec:reviewmajo}

 For later reference we begin with a brief review of {\it vector majorization}. Consider a pair of vectors $\boldsymbol{p},\boldsymbol{q}\in \mathbb{R}^n$. We reorder the components (note that they can also be negative) into descending order and relabel them to define  $\boldsymbol{p}^\downarrow = (p^\downarrow_1,p^\downarrow_2,\ldots ,p^\downarrow_n),\ \boldsymbol{q}^\downarrow = (q^\downarrow_1,q^\downarrow_2,\ldots ,q^\downarrow_n)$ such that $p^\downarrow_1\geq p^\downarrow_2\geq \cdots \geq p^\downarrow_n$ and $q^\downarrow_1 \geq q^\downarrow_2 \geq \cdots \geq q^\downarrow_n$. We then compare partial sums of the components. If the inequalities
\begin{equation}\label{partialsums}
\sum_{i=1}^k p^\downarrow_{i}\geq \sum_{i=1}^k q^\downarrow_{i}\,
\end{equation}
hold for all $k=1,\ldots, n$, then we say that the vector $\boldsymbol{p}$ {\it weakly majorizes} the vector $\boldsymbol{q}$ and denote $\boldsymbol{p}\ggcurly \boldsymbol{q}$. If in addition for $k=n$ we have the equality 
\begin{equation}\label{vectornorm}
\sum_{i=1}^n p^\downarrow_{i} = \sum_{i=1}^n q^\downarrow_{i}\,,
\end{equation}
we say that the vector $\boldsymbol{p}$ {\it majorizes} the vector $\boldsymbol{q}$ and we denote $\boldsymbol{p}\succ \boldsymbol{q}$. 

These definitions can be extended to the discretely infinite case; for majorization the sums $\sum^\infty_ {i=1}$ replacing the finite sums in \eqref{vectornorm} must be finite. A special application is the case when $\boldsymbol{p},\boldsymbol{q}$ are probability vectors associated with a discrete random variable (so that the components are non-negative). In this case \eqref{vectornorm} holds due to the unit normalization of probability vectors, and if $\boldsymbol{p}\succ \boldsymbol{q}$, we can think of the distribution $\boldsymbol{p}$ being "more ordered" or "less random" than
the distribution $\boldsymbol{q}$. This notion can then be implemented to define quantum state majorization  in terms of the eigenvalues of density operators. Given two DV or CV states with density operators $\hat{\rho}_1 $ and $\hat{\rho}_2$, let us call $\lambda_j$ and $\kappa_j$, with $j=1,\dots,n$ or $j=1,2,\ldots, \infty$ their respective eigenvalues. We then assemble the eigenvalues into ordered (probability) vectors $\boldsymbol{\lambda}^\downarrow, \boldsymbol{\kappa}^\downarrow$ and
 say that the state $\hat{\rho}_1$ {\it majorizes} the state $\hat{\rho}_2$,  and denote $\hat{\rho}_1 \succ_{\textrm{\tiny DM}} \hat{\rho}_2 $, if and only if $\boldsymbol{\lambda} \succ \boldsymbol{\kappa}$.
We will henceforth refer to this majorization relation as {\it density matrix majorization} in distinction to the Wigner majorization which will be our main topic. We will also compare density matrix majorization with Wigner majorization to understand whether the two concepts may coincide in some instances. 

Continuous majorization \cite{HLPbook,MaOlAr} is a generalization for functions, introduced in \cite{HLP29} and induces a preorder in the space of positive integrable functions on $\mathbb{R}^{2N}$. 
To avoid clutter, throughout the manuscript, we denote the integrals of these functions on the entire $\mathbb{R}^{2N}$ as
\begin{equation}
\int_{\mathbb{R}^{2N}} f(\boldsymbol{r})d \boldsymbol{r}\equiv\int f(\boldsymbol{r})d \boldsymbol{r}\,,
\end{equation}
namely without specifying any integration domain.

To define the continuous majorization precisely, consider two positive functions $f,g:\mathbb{R}^{2N}\to[0,\infty)$ (we will later address the question what happens if the functions are allowed to have negative values). We first define the generalization of weak majorization of vectors, following \cite{MaOlAr,Chong74} (also called a "spectral order relation"). We say that $f$ {\it weakly majorizes} $g$ and denote $f \ggcurly g$ if one of the following equivalent statements is true \cite{MaOlAr,Chong74}:
\begin{enumerate}
\item 
\label{item:firstcond}
\begin{equation}
\label{eq:firstcond}
\int [f(\boldsymbol{r})-t]_+ d\boldsymbol{r} \geq \int [g(\boldsymbol{r})-t]_+d\boldsymbol{r} \,,
\end{equation}
for any  $t\geq 0$, where 
\begin{equation}
\label{eq:squarebracket}
[y]_+ = \max (y,0)\,.
\end{equation}

\item 
\label{item:second_cond}
\begin{equation}
\label{eq:majocondition2}
\int \Phi (f(\boldsymbol{r})) d\boldsymbol{r} \geq \int \Phi (g(\boldsymbol{r})) d\boldsymbol{r}\,,
\end{equation}
for all non-negative increasing convex functions $\Phi:[0,\infty]\to [0,\infty]$ such that $\Phi (0)=0$.

\item
\begin{equation}
\label{eq:majocondition3}
\int^\infty_t m_f(s) ds \geq \int^\infty_t m_g(s)ds\,,
\end{equation}
  for any  $t\geq 0$, where the {\it level function} $m_f$ is defined as
  \begin{equation}
\label{eq:levelfunctiondef}
m_f(t)\equiv{\rm Vol}(\{\boldsymbol{r}\in\mathbb{R}^{2N}\;|\;f(\boldsymbol{r})\geq t\})  
  \,.
  \end{equation}
  
\item
\label{item:decrrearr}
\begin{equation}
\label{eq:majo_cond_decrrearr}
\int^v_{0} f^{\downarrow}(u)du \geq \int^v_{0} g^{\downarrow} (u)du\,,
\end{equation}
 for any $ v\in[0,\infty) $, where $f^{\downarrow}\,,g^{\downarrow}:[0,\infty]\to[0,\infty]$ are the {\it decreasing rearrangements} of $f$ and $g$ respectively, defined as
\begin{equation}
\label{eq:def_decrrearrang_ext}
f^\downarrow(u)=\inf \big\{ s\in[0,\infty) \; \big|\; m_f(s) \leq u \big\} = m^{-1}_f(u)\,.
\end{equation}
\end{enumerate}
Note in particular that the decreasing rearrangement of a function \eqref{eq:def_decrrearrang_ext} is the counterpart of the decreasing rearrangement $\boldsymbol{p}^\downarrow$ of a vector $\boldsymbol{p}$, but while the function $f$ can be multivariate, its decreasing rearrangement $f^\downarrow$ always has a single variable. 
Note that inequality \eqref{eq:majo_cond_decrrearr} mirrors the partial sum condition in \eqref{partialsums}. Importantly, weak majorization $\ggcurly$ satisfies the following property
\begin{equation}
\label{eq:propweakmajo}
f\ggcurly g\,,\,\, f\ggcurly h
\quad
\Rightarrow
\quad
f\ggcurly t g+(1-t) h\,,\,\,\, t\in[0,1]
\,,
\end{equation}
which can be proved by exploiting the concavity of the functional in \eqref{eq:squarebracket}.

If, in addition to one of the conditions \eqref{eq:firstcond}-\eqref{eq:majo_cond_decrrearr}, we have that
\begin{equation}
\label{eq:normalizationcondition}
\int f(\boldsymbol{r})d \boldsymbol{r}=\int g(\boldsymbol{r})d \boldsymbol{r}\,,
\end{equation}
then we say that $f$ {\it majorizes} $g$ and we denote it as $f\succ g$. The equation \eqref{eq:normalizationcondition} is the counterpart of the equation \eqref{vectornorm}. 

Note that two functions $f$ and $g$ can have the same level function: $m_f=m_g$ (and thus also the same decreasing arrangement $f^\downarrow = g^\downarrow$). In this case we say that $f$ and $g$ are {\it level-equivalent}, $f\sim g$, and define an equivalence class $[f]_\sim = [g]_\sim$. Thus the continuous majorization in fact applies to equivalence classes: $[f]_\sim \succ [g]_\sim \Leftrightarrow f\succ g$.
Note also that continuous majorization can be defined for positive functions on more general measure spaces. However, for this work, we restrict our analysis to Euclidean spaces of even dimensionality.

For later convenience, we recall that, given $f,g:\mathbb{R}^{2N}\to[0,\infty)$, a sufficient condition for the relation $f \succ g$ is the existence of an integral kernel $k:\mathbb{R}^{2N}\times\mathbb{R}^{2N} \to\mathbb{R}$, 
$k(\boldsymbol{r},\boldsymbol{z})\geq 0\ \ \forall \boldsymbol{r},\boldsymbol{z}\in \mathbb{R}^{2N}$, such that \cite{Hickey84,Manjegani23}
\begin{equation}
\label{eq:necessarycondition_wingermajo}
g(\boldsymbol{r})=\int  k(\boldsymbol{r},\boldsymbol{z})f(\boldsymbol{z})d\boldsymbol{z}\,,
\end{equation}
and 
\begin{equation}
\label{eq:bistoch_kernel}
\int d\boldsymbol{r} k(\boldsymbol{r},\boldsymbol{z})=1\,,
\;\;\quad
\int d\boldsymbol{z} k(\boldsymbol{r},\boldsymbol{z})\leq 1\,.
\end{equation}
A kernel with the properties in \eqref{eq:bistoch_kernel} is called semi-doubly stochastic, while, when the second inequality is saturated, $k$ is called doubly stochastic (or bistochastic). Restricting to the doubly stochastic kernels, the condition (\ref{eq:bistoch_kernel}) has a counterpart in the context of discrete probability distributions, where $k$ is replaced by a bistochastic matrix.
\\

The theory of continuous majorization has been recently applied to studying bosonic quantum states by introducing the Wigner majorization (continuous majorization of Wigner functions in quantum phase space) \cite{vanherstraeten2021continuous}; an extension to a fermionic case has been introduced in \cite{Cerf:2024sok}. In this work we focus on further developing the theory of Wigner majorization for bosonic systems \footnote{Technical and conceptual difficulties will arise when we move to consider non-Wigner positive states, with the Wigner function becoming negative in some domain. For that reason, \cite{Garttner:2022axh} considered applying continuous majorization in quantum phase space through Husimi Q distributions which are non-negative and difficulties can be avoided. However, here we are motivated by the familiarity of Wigner functions, the connection between quantum advantage and Wigner negativity, and 
by the resource theories of non-Gaussianity and Wigner negativity. Hence, we focus on Wigner functions.}. 
Consider a system described in terms of a finite set of degrees of freedom represented by pairs of Hermitian canonical operators $\hat{q}_j$ and $\hat{p}_j$, with $j=1,\dots,N$, such that
\be
\label{eq:CCR}
[\hat{q}_i,\hat{p}_j]=\mathrm{i}\delta_{i,j}\,,
\quad
[\hat{q}_i,\hat{q}_j]=[\hat{p}_i,\hat{p}_j]=0\,.
\ee
We refer to these operators as {\it canonical quadrature operators} and, for convenience, we collect them into a vector $\boldsymbol{\hat{r}}\equiv \left(\hat{q}_1,\dots, \hat{q}_N ,\hat{p}_1,\dots, \hat{p}_N\right)^{\textrm{t}}$.
Since the operators $\hat{q}_i$ and $\hat{p}_i$ have continuous eigenvalues on the real line, this framework is known as {\it quantum CV} and we call {\it CV systems} those systems allowing for this description \cite{Serafini17book}.

The states of CV systems are described by density matrices, which are operators on an infinite dimensional Hilbert space. 
Given a CV state with density matrix $\hat{\rho}$, we define the characteristic function $\chi_{\hat{\rho}}$ as
\begin{equation}
\label{eq:characteristicfuntion}
\chi_{\hat{\rho}}(\boldsymbol{r})\equiv\mathrm{Tr}\left(\hat{\rho}e^{\mathrm{i}\boldsymbol{\hat{r}}^{\textrm{t}}J\boldsymbol{r}}\right)\,,
\end{equation}
where $J$ is the standard symplectic matrix
\begin{equation}
\label{eq:def_stand_symp_mat}
J\equiv
\begin{pmatrix}
\boldsymbol{0} & \boldsymbol{1}
\\
-\boldsymbol{1} & \boldsymbol{0}
\end{pmatrix}\,,
\end{equation}
with $\boldsymbol{0}$ representing the $N\times N$ matrix with all the entries equal to zero and $\boldsymbol{1}$ is the $N\times N$ identity matrix.
The vector $\boldsymbol{r}\in\mathbb{R}^{2N}$ can be written as $\boldsymbol{r}=(\boldsymbol{q},\boldsymbol{p})^{\textrm{t}}\equiv(q_1,\dots,q_N,p_1,\dots,p_N)^{\textrm{t}}$, where these $2N$ coordinates parametrize the quantum phase space. To avoid confusion between the quadrature operators and the corresponding phase-space variables, we denote the former with the hat and the latter without it.

The {\it Wigner function} is defined as the symplectic Fourier transform of the characteristic function \cite{Serafini17book}
\begin{equation}
\label{eq:fromCharacteristictoWigner}
W_{\hat{\rho}}(\boldsymbol{r})=\frac{1}{(4\pi^2)^N}\int d\boldsymbol{r'}e^{-\mathrm{i}\boldsymbol{r}^{\textrm{t}}J\boldsymbol{r'}}\chi_{\hat{\rho}}(\boldsymbol{r'})\,.
\end{equation} 
Due to the normalization of the density matrix ${\rm Tr}\hat\rho=1$, the integral of the Wigner function (\ref{eq:fromCharacteristictoWigner}) on the entire phase space is one, namely
\begin{equation}
\label{eq:normalizationW}
\int W_{\hat{\rho}}(\boldsymbol{r})d \boldsymbol{r}=1\,.
\end{equation}
Since $ W_{\hat{\rho}}(\boldsymbol{r})$ is not necessarily a positive function, the condition (\ref{eq:normalizationW}) implies that the Wigner function is a {\it quasi-probability distribution}, which reduces to an ordinary probability distribution when it is positive \footnote{Strictly speaking, even a non-negative Wigner function is  not an ordinary probability distribution, since the random variables are eigenvalues of non-commuting operators \cite{Garttner:2022axh} and therefore it could be more appropriately called a quasi-probability distribution. In this work to simplify the language we use the word probability distribution for a non-negative Wigner function and quasi-probability distribution for non-positive Wigner function. }. We call the CV states for which this happens {\it Wigner-positive states} (see \cite{Broecker95} for a study on the characterization of states belonging to this class demonstrating how difficult this task is).
 We find it worth stressing that Wigner functions are functions defined on the quantum phase space $\mathbb{R}^{2N}$ and uniquely identify a quantum state. 

Here, the theory of continuous majorization 
enters and allows to define a new  notion of majorization between quantum states, that we call in this work "Wigner majorization".  The idea of Wigner majorization between states with positive Wigner functions was introduced in \cite{vanherstraeten2021continuous}, where the authors exploited this tool to study information-theoretic properties of Wigner functions. Given two states $\hat{\rho}_1$ and $\hat{\rho}_2$ with positive Wigner functions $W_1$ and $W_2$ respectively, following \cite{vanherstraeten2021continuous} we say that the state $\hat{\rho}_1$ {\it Wigner-majorizes} $\hat{\rho}_2$, in symbols $\hat{\rho}_1\succ_\textrm{\tiny W}\hat{\rho}_2$, if and only if $W_1 \succ W_2$.
 Notice that conditions \ref{item:firstcond}-\ref{item:decrrearr} are valid for positive integrable functions, and, therefore, the definition of Wigner majorization given above can only be applied between states with positive Wigner functions. 
As we will discuss later, one of the results of this manuscript consists of proposing potential extensions of the definition of Wigner majorization to more general Wigner functions and states.

As we will elaborate more in the forthcoming sections, a widely employed class of quantum operations is the one of {\it Gaussian unitaries}, namely unitary transformations generated by quadratic combinations of the quadrature operators. More precisely, these operations are implemented by a unitary operator of the following form 
\begin{equation}
\hat{U}={\rm exp}\left(-{\rm i} \boldsymbol{\bar{r}}^{\textrm{t}}J\,\boldsymbol{\hat{r}} \right)
{\rm exp}\left({\rm i} \boldsymbol{\hat{r}}^{\textrm{t}}h\,\boldsymbol{\hat{r}} \right)
{\rm exp}\left({\rm i} \boldsymbol{\bar{r}}^{\textrm{t}}J\,\boldsymbol{\hat{r}} \right) \,,
\end{equation}
with $h$ being a $2N\times 2N$ Hermitian matrix and $\bar{r}$ a vector with $2N$ entries \cite{Serafini17book}. 
 We find it worth remarking that applying a Gaussian unitary transformation on a CV state amounts to performing a translation by a vector $\bar{r}$ and a symplectic transformation in the coordinates parameterizing the corresponding Wigner function. The translation and the symplectic transformation are changes of coordinates of the quantum phase space, with a unit determinant. From this fact, through the conditions \ref{item:firstcond}-\ref{item:decrrearr}, one can show that two Wigner functions that differ by a symplectic transformation are level-equivalent. Thus, the corresponding pair of states related by a Gaussian unitary are equivalent in the sense of Wigner majorization \cite{vanherstraeten2021continuous}. In the rest of the manuscript, we will refer to two CV states $\hat\rho_1$ and $\hat\rho_2$ such that $\hat\rho_1 \succ_{\textrm{\tiny W}} \hat\rho_2$ and $\hat\rho_2 \succ_{\textrm{\tiny W}} \hat\rho_1$ as {\it majorization-equivalent}.

\subsection{Majorization for positive Wigner functions}
\label{subsec:results_positiveWigner}

We now begin our summary of the results of this work on Wigner majorization between quantum CV states. We  start with results for states with positive Wigner functions. These outcomes are therefore obtained by exploiting the definition of Wigner majorization through the conditions \ref{item:firstcond}-\ref{item:decrrearr}.

An important class of CV states with positive Wigner functions is the one of {\it Gaussian states}, which are described by Gaussian Wigner functions. 
The Wigner function of a Gaussian state $\hat{\rho}$ takes the general form \cite{Weedbrook12b,Adesso14}
\be
\label{eq:Wigner}
W_{\hat{\rho}}(\boldsymbol{r})=\frac{e^{-\frac{1}{2}(\boldsymbol{r}-\bar{\boldsymbol{r}})^\textrm{t}\gamma^{-1}(\boldsymbol{r}-\bar{\boldsymbol{r}})}}{(2\pi)^N\sqrt{\det\gamma}}\,,
\ee
where
\be
\bar{r}_j\equiv
{\Tr}\left[\hat{\rho}\,\hat{r}_j \right]\,,
\quad
\gamma_{ij}\equiv\frac{1}{2}
{\Tr}\left[ \hat{\rho}\,\{(\hat{r}_i-\bar{r}_i), (\hat{r}_j-\bar{r}_j)\}\right]
\,,
\ee
and $\{,\}$ denotes the anticommutator.
The values $\bar r_j$ are the first moments of the quadrature operators, while $\gamma = (\gamma_{ij})$ is a $2N\times 2N$ real, symmetric, and positive definite matrix, called the {\it covariance matrix}.
Gaussian states are widely employed in various contexts, including quantum information and quantum optics, and their 
study is also relevant for experiments.

To study majorization, one can simplify the Wigner function in Eq.  (\ref{eq:Wigner}) by noting that displacements in phase space—corresponding to Gaussian unitaries—do not affect the relevant integrals in Eqs. (\ref{eq:firstcond}), (\ref{eq:majocondition2}), (\ref{eq:majocondition3}) and (\ref{eq:majo_cond_decrrearr}). Thus, given the Wigner function (\ref{eq:Wigner}), we can always choose $\boldsymbol{\bar{r}}=0$ without altering the majorization properties of Gaussian Wigner functions and the corresponding Gaussian states. For this reason, when not otherwise specified, from now on we consider Gaussian states with vanishing first moments.
Moreover, notice that not all the real, symmetric, positive definite matrices can be covariance matrices of a quantum Gaussian state; the uncertainty principle imposes a constraint on the eigenvalues of $\gamma$, which is encoded in the inequality \cite{Weedbrook12b}
\be
\label{eq:uncertainty}
\gamma+\frac{\mathrm{i}}{2}J\geq 0\,,
\ee
where $J$ is defined in (\ref{eq:def_stand_symp_mat}).
The covariance matrix has $2N(2N-1)$ independent parameters, but not all of these affect the majorization between two Gaussian Wigner functions. 
To show this fact, we remark that any real even-dimensional positive 
definite matrix, and therefore also the covariance matrix, can be decomposed according to the Williamson decomposition \cite{Williamson36}, namely
\be
\label{eq:Williamson}
\gamma=S^{\textrm{t}}\mathcal{D} S\,,
\quad
\mathcal{D}=\textrm{diag}(\sigma_1,\dots,\sigma_N)\oplus \textrm{diag}(\sigma_1,\dots,\sigma_N)
\,,
\ee
where $S$ is a symplectic matrix (sometimes we will refer to it as symplectic basis) and $\sigma_i$ are called symplectic eigenvalues. Given the constraint (\ref{eq:uncertainty}), $\sigma_i\geq 1/2$. If $\gamma$ is the covariance matrix of a pure state, $\sigma_i= 1/2$, $\forall i$ \cite{Weedbrook12b}.
The decomposition (\ref{eq:Williamson}) provides a natural basis for the phase space coordinates.
Changing the coordinates by $\boldsymbol{r}\to S^{-\textrm{t}}\boldsymbol{r}$, we can write the Wigner function (\ref{eq:Wigner}) with  $\boldsymbol{\bar{r}}=0$ as
\be
\label{eq:WignerWilliamson}
W_{\hat{\rho}}(\boldsymbol{r})=\prod_{j=1}^N\frac{e^{-\frac{r_j^2}{2\sigma_j}}}{2\pi \sigma_j}\,,
\ee
where $r_j^2\equiv q_j^2+p_j^2$. Since (\ref{eq:Wigner}) with  $\boldsymbol{\bar{r}}=0$ and (\ref{eq:WignerWilliamson}) differ only by a symplectic transformation in the phase space coordinates, they are level-equivalent. Thus, only the $N$ symplectic eigenvalues affect the Wigner majorization properties of Gaussian states.

To completely characterize the Wigner majorization between Gaussian states, we resort to a result due to Joe \cite{Joe87} that provides a majorization criterion between Gaussian probability distributions. When applied to Gaussian Wigner functions, Joe’s result leads to the first result of this manuscript (a simple proof of Joe's result is sketched in
\footnote{Result 1 can be proven by plugging \eqref{eq:WignerWilliamson} into the left-hand side of \eqref{eq:firstcond} and performing the change of variables $y_j= r_j/\sigma_j$. We obtain
\be
\int \left[\frac{e^{-\frac{1}{2}\boldsymbol{r}^\textrm{t}\boldsymbol{r}}}{(2\pi)^N}-t\sqrt{\det\gamma}\right]_+ d\boldsymbol{r} \equiv \mathcal{I}(t\sqrt{\det\gamma})\,.
 \ee
 The fact that the function $\mathcal{I}(x)$ is monotonically decreasing in $x$ and the condition \eqref{eq:firstcond} implies Result 1.}).
\\

\noindent\fbox{%
    \parbox{\linewidth}{
{\bf Result 1}: Given two Gaussian states $\hat\rho_1$ and $\hat\rho_2$ with Wigner functions $W_1$ and $W_2$ and covariance matrices $\gamma_1$ and $\gamma_2$ respectively, we have
\be
W_1\succ W_2\quad
\Leftrightarrow
\quad \det\gamma_1\leq\det\gamma_2\,,
\ee
which induces the Wigner majorization relation
\be
\label{eq:MajorizationGaussianStates}
\hat{\rho}_1\succ_{\textrm{\tiny W}} \hat{\rho}_2\quad
\Leftrightarrow
\quad \det\gamma_1\leq\det\gamma_2\,.
\ee
}
}
\\

\noindent
This result is very powerful given that it encodes the Wigner majorization properties of a Gaussian state in a single number computable from a $2N\times 2N$ matrix. Notice that the determinant of a covariance matrix is given by the square product of all the symplectic eigenvalues. Thus, as anticipated, the Wigner majorization between Gaussian states only depends on the symplectic spectra of the two corresponding covariance matrices. Note that Result 1 also implies that a Wigner majorization order always exists between an arbitrary pair of $N$-mode Gaussian states. In Sec.\,\ref{subsec:example} we discuss explicit examples based on Result 1. 

One of the main findings of \cite{vanherstraeten2021continuous} is the conjecture on the majorization of the pure state $\hat\rho_0$, defined as the state with Wigner function (\ref{eq:WignerWilliamson}) with all the symplectic eigenvalues $\sigma_j=1/2$, for any $j=1,\dots,N$. The conjecture states that any mixed state with a positive Wigner function is majorized by $\hat{\rho}_0$. In \cite{vanherstraeten2021continuous} this conjecture is demonstrated for some mixtures of the harmonic oscillator eigenstates, in the single-mode ($N=1$) case. Due to Result 1, we can extend the proof of this conjecture to the mixed states belonging to the convex hull of $N$-mode Gaussian states.
\\

\noindent\fbox{%
    \parbox{\linewidth}{%
        {\bf Result 2}:
The $N$-mode pure Gaussian state $\hat{\rho}_0$ Wigner-majorizes every state in the convex hull of $N$-mode Gaussian states, namely
\begin{equation}
\label{eq:result2}
\hat{\rho}_{0}\succ_{\textrm{\tiny W}}\hat{\rho},\;\;\quad \forall\,
\hat{\rho}=\sum_{i}p_i\hat{\sigma}_i\,,
\end{equation}
where $\hat{\sigma}_i$ are Gaussian states and $\sum_i p_i=1$, $p_i \geq 0$.
    }%
}
\\

This result is proved by using \eqref{eq:propweakmajo}, \eqref{eq:MajorizationGaussianStates} and the fact that, among the Gaussian states, the covariance matrix of pure states has the smallest determinant allowed.
In Sec.\,\ref{subsec:majorizationGaussian}, we provide an alternative proof of \eqref{eq:result2}.
Extending Result 2 to linear combinations of Gaussian states that allow for negative coefficients would include cat states and Gottesman-Kitaev-Preskill states, widely used in quantum information theory \cite{Bourassa:2021ezn}. This generalization deserves a more careful analysis that we postpone to future works.

In Sec.\,\ref{subsec:majorizationGaussian} we derive also the following statement.
\\

\noindent\fbox{%
    \parbox{\linewidth}{
{\bf Result 3}:
Consider a one-parameter family of Gaussian density matrices $\hat{\rho}(\tau)$, where $\tau$ can be a coupling constant, the temperature, or other physical parameters. As a consequence, $\gamma$ and the symplectic eigenvalues $\sigma_k$ depend on $\tau$. The following majorization criterion as a function of the parameter $\tau$ holds for any pair of states with $\tau_2 >\tau_1$:
\begin{equation}
\label{eq:Wignermajo_1parameter}
\
\partial_\tau\sigma_k(\tau)>0\,\;\forall k,\ \forall \tau \in [\tau_1,\tau_2]\,\;\Rightarrow\,\;
\hat{\rho}_{\tau_1}\succ_{\textrm{\tiny W}} \hat{\rho}_{\tau_2}\, 
\,.
\end{equation}
}
}
\\

\noindent

This result is a special case of a more general one: if $\det \gamma (\tau)$ is monotonically decreasing in the interval $[\tau_1, \tau_2]$, the right-hand side of \eqref{eq:Wignermajo_1parameter} follows from Result 1. The specific case of Result 3 helps to compare Wigner majorization and density matrix majorization. It would be
insightful to understand whether they coincide in some instances; this question is also related to the interconversibility of Gaussian states \cite{Jabbour15PRA}.
In Sec.\,\ref{subsec:WignervsDM}, we show that Wigner majorization is equivalent to the density matrix majorization if we restrict to single-mode Gaussian states. 
\\

\noindent\fbox{%
    \parbox{\linewidth}{
{\bf Result 4}: Given two single-mode Gaussian states with density matrices $\hat{\rho}_1 $ and $\hat{\rho}_2$, then
\be
\label{eq:majoequivalence}
\hat{\rho}_1 \succ_{\textrm{\tiny DM}} \hat{\rho}_2 \quad
\Leftrightarrow
\quad
\hat{\rho}_1 \succ_{\textrm{\tiny W}} \hat{\rho}_2\,.
\ee
}
}
\\

This statement ceases to be true in general when $N>1$. We show this fact in Fig.\,\ref{fig:partialsums}, providing explicit examples of inequivalence between the two majorizations when two-mode Gaussian states are considered.
We also identify a subset of pairs of multi-mode states where the two majorizations are equivalent, as discussed in Sec.\,\ref{subsec:DMvsWigner_onemode}.

To conclude our analysis on Wigner majorization among Gaussian states, in Sec.\,\ref{subsec:example}, Result 3 is applied to examples from the context of harmonic lattices. In particular, we show that any $N$-mode Gaussian thermal state at a given temperature always majorizes another thermal state of the same system at higher temperature. This is consistent
with the idea that Wigner majorization orders quantum states based on their mixedness \cite{vanherstraeten2021continuous}. In addition, we also show that the thermal state of a harmonic chain made by $N$ sites Wigner-majorizes any other thermal state of the same system with a smaller frequency parameter and the same temperature.
\\

Wigner functions play a prominent role in the resource theory of non-Gaussianity and the resource theory of Wigner negativity \cite{parisferraro18, Takagi:2018rqp}. In the former case, the free states are given by CV Gaussian states, while, in the latter case, they are given by CV states with positive Wigner functions. In both resource theories, the free operations are given by the so-called Gaussian protocols, including Gaussian measurements and Gaussian channels. A Gaussian channel is a completely positive and trace-preserving operation that maps Gaussian states into Gaussian states. When acting on a Gaussian state, its action on the input covariance matrix $\gamma$ is given by \cite{Giedke:2002lqi, Serafini17book}
\be
\label{eq:gaussian channel def}
\gamma\to X\gamma X^{\textrm{t}}+Y
\,,
\qquad
Y+\mathrm{i}\frac{J}{2}\geq \mathrm{i} X \frac{J}{2} X^{\textrm{t}}\,,
\ee
where $X$ and $Y$ are $2N\times 2N$ matrices and $Y$ is additionally symmetric and positive definite. Gaussian channels map Gaussian states into Gaussian states according to (\ref{eq:gaussian channel def}), but can also be applied to non-Gaussian states. In the latter case, their action is more complicated, as discussed later in this manuscript.
To understand whether Wigner majorization plays any role in the context of the aforementioned resource theories, we find it insightful to study the majorization relation between a given input state and the output state obtained after applying a Gaussian channel.
Considering first a simple example of single-mode Gaussian channels, we can gain the following insight for $N$-mode Gaussian channels, discussed in Sec.\,\ref{subsec:Wignermajo_thermonoiseChannels}.
\\

\noindent\fbox{%
    \parbox{\linewidth}{{\bf Result 5}: 
    It is not true that a CV input state always Wigner-majorizes the output state obtained by applying a Gaussian channel or that it is always Wigner-majorized by it. The direction of a Wigner majorization relation depends on the choice of the CV input state and a Gaussian channel acting on it.
}}
\\

\noindent
This result implies that the Wigner majorization among CV Wigner-positive states does not allow any counterpart of Nielsen's theorem, where the role of LOCC would be replaced by Gaussian protocols. 
Result 6 below will explain why: the convolution kernels are not in general semi-doubly stochastic. 

Given the discussion above, it is natural to try to understand whether we can identify a subclass of {\it Wigner-majorizing Gaussian channels}. A Gaussian channel $\mathcal{E}$ belongs to this class if, given an input state $\hat\rho_{\textrm{\tiny in}}$, we have that $\hat\rho_{\textrm{\tiny in}}\succ_{\textrm{\tiny W}}\mathcal{E}(\hat\rho_{\textrm{\tiny in}})$. To find this class of channels when $\hat\rho_{\textrm{\tiny in}}$ is a CV positive-Wigner state, we reinterpret a result found in \cite{Walschaers:2021zvx} in view of exploiting it in the context of Wigner majorization.
\\

\noindent\fbox{%
    \parbox{\linewidth}{
{\bf Result 6:}
Given the Wigner function $W_{\textrm{\tiny in}}$ of an input state, the Wigner function $W_{\textrm{\tiny out}}$ of the output state $\hat\rho_{\textrm{\tiny out}}\equiv\mathcal{E}(\hat\rho_{\textrm{\tiny in}})$ is given by
\begin{equation}
\label{eq:inputoutputNewW}
W_{\textrm{\tiny out}}(\boldsymbol{r})=\int d\boldsymbol{z} k(\boldsymbol{r},\boldsymbol{z})W_{\textrm{\tiny in}}(\boldsymbol{z})
\,,
\end{equation}
where the integral kernel is defined as
\begin{equation}
\label{eq:k_kerneldef}
k(\boldsymbol{r},\boldsymbol{z})\equiv \frac{e^{-\frac{1}{2}\boldsymbol{r}^{\textrm{t}}Y^{-1}\boldsymbol{r}-\frac{1}{2}\boldsymbol{z}^{\textrm{t}}X^{\textrm{t}}Y^{-1}X\boldsymbol{z}+\boldsymbol{z}^{\textrm{t}}X^{\textrm{t}}Y^{-1}\boldsymbol{r}}}{(2\pi)^N\sqrt{\det Y}}\,,
\end{equation}
in terms of the matrices $X$ and $Y$ characterizing the channel $\mathcal{E}$. The kernel $k$ satisfies the properties
\begin{equation}
\label{eq:propertykernel1}
\int d\boldsymbol{r} k(\boldsymbol{r},\boldsymbol{z})=1\,,
\;\;\quad
\int d\boldsymbol{z} k(\boldsymbol{r},\boldsymbol{z})=\frac{1}{\det X}\,.
\end{equation}
}
}
\\

A proof of the formulas (\ref{eq:inputoutputNewW})-(\ref{eq:k_kerneldef}) is reported for completeness in Appendix \ref{app:proofinputoutputformula}. Importantly, note that the convolution formula \eqref{eq:inputoutputNewW} with the unit normalization of the integral over $\boldsymbol{r}$ \eqref{eq:propertykernel1} plays a central role in the proof of the  Eisert-Mari theorem \cite{Mari:2012ypq}. In the DV case, a convolution formula follows from the Choi-Jamiolkowski dual representation of the action of the channel. However, as we discuss in more detail in Sec.\,\ref{subsec:Wignermajo_thermonoiseChannels}, in the CV case the dual representation is ill-defined, since the C-J dual mapping involves the two-mode squeezed state in the infinite squeezing limit. The formulas \eqref{eq:inputoutputNewW}- \eqref{eq:k_kerneldef} do not rely on the C-J dual mapping and, at least for Gaussian channels, 
give a well-defined convolution for the Eisert-Mari theorem in the CV case. While we believe that similar convolution formulas can be derived for more general positive channels, it appears to be an important problem to find such generalizations, to strengthen the proof of the Eisert-Mari theorem in the CV case. 
Comparing (\ref{eq:propertykernel1}) with (\ref{eq:bistoch_kernel}), we notice that $k$ is not a semi-doubly bistochastic kernel, unlike in the convolution formula in the DV case.  However, using the properties of $k$ and the necessary condition (\ref{eq:necessarycondition_wingermajo}) for Wigner majorization for states with positive Wigner functions, we prove the following statement.
\\

\noindent\fbox{%
    \parbox{\linewidth}{
{\bf Result 7:}
The class of Gaussian channels $\mathcal{E}_{X,Y}$ characterized by the matrices $X$ and $Y$ such that $\det X\geq 1$ is a class of Wigner-majorizing channels when applied to CV states with positive Wigner functions, namely, for any CV state $\hat\rho$ with positive Wigner function,
\begin{equation}
\label{eq:majorizingchannel}
\hat\rho\succ_{\textrm{\tiny W}}\mathcal{E}_{X,Y}(\hat\rho)\,.
\end{equation}
}
}
\\

\noindent
Within this class, important examples arise when we further restrict to channels with $\det X= 1$. These represent special cases where the associated kernel becomes doubly stochastic.
 First, notice that the Gaussian unitaries correspond to Gaussian channels with $Y=\boldsymbol{0}$ and $X$ symplectic, the latter conditions implying $\det X= 1$. This is a consistency check of the fact that Gaussian unitaries are majorizing channels, despite in a trivial way. Indeed, the output state after a Gaussian unitary is majorization-equivalent to the input state and therefore is majorized by it. In addition to this trivial example, the class of majorizing channels with $\det X= 1$ includes {\it classical mixing channels}, namely Gaussian channels which implement displacements on the input state randomly distributed according to a given Gaussian probability distribution and are models of Gaussian noise \cite{Serafini17book}. An example of the case $\det X >1$ is the amplification channel. 
We elaborate more on this class of Wigner-majorizing channels in Sec.\,\ref{subsec:majorizing channel positive}.

\subsection{Majorization for generic Wigner functions}
\label{subsec:results_genericWigner}

The results described in the previous subsection apply to CV states with positive Wigner functions. 
It is desirable to extend the concept of Wigner majorization to a more general class of CV states including Wigner-negativity,  and understand the interplay of these extensions with Gaussian channels. The findings reported in this section aim at filling this gap.  

As a first step, we introduce a proposal to extend Wigner majorization to CV states whose Wigner functions are not necessarily positive. Physically motivated by resource theory arguments, as discussed later in this section, we also require that the Wigner functions of the considered states are absolutely integrable on the entire phase space, i.e. they belong to $L^1(\mathbb{R}^{2N})$. 
\\

\noindent\fbox{%
    \parbox{\linewidth}{{
 {\bf Result 8 (Proposal 1):} Consider  two {\it generic} CV states with density matrices $\hat{\rho}_1$ and $\hat{\rho}_2$ and Wigner functions $W_1$ and $W_2$ respectively  and assume that $W_1,\, W_2\in L^1(\mathbb{R}^{2N})$.
 We propose to extend the majorization relation $\succ_{\textrm{\tiny W}}$ given in Sec.\,\ref{subsec:reviewmajo} to this more general set of states. We define that
$\hat{\rho}_1\succ_{\textrm{\tiny W}} \hat{\rho}_2$ if and only if $|W_1|\ggcurly |W_2|$, where the relation $\ggcurly$ is defined by \eqref{eq:firstcond}-\eqref{eq:def_decrrearrang_ext}.
 }
 }}
 \\
 
\noindent Let us comment on this proposal.  Since for Wigner functions which are not generically positive the integral of $|W|$ over the entire phase space is not one, we do not impose the condition (\ref{eq:normalizationcondition}). Thus, the Wigner majorization (in the non-positive case) is established by the weak majorization $\ggcurly$ defined in Sec.\,\ref{subsec:reviewmajo}. In the positive case, this becomes the majorization $\succ$ due to $|W|=W$ and unit normalization. The main virtue of Proposal 1 is that it is convenient to work with. Testing the weak majorization of absolute values of Wigner functions by condition 1 (see the inequality \eqref{eq:firstcond}) is simple to do with Mathematica or an equivalent program.  In Sec.\,\ref{subsec:proposal1}, we test this general proposal for Wigner majorization on some exemplary CV states (see Fig.\,\ref{fig:proposal1}).

We also provide an alternative proposal for Wigner majorization and a further new notion of preorder between generic CV states.
\\ 

\noindent\fbox{%
    \parbox{\linewidth}{{
    {\bf Alternative proposal (Proposal 2):}
    Consider  two {\it generic} CV states with density matrices $\hat{\rho}_1$ and $\hat{\rho}_2$ and Wigner functions $W_1$ and $W_2$ respectively. We propose an alternative extension of the majorization relation $\succ_{\textrm{\tiny W}}$ given in Sec.\,\ref{subsec:reviewmajo} to this more general set of states. To distinguish it from Proposal 1, we use the symbol $\succ_{\textrm{\tiny W}_\textrm{\tiny 2}}$ for this second proposal. We define that
    $\hat{\rho}_1\succ_{\textrm{\tiny W}_\textrm{\tiny 2}} \hat{\rho}_2$ if and only if one of the two (equivalent) conditions is verified
\begin{equation}
\label{eq:final_negMajo1}
\int [W_1-t]_+ d\boldsymbol{r} - \int [W_2-t]_+d\boldsymbol{r}\geq 0\,,
\quad
\forall t\in\mathbb{R}\,,
\end{equation}
where $[\cdot]_+$ is defined in (\ref{eq:squarebracket}),
or
\begin{equation}
\label{eq:final_negMajo2}
\int^\infty_t m_{W_1}(s) ds - \int^\infty_t m_{W_2}(s)ds\geq0\,,
\quad
\forall t\in\mathbb{R}\,,
\end{equation}
with $m_W$ given by (\ref{eq:levelfunctiondef}).
     }
 }}
  \\
  
\noindent Proposal 2 is motivated by the DV case investigated by Koukoulekidis and Jennings in \cite{Koukoulekidis:2021ppu}.
It consists of extending conditions \eqref{eq:firstcond} and \eqref{eq:majocondition3} to non-positive functions. This leads to divergent integrals.
 However, the integrals appearing in the definition are divergent separately, but in Sec.\,\ref{subsec:proposal2} we show that, under certain (quite natural) circumstances, the divergences are crucially canceled out once we consider the differences in (\ref{eq:final_negMajo1}) and (\ref{eq:final_negMajo2}) (see also Fig.\,\ref{fig:proposal2}).

In principle, majorization relations can order quantum states following various criteria. Proposals 1 and 2 aim to generalize the preorder from  Sec.\,\ref{subsec:reviewmajo},  extending the notion of uncertainty—quantified via the shape and spread of Wigner functions—to states with Wigner negativity. As we will detail in Sec.\,\ref{subsec:proposal3}, the following relation aims to introduce an order which will be promising for the resource theory of non-Gaussianity.
\\

\noindent\fbox{
    \parbox{\linewidth}{{
     {\bf Alternative proposal (Tautological preorder):}
     Consider  two {\it generic} CV states with density matrices $\hat{\rho}_1$ and $\hat{\rho}_2$ and Wigner functions $W_1$ and $W_2$ respectively. We provide a new notion of preorder among these states. This relation differs from Propositions 1 and 2 and is denoted by the symbol $\succ_{\textrm{\tiny W}_\textrm{\tiny t}}$. We define that
    $\hat{\rho}_1\succ_{\textrm{\tiny W}_\textrm{\tiny t}} \hat{\rho}_2$ if and only if $W_1 \succ_{\textrm{t}} W_2$, namely
if there exists a kernel $k$ of the form (\ref{eq:k_kerneldef}) such that 
\begin{equation}
\label{eq:tautologicalcondition}
W_{2}(\boldsymbol{r})=\int d\boldsymbol{z} k(\boldsymbol{r},\boldsymbol{z})W_{1}(\boldsymbol{z})
\, .
\end{equation}
     }
 }}
  \\

\noindent The tautological preorder is different from the previous two proposals since it establishes that a CV state $\hat{\rho}_1$ majorizes $\hat{\rho}_2$ whenever the latter is obtained from the former via a Gaussian channel. Thus, this definition tautologically induces a counterpart of Nielsen's theorem for the new preorder and Gaussian channels. 
This property explains the choice of the name for this proposal.
To show that this relation can be legitimately called preorder, in Appendix \ref{app:preorder} we show that it satisfies all the required properties
The tautological preorder describes a different preorder compared to the Wigner majorization. Indeed, it is not equivalent to Proposals 1 and 2 when restricted to Wigner-positive states and, as discussed in Sec.\,\ref{subsec:proposal3}, it orders states according to their Gaussianity.
\\

\noindent\fbox{%
    \parbox{\linewidth}{
{\bf Remark on tautological preorder}: 
All the Gaussian states are equivalent with respect to the tautological preorder.
Moreover, according to the tautological preorder, every Gaussian state is less ordered than any other CV states. This implies that Gaussian states are at the bottom of the preorder induced by $\succ_{\textrm{ W}_\textrm{t}} $, which provides a legitimate candidate as a preorder for non-Gaussianity.
}}
\\

\noindent
Note that while Proposals 1 and tautological preorder hold for any CV state with absolutely integrable Wigner function, Proposal 2 has a more restricted range of validity due to an additional technical condition on the Wigner functions required for the integrals in (\ref{eq:final_negMajo1}) and (\ref{eq:final_negMajo2}) to be well-defined (see Sec.\,\ref{subsec:proposal2} for more details). Note further that Proposals 1 and 2 reduce to the Wigner majorization discussed in Sec.\,\ref{subsec:reviewmajo} when applied to CV states with positive Wigner functions. This does not happen for the tautological preorder, which has a different behavior also when considered on Wigner-positive states, as discussed in Sec.\,\ref{subsec:proposal3}.

The three preorders listed above have an important common feature concerning the {\it Wigner logarithmic negativity}. Given a CV state $\hat{\rho}$ with Wigner function $W_{\hat{\rho}}$, the Wigner logarithmic negativity is defined as
\begin{equation}
\label{eq:Wlogneg_def}
\mathcal{N}_{W_{\hat{\rho}}}\equiv
\ln\left(\int |W_{\hat{\rho}}(\boldsymbol{r})|d\boldsymbol{r}\right)\,.
\end{equation}
Due to the normalization (\ref{eq:normalizationW}), $\mathcal{N}_{W_{\hat{\rho}}}$ is vanishing when $W_{\hat{\rho}}$ is positive and non-vanishing otherwise. 
In other words, the Wigner logarithmic negativity measures how much a CV state is not Wigner-positive. 
This quantity is tightly related to the resource theories of non-Gaussianity and Wigner negativity, and, indeed, it has been proven monotonic under Gaussian protocols \cite{parisferraro18, Takagi:2018rqp}. In Sec.\,\ref{subsec:relationWLN}, we prove the following result, which shows $\mathcal{N}_{W_{\hat{\rho}}}$ to be an insightful quantity in the context of Wigner majorization.
\\

\noindent\fbox{%
    \parbox{\linewidth}{{
{\bf Result 9}: Given two CV states $\hat{\rho}_A$ and $\hat{\rho}_B$ with Wigner functions $W_A$ and $W_B$ respectively, using any of the three preorder proposals reported above, we have that, for $*\in \{\textrm{W},\textrm{ W}_\textrm{2},\textrm{ W}_\textrm{t}\}$,
\begin{equation}
\hat{\rho}_A\succ_{*} \hat{\rho}_B\;\quad \Rightarrow \;\quad
\mathcal{N}_{W_{A}}\geq \mathcal{N}_{W_{B}}\,.
\end{equation}
}
}
}
\\

\noindent
In other words, each of the proposals introduced above for preorder between CV states implies a relation between their Wigner logarithmic negativity.
At this point, the exclusion of states with non-absolutely integrable Wigner functions from our analysis can be justified on physical grounds. Indeed, from (\ref{eq:Wlogneg_def}), we see that, if $W_{\hat\rho}\notin L^1(\mathbb{R}^{2N})$, then the Wigner logarithmic negativity diverges, rendering it unsuitable as a resource quantifier in this context.

The Wigner Rényi entropies are other quantities that can be defined from the Wigner function $W_{\hat{\rho}}$ of given state $\hat{\rho}$ and are defined as
\begin{equation}
\label{eq:RenyiWigner}
S_{W_{\hat{\rho}}}^{(\alpha)}=\frac{1}{1-\alpha}\ln\left(\int [W_{\hat{\rho}}(\boldsymbol{r})]^\alpha d\boldsymbol{r}\right)\,.
\end{equation}
While these quantities are finite and real-valued for any $\alpha>0$ when $\hat{\rho}$ is a Wigner-positive state, issues arise in the more general case. 
In Sec.\,\ref{subsec:relationWLN} we address the question of whether $S_{W_{\hat{\rho}}}^{(\alpha)}$ is monotonic under Gaussian channels, deriving the following result.
\\

\noindent\fbox{%
    \parbox{\linewidth}{{
{\bf Result 10:}
Consider a CV state $\hat\rho_{\textrm{\tiny in}}$ with Wigner function $W_{\textrm{\tiny in}}$ and act on it with a Gaussian channels $\mathcal{E}$ determined by the matrices $X$ and $Y$. If we denote the output state as
$ \hat\rho_{\textrm{\tiny out}}=\mathcal{E}(\hat\rho_{\textrm{\tiny in}})$ and its Wigner function as $W_{\textrm{\tiny out}}$, the following inequality holds $\forall~\alpha\geq 1 $:
 \begin{equation}
 \label{eq:ineq_quasi_RenyiNegativity}
\  (\det X)^{\alpha-1} \int d\boldsymbol{r}~|W_{\textrm{\tiny out}}(\boldsymbol{r})|^{\alpha} \leq \int d\boldsymbol{r}~|W_{\textrm{\tiny in}}(\boldsymbol{r})|^\alpha
 \,.
 \end{equation}
 }}}
 \\
 
 \noindent
When $\alpha=1$, (\ref{eq:ineq_quasi_RenyiNegativity}) is consistent with the monotonicity of Wigner logarithmic negativity under Gaussian protocols proven in \cite{parisferraro18, Takagi:2018rqp}. On the other hand, when $\alpha> 1$, the channel-dependent term with the matrix $X$  of the left-hand side of (\ref{eq:ineq_quasi_RenyiNegativity}) spoils the monotonicity of Wigner R\'enyi entropies (\ref{eq:RenyiWigner}) under Gaussian channels: instead we find the inequality
\be\label{inequalities2}
  -\ln (\det X) +  S^{(\alpha)}_{W_{\textrm{\tiny out}}} \geq S^{(\alpha )}_{W_{\textrm{\tiny in}}}\ \ , \ \ \forall~\alpha> 1  \ .
\ee
Thus, there is no monotonicity under Gaussian channels. In Fig.\,\ref{fig:WignerRenyi} we illustrate with some examples that the Wigner R\'enyi entropies may increase or decrease under a Gaussian channel, depending
on choices of input states and channel parameters. 

To gain more insights on the Wigner majorization between a generic input CV state and the corresponding output state obtained from applying a Gaussian channel, we focus on Proposal 1 described in Result 8, which proves to be the most versatile and also the easiest to evaluate among the three proposals. In Sec.\,\ref{subsec:ChannelsGenericWigner} we prove that, when employing this Wigner majorization proposal, Result 7 can be extended to generic, not necessarily Wigner positive, CV states with an absolutely integrable Wigner function.
\\

\noindent\fbox{%
    \parbox{\linewidth}{
{\bf Result 11:}
The class of Gaussian channels $\mathcal{E}_{X,Y}$ characterized by the matrices $X$ and $Y$ such that $\det X\geq 1$ is a class of Wigner-majorizing channels when applied to generic CV states. Thus, for any CV state $\hat\rho$, we have that,
according to Proposal 1 in Result 8,
\begin{equation}
\hat\rho\succ_{\textrm{\tiny W}}\mathcal{E}_{X,Y}(\hat\rho)\,.
\end{equation}
}
}
\\

\noindent So far, when considering quantum operations, we have focused on Gaussian channels. To discuss different instances, we consider the quantum operation defined by the following convex combination of Gaussian unitary channels applied on a density matrix $\hat\rho$ of a CV state
\begin{equation}
\label{eq:random Gaussian unitary channel}
\mathcal{E}_p(\hat{\rho})=\int dS\int d\boldsymbol{\bar{r}}\, p(S,\boldsymbol{\bar{r}}) U_{S,\boldsymbol{\bar{r}}} \hat{\rho} U_{S,\boldsymbol{\bar{r}}}^\dagger\,,
\end{equation}
where the integral over $\boldsymbol{\bar{r}}$ is performed over the entire phase space, while the integral over $S$ is performed over the set of symplectic matrices corresponding to the Gaussian unitaries applied to $\hat\rho$. Examples of channels in this class have been considered in \cite{Mancini20Capacity,Lami:2022bjh} and their quantum capacity has been studied. The transformation in (\ref{eq:random Gaussian unitary channel}) amounts to apply on the input state random combinations $U_{S,\boldsymbol{\bar r}}$  of Gaussian unitary transformations and displacements distributed according to the probability density function $p(S,\boldsymbol{\bar r})$. For this reason, we call the channel $\mathcal{E}_p$ {\it random Gaussian unitary channel}. 
 By focusing again on the Wigner majorization Proposal 1 in Result 8, in Sec.\,\ref{subsec:ChannelsGenericWigner}, we prove that any CV state with an absolutely integrable Wigner function after the application of a random Gaussian unitary channel is majorized by the corresponding input state. 
\\

\noindent\fbox{%
    \parbox{\linewidth}{
{\bf Result 12:} For any CV state and any choice of channel of the form (\ref{eq:random Gaussian unitary channel}) we have that, according to Proposal 1 in Result 8,
\begin{equation}
\label{eq:inputmajorization_RGUC}
\hat{\rho}\succ_{\textrm{\tiny W}}\mathcal{E}_p(\hat{\rho})\,.
\end{equation}
}}
\\

\noindent
Note that this result can be viewed as a quantum phase space counterpart of Uhlmann's theorem of (density matrix) majorization. The latter states that the output state of any unital channel, i.e. a quantum channel that maps the identity into itself, is always density matrix-majorized by the corresponding input state \cite{Uhlmann-thm}. Furthermore, since random Gaussian unitary channels $\mathcal{E}_p$ in (\ref{eq:random Gaussian unitary channel}) are also examples of unital channels,  Uhlmann's theorem of majorization should also be applicable along with Result 12 ({\it i.e.}, also $\hat{\rho}\succ_{\textrm{\tiny DM}}\mathcal{E}_p(\hat{\rho})$ holds). A heuristic interpretation of Result 12 is that random Gaussian unitary channels (at least, random displacement channels) are models of bosonic noise, and, according to Result 12, the Wigner function of the output state is a "more random" distribution in quantum phase space than that of the input state.

\section{Continuous majorization and positive Wigner functions}
\label{sec:MajorizationPositive}

In this section, we study the Wigner majorization between Wigner positive states, focusing on Gaussian states. 
Exploiting known results in the theory of majorization between probability distributions, we provide a criterion for Wigner majorization of Gaussian states. This condition allows analytic control on the Wigner majorization properties of a large class of continuously parametrized families of Gaussian states, which includes examples of physical relevance, such as thermal states of harmonic chains. We finally discuss a quantitative comparison between the Wigner majorization and the density matrix majorization for single- and two-mode Gaussian states.

\subsection{Wigner majorization of Gaussian states}
\label{subsec:majorizationGaussian}

The theory of majorization between Gaussian probability distributions has been discussed in detail in \cite{Joe87}, where it has been found that, given two Gaussian distributions $f_{\Sigma_1}$ and $f_{\Sigma_2}$ with covariance matrices $\Sigma_1$ and $\Sigma_2$ respectively, $f_{\Sigma_1}\succ f_{\Sigma_2}$ according to the definition reviewed in Sec.\,\ref{subsec:reviewmajo} if and only if $\det \Sigma_1\leq \det \Sigma_2$. This result immediately applies to the Wigner majorization between two Gaussian states $\hat{\rho}_1 $ and $\hat{\rho}_2 $ with covariance matrices $\gamma_1$ and $\gamma_2$ constrained by (\ref{eq:uncertainty}). Indeed, recalling that the Wigner function of an $N$-mode Gaussian state is a Gaussian distribution with $2N$ variables, recalling the definition of Wigner majorization in Sec.\,\ref{subsec:reviewmajo}, we straightforwardly obtain.
\\

\noindent\fbox{%
    \parbox{\linewidth}{
{\bf Result 1}: Given two Gaussian states $\hat\rho_1$ and $\hat\rho_2$ with Wigner functions $W_1$ and $W_2$ and covariance matrices $\gamma_1$ and $\gamma_2$ respectively, we have
\be
W_1\succ W_2\quad
\Leftrightarrow
\quad \det\gamma_1\leq\det\gamma_2\,,
\ee
which induces the Wigner majorization relation
\be
\label{eq:MajorizationGaussianStates_mt}
\hat{\rho}_1\succ_{\textrm{\tiny W}} \hat{\rho}_2\quad
\Leftrightarrow
\quad \det\gamma_1\leq\det\gamma_2\,.
\ee
}
}
\\

\noindent
Let us discuss some of the implications of this result.
Consider two generic Gaussian states characterised by their covariance matrices $\gamma_1$ and $\gamma_2$
and whose corresponding Wigner functions $W_1$ and $W_2$ are obtained from (\ref{eq:Wigner}).
Result 1 implies that $W_1\succ W_2$ when $\det \gamma_1\leqslant \det\gamma_2$,
where 
\be
\label{det-gamma-from-symp-spect}
\det\gamma=\prod_{j=1}^N\sigma_j^2\geq\frac{1}{4^N}\,,
\ee
which can be obtained  by combining (\ref{eq:Williamson})
with the fact that $\textrm{det} \,S = 1$ for any real symplectic matrix $S$.
The determinant of the covariance matrix is largely studied given its relation with various quantities such as the purity \cite{Adesso14}, the Shannon entropy of Wigner functions \cite{Hertz_2017,Adesso:2012ni,VanHerstraeten:2021nce} and the entropy of outcomes of Gaussian measurements \cite{Lami17Logdetinequalities}.
In the context of Wigner majorization, $\det \gamma$ allows to conclude that the preorder between
two Gaussian states is determined only by 
the symplectic spectra of the corresponding covariance matrices
and the Wigner function majorization between Gaussian states 
provides a {\it total} preorder, 
namely either $W_1\succ W_2$ or $W_2 \succ W_1$
for any given pair of Wigner functions $W_1$ and $W_2$.

The above mentioned sufficient condition for the Wigner majorization 
implies that any monotonically decreasing function of $\det\gamma$ preserves the majorization relation.
i.e. it is Schur convex from the point of view of the Wigner function majorization.
An interesting example is the purity $\mathrm{Tr}(\hat{\rho}^2)$,
satisfying $\mathrm{Tr}(\hat{\rho}^2) \leq 1$ for any normalised quantum state, 
which is saturated only by the pure states. 
The purity of a Gaussian state is determined by $\det\gamma$ as follows \cite{Adesso14}
\be
\label{eq:purity_def}
\mathrm{Tr}(\hat{\rho}^2)=\frac{1}{2^N\sqrt{\det\gamma}}\,,
\ee
where the prefactor depends on the convention adopted 
in the definition of the covariance matrix.
This is consistent with the interpretation of the majorization in terms of the mixedness of a state;
indeed, if a state majorizes another state, it is less mixed and, correspondingly, its purity is larger.

We find it instructive to consider also the second R\'enyi entropy $S^{(2)}(\hat{\rho}) \equiv -\ln\mathrm{Tr}(\hat{\rho}^2)$.
In the case of a bosonic Gaussian state, from (\ref{eq:purity_def}) one finds
\be
S^{(2)}(\hat{\rho})
=N\ln 2+\frac{1}{2}\ln \det\gamma=\sum_{j=1}^N\ln(2\sigma_j)\,,
\ee
which is an increasing function of $\det\gamma$.
Thus, by applying the conclusions of discussion above,
for two bosonic Gaussian states $\hat{\rho}_1$ and $\hat{\rho}_2$
we have that 
\be
\label{eq:majorizationRenyi2}
\hat{\rho}_1 \succ_{\textrm{\tiny W}} \hat{\rho}_2 \;\quad\; \Leftrightarrow \quad S^{(2)}(\hat{\rho}_1)\leq S^{(2)}(\hat{\rho}_2)\,,
\ee
meaning that also the second R\'enyi entropy $S^{(2)}(\hat{\rho})$ characterises 
the Wigner majorization relation between two Gaussian states in a unique way.

As reviewed in Sec.\,\ref{subsec:results_positiveWigner}, the authors of \cite{vanherstraeten2021continuous} conjectured that the pure state $\hat\rho_0$ given by (\ref{eq:WignerWilliamson}) with $\sigma_j=1/2$, for any $j=1,\dots,N$, Wigner majorizes any other CV Wigner-positive state.
This conjecture was verified in \cite{vanbever2021} for Wigner positive mixtures of Fock states.
 Aiming to find more evidences for this conjecture, Result 1 allows to prove this conjecture for the states contained in the convex-hull of $N$-mode Gaussian states.
 \\

\noindent\fbox{%
    \parbox{\linewidth}{%
       {\bf Result 2}:
The $N$-mode pure Gaussian state $\hat{\rho}_0$ Wigner-majorizes every state in the convex hull of $N$-mode Gaussian states, namely
\begin{equation}
\label{eq:result2_mt}
\hat{\rho}_{0}\succ_{\textrm{\tiny W}}\hat{\rho},\;\;\quad \forall\,
\hat{\rho}=\sum_{i}p_i\hat{\sigma}_i\,,
\end{equation}
where $\hat{\sigma}_i$ are Gaussian states and $\sum_i p_i=1$, $p_i \geq 0$.
    }%
}
\\

\noindent
We prove this result in the following.
 For this purpose, we need to show that $W_{\hat\rho_0}\succ W_{\hat\rho}$, where $W_{\hat\rho_0}$ is the Wigner function of the pure state $\hat\rho_0$ and $W_{\hat\rho}$ the one of the generic state in the convex hull defined in (\ref{eq:result2_mt}).
First we observe that $\hat\rho_0$ Wigner majorizes any other Gaussian states, given that the determinant of the covariance matrices of pure states is the smallest possible. Thus, since the states $\hat\sigma_i$ are Gaussian, $\hat\rho_0\succ_{\textrm{\tiny W}} \hat\sigma_i$. At this point, we only have to show that every convex combination of $\hat\sigma_i$ is Wigner majorized by $\hat\rho_0$.
Given the state $\hat\rho$ defined in (\ref{eq:result2_mt}), its Wigner function can be written in terms of the Wigner functions of the states $\hat\sigma_i$ as
\begin{equation}
W_{\hat\rho}=\sum_{i}p_i W_{\hat\sigma_i}\,.
\end{equation}
 Thus, to prove the Wigner majorization, we have to prove that
\begin{equation}
\label{eq:condition}
\int \Phi\left(\sum_{i}p_i W_{\hat\sigma_i}\right) d\boldsymbol{r}\leq \int \Phi(W_{\hat\rho_0}) d\boldsymbol{r}
\end{equation}
for any positive function $\Phi:[0,\infty)\to [0,\infty)$
which is also convex and non-increasing.
Since $\Phi$ is convex, we can apply Jensen's inequality, obtaining
\begin{equation}
\label{eq:jensencondition}
\int \Phi\left(\sum_{i}p_i W_{\hat\sigma_i}\right) d\boldsymbol{r}\leq \sum_i p_i \int \Phi\left( W_{\hat\sigma_i}\right) d\boldsymbol{r}\,.
\end{equation}
Crucially, since $\hat\rho_0\succ_{\textrm{\tiny W}} \hat\sigma_i$ for any $\hat\sigma_i$, we have that
\begin{equation}
\int \Phi\left( W_{\hat\sigma_i}\right) d\boldsymbol{r}\leq \int \Phi(W_{\hat\rho_0}) d\boldsymbol{r}\,,
\end{equation}
which, combined with the fact that $\sum_i p_i=1$ and (\ref{eq:jensencondition}), leads to
(\ref{eq:condition}), concluding the proof of Result 2.

Finally, to showcase the applicability of Result 1, we derive Result 3, which provides a criterion to identify a Wigner majorization order along certain continuously parametrized families of Gaussian states.
\\

\noindent\fbox{%
    \parbox{\linewidth}{
{\bf Result 3}:
Consider a one-parameter family of Gaussian density matrices $\hat{\rho}(\tau)$, where $\tau$ can be a coupling constant, the temperature, or other physical parameters. As a consequence, $\gamma$ and the symplectic eigenvalues $\sigma_k$ depend on $\tau$. The following majorization criterion as a function of the parameter $\tau$ holds for any pair of states with $\tau_2 >\tau_1$:
\begin{equation}
\label{eq:Wignermajo_1parameter_mt}
\
\partial_\tau\sigma_k(\tau)>0\,\;\forall k,\ \forall \tau \in [\tau_1,\tau_2]\,\;\Rightarrow\,\;
\hat{\rho}_{\tau_1}\succ_{\textrm{\tiny W}} \hat{\rho}_{\tau_2}\, 
\,.
\end{equation}
}
}
\\

\noindent
Exploiting this finding, in the next subsection we discuss concrete physical examples for systems of coupled oscillators.

Considering a one-parameter family of Gaussian density matrices $\hat{\rho}(\tau)$,
parameterised by the real parameter $\tau$,
from (\ref{det-gamma-from-symp-spect}) it is straightforward to prove Result 3.
Indeed, taking first the logarithm of (\ref{det-gamma-from-symp-spect}) and then its derivative w.r.t. $\tau$,
we find
\be
\label{eq:der_detgamma}
\partial_\tau\det\gamma(\tau)=2\det\gamma(\tau)\sum_{k=1}\frac{\partial_\tau\sigma_k(\tau)}{\sigma_k(\tau)}\,.
\ee
Hence the condition $\partial_\tau\sigma_k(\tau)> 0$ for all values of $k$
implies that $\det\gamma_A(\tau_1)>\det\gamma_A(\tau_2)$ when $\tau_1> \tau_2$.
Thus, the above majorization criterion leads to 
\be
\label{eq:WmajorizationGaussianRDM_2}
\partial_\tau\sigma_k(\tau)> 0\,,\,\forall k
\;\;\Rightarrow\;\;
W_A(\tau_2)\succ W_A(\tau_1)\,,
\ee
when $\tau_1> \tau_2$, which is equivalent to Result 3.

\subsection{Examples}
\label{subsec:example}

The model considered in our examples is the harmonic chain 
made by $N$ sites, whose Hamiltonian is
\be
\label{eq:HamHL}
\widehat{H}
=
\sum_{i=1}^N\left(
\frac{1}{2 m}\hat{p}_i^2+\frac{m \omega^2}{ 2}\hat{p}_i^2
\right)+\sum_{\langle i,j \rangle}\frac{\kappa}{2}(\hat{q}_i-\hat{q}_j)^2
\,,
\ee
we can set $\kappa=1$ and $m=1$ without loss of generality 
and periodic boundary conditions are imposed. 
The Hamiltonian (\ref{eq:HamHL}) can be diagonalised through the standard procedure, 
which requires to introduce the bosonic creation and annihilation operators $\hat{b}_k^\dagger $ and $\hat{b}_k$,
and the result is 
\be
\label{eq:HamHLdiag}
\widehat{H}
=\sum_{k=1}^N\Omega_k\left(\hat{b}^\dagger_k\hat{b}_k +\frac{1}{2}\right)
\,,
\ee
where the dispersion relation $\Omega_k>0$ 
for the periodic boundary conditions reads
\be
\label{dispersion-relation-HC}
\Omega_k=\sqrt{\omega^2+4[\sin(\pi k/N)]^2}
\,,
\ee
with $k=1,\dots,N$.

When the harmonic chain (\ref{eq:HamHL}) is at finite inverse temperature $\beta$,
its density matrix is 
\be
\label{eq:thermalDM}
\hat{\rho}_{\textrm{\tiny th}}\propto e^{-\beta \widehat{H}}\,.
\ee
In this case the symplectic spectrum is given by 
\be
\label{eq:thermalsymplecticspect}
\sigma_{\textrm{\tiny th},k}=\frac{1}{2}\coth(\beta\Omega_k/2)
\,,
\ee
where $k=1,\dots,N$,
which provides the determinant of the corresponding 
covariance matrix through (\ref{det-gamma-from-symp-spect}).

For any given value of $k=1,\dots,N$, the symplectic eigenvalue $\sigma_{\textrm{\tiny th},k}$ is a decreasing function of $\beta$, hence
\be
\sigma_{\textrm{\tiny th},k}(T_1)>\sigma_{\textrm{\tiny th},k}(T_2)\,,
\qquad
\textrm{if}\;\;T_1>T_2\,,
\ee
Thus, Result 3 implies that
\be
\hat{\rho}_{T_2}\succ_{\textrm{\tiny W}} \hat{\rho}_{T_1}\,,
\qquad
\textrm{if}\;\;T_1>T_2\,,
\ee
in agreement 
with the fact that the interpretation of the majorization order is related to the degree of mixedness of the states 
and that the pure states majorize all the other states.

Another majorization relation is obtained by 
considering two states at the same temperature 
and different values of $\omega$ in their dispersion relation (\ref{dispersion-relation-HC}).
Since
\be
\sigma_{\textrm{\tiny th},k}(\omega_1)<\sigma_{\textrm{\tiny th},k}(\omega_2)\,,
\qquad
\textrm{if}\,\,\,\omega_1>\omega_2\,,
\ee
the same analysis leads to 
\be
\hat{\rho}_{\omega_1}\succ_{\textrm{\tiny W}} \hat{\rho}_{\omega_2}\,,
\qquad
\textrm{if}\,\,\,\,\omega_1>\omega_2\,.
\ee
This  agrees with the result discussed in the Appendix\;C of \cite{Arias:2023kni},
where the density matrix majorization analysis has been performed for the mixed states 
given by the reduced density matrices of half infinite chains 
having different frequencies in their dispersion relation.

\subsection{Comparison with density matrix majorization}
\label{subsec:WignervsDM}

Wigner majorization is not the only majorization relation between quantum states introduced in the literature. The density matrix majorization defined in Sec.\,\ref{subsec:reviewmajo} has been widely employed in the study of entanglement and, more generally, in the context of resource theory. In this section, we discuss a comparison between the density matrix majorization and the Wigner majorization. Although not fully general, the analysis in the context of Gaussian states allows to draw insightful conclusions.

\subsubsection{Density matrix majorization between Gaussian states}
\label{subsec:DMvsWigner_onemode}

To discuss the density matrix majorization properties, we recall that the density matrix of any $N$-mode Gaussian state can be written in the following form
\begin{equation}
\label{eq:GaussianDM}
\hat{\rho}=\frac{e^{-\sum_{k=1}^N \varepsilon_k\hat{n}_k }}{Z}\,,
\end{equation}
where $\hat{n}_k$ are bosonic number operators and $\varepsilon_k>0$ are called single-particle energies. The $\varepsilon_k$s can be expressed in terms of the symplectic eigenvalues $\sigma_k$ of the same Gaussian state as
\begin{equation}
\label{eq:relationsigma_epsilon}
\varepsilon_k=\log\left(\frac{\sigma_k+\frac{1}{2}}{\sigma_k-\frac{1}{2}}\right)
\,.
\end{equation}
The normalization factor $Z$ in (\ref{eq:GaussianDM}) can be written in terms of  single-particle energies as
\begin{equation}
Z=\prod_{k=1}^N\frac{1}{1-e^{-\varepsilon_k}}\,.
\end{equation}
From (\ref{eq:GaussianDM}), the eigenvalues of $\hat{\rho}$ are
\begin{equation}
\label{eq:eigenvaluesGaussianDM}
\lambda(n_1,n_2,\dots,n_N)=\prod_{k=1}^N (1-e^{-\varepsilon_k})e^{-\varepsilon_k n_k} \equiv\prod_{k=1}^N\lambda_k(n_k)\,,
\end{equation}
where $n_k$ are the non-negative integer occupation numbers of the different modes.
If $\hat\rho$ in (\ref{eq:GaussianDM}) depends on a continuous parameter $\tau$, i.e. $\hat\rho(\tau)$, through the single-particle energies $\varepsilon_k(\tau)$, a dependence on $\tau$ is induced also in the symplectic eigenvalues $\sigma_k(\tau)$, in the density matrix eigenvalues $\lambda(n_1,n_2,\dots,n_N;\tau)$ and in the $k$-th mode contribution $\lambda_k(n_k;\tau)$. 
Notice that each eigenvalue in (\ref{eq:eigenvaluesGaussianDM}) depends on $N$ occupation numbers; hence it is difficult to order the elements of the spectrum of $\hat\rho$ to directly check the density matrix majorization. To overcome this difficulty, we exploit the factorized structure of $\lambda(n_1,n_2,\dots,n_N;\tau)$. Indeed, a result proved in \cite{Orus:2005jq} claims
that, if, for any pair $\tau_1> \tau_2$, we have
\begin{equation}
\label{eq:Orus_thm_if}
\sum_{n_k=0}^m \lambda_k (n_k,\tau_1) \geq \sum_{n_k=0}^m
\lambda_k (n_k,\tau_2)\,,\quad
\forall\, m\geq 0\,,\, 1\leq k\leq N\,,
\end{equation}
then
\begin{equation}
\label{eq:Orus_thm_then}
\hat{\rho}({\tau_1})\succ_{\textrm{\tiny DM}}\hat{\rho}({\tau_2})\,.
\end{equation}
In other words, the density matrix majorization can be proved by checking the condition (\ref{partialsums}) for every mode.
Exploiting the exponential form in (\ref{eq:eigenvaluesGaussianDM}), after a bit of algebra, we can show that (\ref{eq:Orus_thm_if}) is satisfied if $\partial_\tau \varepsilon_k(\tau)>0$. Using that (\ref{eq:relationsigma_epsilon}) is a monotonically decreasing function and (\ref{eq:Orus_thm_if})-(\ref{eq:Orus_thm_then}), we conclude that
\be
\label{eq:criterion_DMmajo}
\partial_\tau\sigma_k(\tau)< 0\,,\;\forall k\,,\; \forall \tau_1\geq \tau\geq\tau_2
\;\;\Rightarrow\;\;
\hat{\rho}_{\tau_1}\succ_{\textrm{\tiny DM}}\hat{\rho}_{\tau_2}\,.
\ee
This result generalizes the one derived in Appendix C of \cite{Arias:2023kni}, where this analysis was restricted to the reduced density matrix of half harmonic chain and the parameter $\tau$ was the frequency parameter. The result (\ref{eq:criterion_DMmajo}) holds more generally for any family of Gaussian states continuously parameterized.
As a corollary of (\ref{eq:criterion_DMmajo}), given a pair of states $\hat\rho_1$ and $\hat\rho_2$ with symplectic eigenvalues $\sigma_k^{(1)}$ and $\sigma_k^{(2)}$ respectively, with label $k$ enumerating the eigenvalues in descending order, if $\sigma_k^{(1)}<\sigma_k^{(2)}$ for any $k$, then $\hat\rho_1\succ_{\textrm{\tiny W}}\hat\rho_2$ and $\hat\rho_1\succ_{\textrm{\tiny DM}}\hat\rho_2$.

Our conclusion allows a first comparison between Wigner majorization and density matrix majorization. Indeed, comparing (\ref{eq:Wignermajo_1parameter_mt}) and (\ref{eq:criterion_DMmajo}), we notice that, given a family of Gaussian states parametrized by $\tau$, if $\partial_\tau\sigma_k(\tau)< 0$ for any $k$, then we have both Wigner and density matrix majorization order along the line parameterized by $\tau$. 
Since the single-mode Gaussian states are completely specified by their unique symplectic eigenvalue, when applied to this class of states,  this comparison establishes the equivalence between density matrix majorization and Wigner majorization, which corresponds to
\\

\noindent\fbox{%
    \parbox{\linewidth}{
{\bf Result 4}: Given two single-mode Gaussian states with density matrices $\hat{\rho}_1 $ and $\hat{\rho}_2$, then
\be
\label{eq:majoequivalence_mt}
\hat{\rho}_1 \succ_{\textrm{\tiny DM}} \hat{\rho}_2 \quad
\Leftrightarrow
\quad
\hat{\rho}_1 \succ_{\textrm{\tiny W}} \hat{\rho}_2\,.
\ee
}
}
\\

\subsubsection{Majorization analysis of two-mode Gaussian states}

\begin{figure}[t!]
\vspace{.2cm}
\hspace{-0.827cm}
\includegraphics[width=0.48\textwidth]{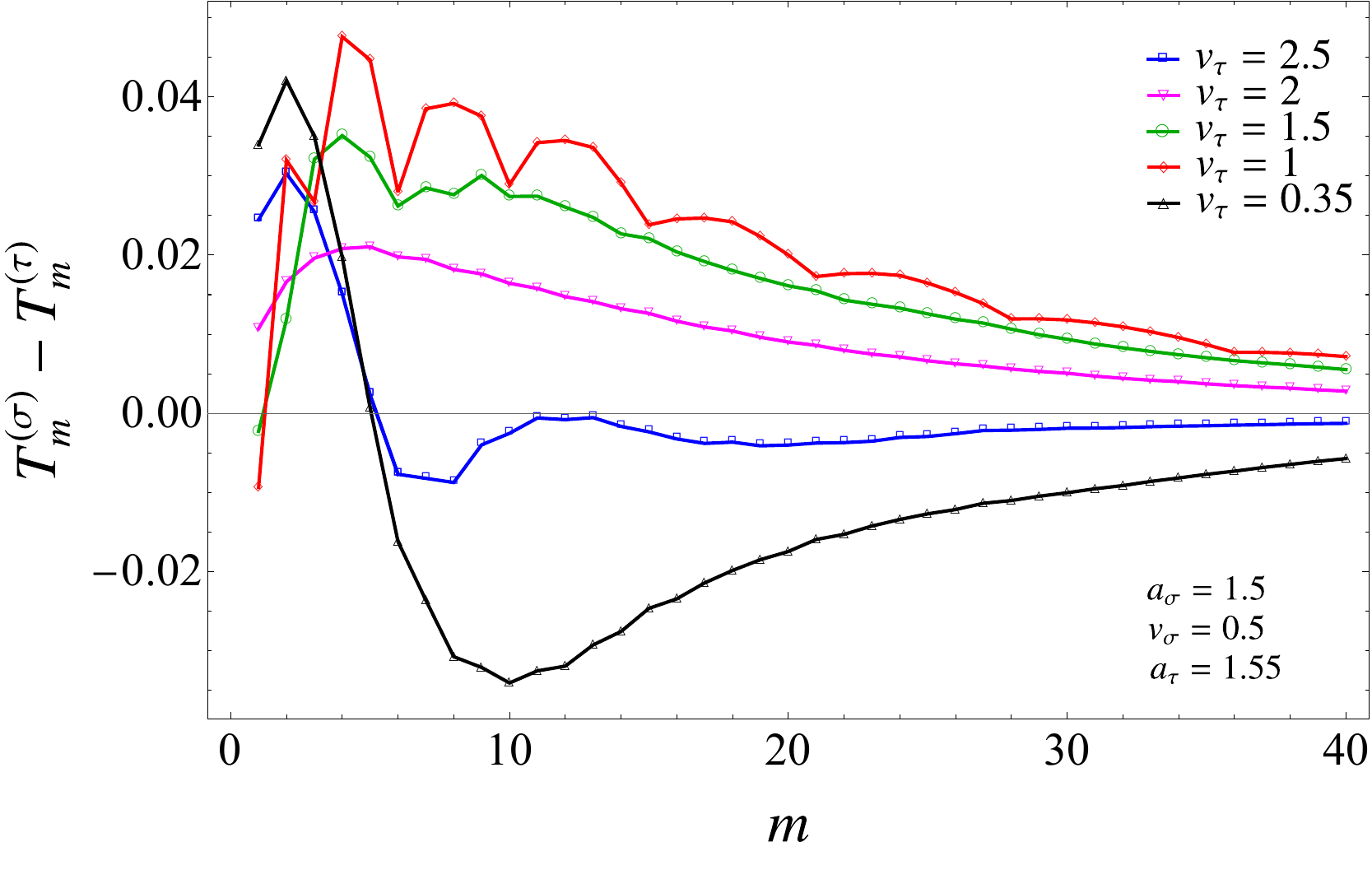}
\vspace{-0.3cm}
\caption{
Differences of the partial sums (\ref{eq:partialsums}) of eigenvalues of density matrices representing two-mode Gaussian states as functions of the number $m$ of terms involved in partial sums. The data are reported for five pairs of Gaussian states, whose symplectic eigenvalues are displayed in the parametrization (\ref{eq:twomode_par}). 
When one of the curves crosses the horizontal axis, the density matrix majorization between the two corresponding Gaussian states is ruled out.
}
\label{fig:partialsums}
\end{figure}

To understand whether the equivalence between Wigner majorization and density matrix majorization holds for  $N$-mode Gaussian states with $N>1$, it is insightful to examine the case of two-mode Gaussian states.
Since Gaussian states are considered, the Wigner majorization is established through the determinants of their covariance matrices. As shown via (\ref{eq:majorizationRenyi2}), this criterion can be rephrased in terms of the second Renyi entropies. In this section, we will exploit this formulation of the Wigner majorization criterion. 
As for the density matrix majorization, the eigenvalues (\ref{eq:eigenvaluesGaussianDM}) must be considered. 
When $N=2$, these eigenvalues depend on two occupation numbers and read
\begin{equation}
\label{eq:spectrumtwomodes}
\lambda(n_1,n_2)=
\frac{1}{\sigma_1+\frac{1}{2}}\frac{1}{\sigma_2+\frac{1}{2}}
\left(\frac{\sigma_1-\frac{1}{2}}{\sigma_1+\frac{1}{2}}\right)^{n_1}
\left(\frac{\sigma_2-\frac{1}{2}}{\sigma_2+\frac{1}{2}}\right)^{n_2}
\,,
\end{equation}
where also (\ref{eq:relationsigma_epsilon}) has been exploited. 
For the forthcoming analysis, 
a convenient parametrization of the two symplectic eigenvalues is
\begin{equation}
\label{eq:twomode_par}
\sigma_1= a v\,,
\qquad
\sigma_2= \frac{a}{v}\,.
\end{equation} 
where $a$ and $v$ are positive and such that $\sigma_1,\sigma_2>1/2$. In this parametrization, $S^{(2)}$ reads
\begin{equation}
S^{(2)}(\hat\rho)=\log 4+\log (\sigma_1\sigma_2)=2\log (2 a)\,,
\end{equation}
namely $a$ parametrizes the contribution of the symplectic spectrum to $S^{(2)}$. In particular, given two states $\hat\rho_\sigma$ and $\hat\rho_\tau$ with symplectic spectra parameterized by $(a_\sigma,v_\sigma)$ and $(a_\tau,v_\tau)$ respectively, from (\ref{eq:majorizationRenyi2}) we have that $a_\tau\geq a_\sigma$ implies $\hat\rho_\sigma\succ_{\textrm{\tiny W}} \hat\rho_\tau$.
Checking whether there is a density matrix majorization relation between $\hat\rho_\sigma$ and $\hat\rho_\tau$ is more complicated because 
the eigenvalues \eqref{eq:spectrumtwomodes} and their partial sums have to be studied. 
The density matrix majorization between two states can be ruled out by applying the following procedure.
\begin{enumerate}
\item
Compute the first $\mathcal{N}$ elements of the spectra of $\hat\rho_{\sigma}$ and $\hat\rho_{\tau}$ by using \eqref{eq:spectrumtwomodes} and order them in a decreasing way $\left\lbrace\lambda^{(\sigma)}_i\right\rbrace_{i=1}^\mathcal{N}$ and $\left\lbrace\lambda^{(\tau)}_i\right\rbrace_{i=1}^\mathcal{N}$.
\item
Compute the first $\mathcal{N}$ partial sums for the two states, i.e.
\begin{equation}
\label{eq:partialsums}
T^{(\sigma)}_m=\sum_{j=1}^m \kappa^{(\sigma)}_j\,, \quad
T^{(\tau)}_m=\sum_{j=1}^m \kappa^{(\tau)}_j\,,
\quad m=1,\dots,\mathcal{N}\,.
\end{equation}
\item
Compute $T^{(\sigma)}_m-T^{(\tau)}_m$ and study its sign as a function of $m$.
\end{enumerate}
The outcome of an analysis performed following these points
is reported in Fig.\,\ref{fig:partialsums}, where the differences between the partial sums $T^{(\sigma)}_m$ and $T^{(\tau)}_m$ in (\ref{eq:partialsums}) are reported for five pairs of two-mode Gaussian states $\hat\rho_\sigma$ and $\hat\rho_\tau$ as a function of the number $m$ of eigenvalues included. All the pairs of states are chosen in such a way that $a_\tau>a_\sigma$, which implies $\hat\rho_\sigma\succ_{\textrm{\tiny W}}\hat\rho_\tau$. On the other hand, we see that four of the five curves in the figure cross the horizontal axis, meaning that $T^{(\sigma)}_m-T^{(\tau)}_m$ has no definite sign. Thus, we can rule out the density matrix majorization between the corresponding pairs of states and, consequently, the equivalence between Wigner majorization and density matrix majorization in the case of $N$-mode Gaussian states with $N>1$.
We find it worth noticing that a detailed analysis of the density matrix majorization between two-mode pure Gaussian states was reported in \cite{Jabbour15PRA}. The results of \cite{Jabbour15PRA} are a promising starting point for classifying the $N$-mode Gaussian states for which density matrix and Wigner majorizations are equivalent, as it happens for the entire set of single-mode Gaussian states.

\section{Wigner majorization and quantum channels}
\label{sec:Wignermajo_Channels}

Density matrix majorization is relevant for establishing the existence of quantum channels connecting two quantum states. Motivated by this application, we want to understand whether similar insights can be drawn also through the Wigner majorization. For this purpose, in this section we study the Wigner majorization order between two states related by a Gaussian channel and whether the occurrence of a Wigner majorization relation implies the existence of Gaussian channels connecting two states. 
These investigations could be of interest in view of applying Wigner majorization to the resource theories of Wigner negativity and non-Gaussianity \cite{parisferraro18, Takagi:2018rqp}. 

\subsection{Wigner majorization between input and output states}
\label{subsec:Wignermajo_thermonoiseChannels}

We begin our analysis with a single-mode Gaussian state example, which teaches us an interesting lesson on Wigner majorization.
Consider the single-mode Gaussian channel whose action on the input single-mode Gaussian state is given by (\ref{eq:gaussian channel def}) with \cite{Holevo:2001zsx,Giovannetti2004}
\begin{equation}
\label{eq:noisechannel}
X=\sqrt{1-s}\boldsymbol{1}_2\,,\qquad\qquad Y=sc\boldsymbol{1}_2\,,
\end{equation}
where $s\in [0,1]$, $c\geq 1/2$ and $\boldsymbol{1}_2$ is the $2\times 2$ identity matrix. The channel defined by \eqref{eq:noisechannel} is called {\it thermal noise channel} and physically describes an operation that 
mixes the initial state and a thermal state with symplectic eigenvalue $c$ according to the mixing parameter $s$.
When $c=1/2$, the channel \eqref{eq:noisechannel} is called {\it loss channel}. This channel has the pure Gaussian state with covariance matrix $\gamma_0=\frac{1}{2}\boldsymbol{1}_2$, i.e. the vacuum, as a fixed point.
Given an input single-mode Gaussian state with covariance matrix $\gamma_{\textrm{\tiny in}} $ and unique symplectic eigenvalue $\sigma_{\textrm{\tiny in}}$, after applying a thermal noise channel we have
\be
\gamma_{\textrm{\tiny in}} \mapsto \gamma_{\textrm{\tiny out}}\equiv (1-s)\gamma_{\textrm{\tiny in}} + s c \boldsymbol{1}_2  
\;\; \Rightarrow \;\; \sigma_{\textrm{\tiny out}} = (1-s)\sigma_{\textrm{\tiny in}} + s c\,,
\ee
where $\gamma_{\textrm{\tiny out}}$ is the covariance matrix of the output state and $\sigma_{\textrm{\tiny out}}$ its symplectic eigenvalue.
It is straghtforward to notice that $\sigma_{\textrm{\tiny out}}\geq \sigma_{\textrm{\tiny in}}$ if $c\geq \sigma_{\textrm{\tiny in}}$ and, otherwise, we have $\sigma_{\textrm{\tiny in}} > \sigma_{\textrm{\tiny out}}$.
By (\ref{eq:MajorizationGaussianStates_mt}) we have that the input state Wigner majorizes the output if $\sigma_{\textrm{\tiny out}}\geq \sigma_{\textrm{\tiny in}}$ and, vice-versa, the output Wigner majorizes the input when $\sigma_{\textrm{\tiny in}}\geq \sigma_{\textrm{\tiny out}}$.
For instance, if the input state is thermal with $\sigma_{\textrm{\tiny in}} =\coth (\beta \Omega /2)$ (see (\ref{eq:thermalsymplecticspect})), 
it is necessary to mix it with a hotter state in order to have that the input state Wigner majorizes the output state, moving towards more disorder, as intuitively expected. 
On the other hand, if we choose to mix the input state with a colder state, the opposite situation occurs. 
Notice that, when considering a loss channel ($c=1/2$), $c\leq \sigma_{\textrm{\tiny in}}$ and, therefore, the output certainly Wigner majorizes the input.
 We conclude that, in general, there is no fixed majorization order between a given state and the output state after applying a Gaussian channel.
 This counterexample proves
 \\

\noindent\fbox{%
    \parbox{\linewidth}{
   {\bf Result 5}: 
    It is not true that a CV input state always Wigner-majorizes the output state obtained by applying a Gaussian channel or that it is always Wigner-majorized by it. The direction of a Wigner majorization relation depends on the choice of the CV input state and a Gaussian channel acting on it.
    }
}
\\

\noindent

In the remainder of this subsection, we contrast the kernel (\ref{eq:inputoutputNewW}) which connects input and output states with the kernel that one would obtain via a naive application of channel-state duality. We remind the reader that the duality between channels and quantum states is well-defined in the finite dimensional case by means of the so-called C-J isomorphism \cite{JAMIOLKOWSKI1972275,CHOI1975285}. The state associated to a channel is defined on an extended Hilbert space (two copies of the original Hilbert space), and in the finite dimensional case the kernel that relates the input and output states can be identified with the Wigner phase space density of the state which is dual to the channel. In \cite{Koukoulekidis:2021ppu} this result was used to show that output states are always Wigner majorized by their input states for the case of positivity-preserving channels (channels which map discrete Wigner positive states to discrete Wigner positive states). 

In the infinite dimensional CV setting, the C-J isomorphism is no longer well-defined and the map between quantum channels and states on extended Hilbert spaces breaks down. Technically, this is because the C-J isomorphism is defined using the maximally mixed state, which ceases to exist in the infinite dimensional case. One can however construct an approximate version of the C-J isomorphism using a regulated version of the maximally mixed state. As we will now show, the phase space density of the state obtained using this approximate channel-state map does not agree with the kernel in (\ref{eq:inputoutputNewW}). 

We begin by rewriting the kernel $k$, 
which connects input and output Wigner functions, as a function on a doubled phase space parametrized by $\boldsymbol{\zeta}\equiv(\boldsymbol{r},\boldsymbol{z})^{\textrm{t}}$. It is straightforward to find
\begin{equation}
\label{eq:k_kerneldef_extended}
k(\boldsymbol{\zeta})\equiv \frac{e^{-\frac{1}{2}\boldsymbol{\zeta}^{\textrm{t}}\Gamma^{-1}\boldsymbol{\zeta}}}{(2\pi)^N\sqrt{\det Y}}
\,,\quad\,\,
\Gamma^{-1}=\begin{pmatrix}
Y^{-1} & -Y^{-1}X
\\
-X^{\textrm{t}}Y^{-1} & X^{\textrm{t}}Y^{-1}X
\end{pmatrix}\,.
\end{equation} 
Notice  that $\Gamma^{-1}$ is well-defined, but $\Gamma=(\Gamma^{-1})^{-1}$ is not because $\det\Gamma^{-1}=0$. This hampers the interpretation of $k$ as a Gaussian Wigner function on a doubled phase space.
 At this point, we apply the C-J isomorphism to the Gaussian channel characterized by the matrices $X$ and $Y$. The rough idea is to associate with this channel a Choi state defined on two copies of the Hilbert space where the channel acts. A more detailed description is reported in Appendix \ref{app:CJapplication to Majorization}. 
Once the CV Choi state is determined, it can be described in terms of its Wigner function, 
which is defined on a doubled version of the original quantum phase space.
For the considered Gaussian channel (see the definition in (\ref{eq:gaussian channel def})), the Wigner function of the corresponding Choi state reads \cite{Serafini17book}
\be
\label{eq:CJ-GaussianWinger}
W_C(\boldsymbol{R})=\frac{e^{-\frac{1}{2}\boldsymbol{R}^\textrm{t}\gamma_C^{-1}\boldsymbol{R}}}{(2\pi)^{2N}\sqrt{\det\gamma_C}}\,,
\qquad\qquad
\,\,
\boldsymbol{R}\equiv(\boldsymbol{r},\boldsymbol{r}^{\textrm{\tiny aux}})^{\mathrm{t}}\,,
\ee
where
\begin{equation}
\label{eq:gammaChoi}
\gamma_C=\lim_{\nu\to\infty}
\begin{pmatrix}
\cosh(2\nu)X X^{\textrm{t}}+Y && \sinh(2\nu) X \Sigma_n\\
\sinh(2\nu) \Sigma_n X^{\textrm{t}} && \cosh(2\nu)\boldsymbol{1}_{2N}
\end{pmatrix}\,,
\end{equation}
and 
\begin{equation}
\Sigma_n\equiv\bigoplus_{j=1}^N
\begin{pmatrix}
1 &&0\\
0 && -1
\end{pmatrix}
\,.
\end{equation}
The vector $\boldsymbol{R}$ has $4N$ entries and parametrizes the doubled quantum phase space where the Wigner function of the Choi state is defined.
The parameter $\nu$ in (\ref{eq:gammaChoi}) plays the role of a regulator that must be introduced because
the maximally mixed state in CV systems entering the definition of the C-J map is not normalizable. The fact that the limit in (\ref{eq:gammaChoi}) does not exist is indeed a manifestation of the non-normalizability of the maximally mixed state.
We can now conclude that the regularized Choi Wigner function (\ref{eq:CJ-GaussianWinger}) is different from (\ref{eq:k_kerneldef_extended}).

The failure of the C-J isomorphism in representing a Gaussian channel as a Wigner function on a doubled phase space is a further manifestation of the difference between CV and DV systems. This fact is also reflected in the different behaviour described in this section of positivity-preserving channels with respect to Wigner majorization.

\subsection{A class of majorizing Gaussian channels }
\label{subsec:majorizing channel positive}

In the previous subsection, we discussed the absence of a definite Wigner majorization order between input and output states obtained by applying Gaussian channels on CV states. 
At this point, it is a natural question to ask whether we can identify a class of  {\it Wigner-majorizing} Gaussian channels. 
A Gaussian channel $\mathcal{E}$ belongs to this class if
$\hat\rho_{\textrm{\tiny in}}\succ_{\textrm{\tiny W}}\mathcal{E}(\hat\rho_{\textrm{\tiny in}})$ for any input state $\hat\rho_{\textrm{\tiny in}}$.
For this purpose, we exploit (\ref{eq:inputoutputNewW})-(\ref{eq:k_kerneldef}), which tell how the Wigner function of a given input state (not necessarily Gaussian) transforms under Gaussian channels.

Consider a Wigner-positive input state $\hat\rho_{\textrm{\tiny in}}$ and the output state $\hat\rho_{\textrm{\tiny out}}\equiv \mathcal{E}(\hat\rho_{\textrm{\tiny in}})$ obtained after the application of the Gaussian channel $\mathcal{E}$. Let us call $W_{\textrm{\tiny in}}$ and $W_{\textrm{\tiny out}}$  the corresponding positive Wigner functions.
As discussed in Sec.\,\ref{subsec:results_positiveWigner}, if $W_{\textrm{\tiny in}}$ and $W_{\textrm{\tiny out}}$ are related as in (\ref{eq:necessarycondition_wingermajo}), then $W_{\textrm{\tiny in}}\succ W_{\textrm{\tiny out}}$, and therefore $\hat\rho_{\textrm{\tiny in}}\succ_{\textrm{\tiny W}}\hat\rho_{\textrm{\tiny out}}$ 
whenever the kernel $k$ in (\ref{eq:necessarycondition_wingermajo}) is semi-doubly stochastic, according to the condition (\ref{eq:bistoch_kernel}).
The following result shows that the kernel (\ref{eq:propertykernel1}) is not in general semi-doubly stochastic.
\\

\noindent\fbox{%
    \parbox{\linewidth}{
{\bf Result 6:}
Given the Wigner function $W_{\textrm{\tiny in}}$ of an input state, the Wigner function $W_{\textrm{\tiny out}}$ of the output state $\hat\rho_{\textrm{\tiny out}}\equiv\mathcal{E}(\hat\rho_{\textrm{\tiny in}})$ is given by
\begin{equation}
\label{eq:inputoutputNewW}
W_{\textrm{\tiny out}}(\boldsymbol{r})=\int d\boldsymbol{z} k(\boldsymbol{r},\boldsymbol{z})W_{\textrm{\tiny in}}(\boldsymbol{z})
\,,
\end{equation}
where the integral kernel is defined as
\begin{equation}
\label{eq:k_kerneldef_mt}
k(\boldsymbol{r},\boldsymbol{z})\equiv \frac{e^{-\frac{1}{2}\boldsymbol{r}^{\textrm{t}}Y^{-1}\boldsymbol{r}-\frac{1}{2}\boldsymbol{z}^{\textrm{t}}X^{\textrm{t}}Y^{-1}X\boldsymbol{z}+\boldsymbol{z}^{\textrm{t}}X^{\textrm{t}}Y^{-1}\boldsymbol{r}}}{(2\pi)^N\sqrt{\det Y}}\,,
\end{equation}
in terms of the matrices $X$ and $Y$ characterizing the channel $\mathcal{E}$. The kernel $k$ satisfies the properties
\begin{equation}
\label{eq:propertykernel1_mt}
\int d\boldsymbol{r} k(\boldsymbol{r},\boldsymbol{z})=1\,,
\;\;\quad
\int d\boldsymbol{z} k(\boldsymbol{r},\boldsymbol{z})=\frac{1}{\det X}\,.
\end{equation}
}
}
\\

\noindent
However, by choosing Gaussian channels with $\det X\geq 1$,  (\ref{eq:propertykernel1_mt}) becomes semi-doubly stochastic, and the relation (\ref{eq:inputoutputNewW}) between input and output Wigner functions is consistent with the condition (\ref{eq:necessarycondition_wingermajo}). Exploiting this observation, we obtain
\\

\noindent\fbox{%
    \parbox{\linewidth}{
{\bf Result 7:}
The class of Gaussian channels $\mathcal{E}_{X,Y}$ characterized by the matrices $X$ and $Y$ such that $\det X\geq 1$ is a class of Wigner-majorizing channels when applied to CV states with positive Wigner functions, namely, for any CV state $\hat\rho$ with positive Wigner function,
\begin{equation}
\label{eq:majorizingchannel_mt}
\hat\rho\succ_{\textrm{\tiny W}}\mathcal{E}_{X,Y}(\hat\rho)\,.
\end{equation}
}
}
\\

\noindent
In Sec.\,\ref{subsec:ChannelsGenericWigner}, we generalize Result 7 to any input state, not necessarily Wigner-positive.

As a first example of a Wigner-majorizing channel, consider the {\it amplification channel} \cite{EisertWolf05}, defined by
\begin{equation}
X=\sqrt{\eta}\boldsymbol{1}_2\,,\qquad\qquad Y=(\eta-1)\boldsymbol{1}_2\,,
\end{equation}
where $\eta\in [1,\infty)$\,.   
Since, $\det X=\eta \geq 1$, these channels satisfy \eqref{eq:majorizingchannel_mt}.
We then move to examples of Gaussian channels with $\det X=1$, i.e. with an associated doubly stochastic kernel.
We first note that the single-mode thermal noise channel defined (\ref{eq:noisechannel}) and discussed in the previous section does not enter this category unless it is the trivial channel. Indeed, in that case, $\det X=1-s$, which is equal to one only when $s=0$, i.e. for the identity channel.
Consider then the Gaussian unitaries. These channels are characterized by $Y=0$ and $X$ symplectic and, therefore, with $\det X=1$. Thus, the Gaussian unitaries are Wigner-majorizing channels.
 This is consistent with our previous remarks since we have stressed that states that differ by symplectic transformations are majorization-equivalent and mutually majorize each other.
Another non-trivial subclass of Wigner-majorizing Gaussian channels contains the {\it classical mixing channels}.
As mentioned in Sec.\,\ref{subsec:results_positiveWigner}, the classical mixing channels implement on the input state a random displacement of the first moments distributed according to a Gaussian probability density function with a certain covariance matrix $Y$. Thus, they can be thought of as a special case of (\ref{eq:random Gaussian unitary channel}) with $U_{S,\boldsymbol{\bar r}}=U_{\boldsymbol{\bar r}}$ and
\begin{equation}
\label{eq:pdfGaussian_classicalmixing}
p(\boldsymbol{\bar{r}})
=
\frac{e^{-\frac{1}{2}\bar{\boldsymbol{r}}^\textrm{t}Y^{-1}\bar{\boldsymbol{r}}}}{(2\pi)^N\sqrt{\det Y}}
\,.
\end{equation}
One can prove that the resulting channel is Gaussian and characterized by the $2N\times 2N$ matrices $X=\boldsymbol{1}$ and $Y$ given by the covariance matrix in (\ref{eq:pdfGaussian_classicalmixing}) \cite{Holevo:2001zsx,Giovannetti2004,Serafini17book}. From these properties, we find that $\det X=1$ and, therefore, the classical mixing channels are Wigner majorizing channels. 
An equivalent argument valid for Gaussian input states is the following. 
In these cases, from (\ref{eq:gaussian channel def}), the covariance matrix of the input state changes as $\gamma\mapsto \gamma +Y$
under the action of the classical mixing channel.
A straightforward linear algebra analysis allows to check that, 
since both $\gamma$ and $Y$ are real, symmetric, and positive definite, we have (see the exercise on page 511 of \cite{HornJohnsonBook})
\begin{equation}
\label{eq:detsummatrix}
\det (\gamma+Y)\geq\det \gamma\,.
\end{equation}
According to the criterion (\ref{eq:MajorizationGaussianStates_mt}), this inequality implies that a Gaussian input state always Wigner majorizes the corresponding output state after a classical mixing channel.

\section{Continuous majorization for generic Wigner functions}
\label{sec:MajorizationNegative}

In the previous sections we have discussed
Wigner majorization relations between Wigner-positive states.
In this section, we explore possible extensions of the previous analysis to generic CV states by introducing three proposals to establish a preorder among states that are not generically Wigner-positive. Two of these provide a generalization of the Wigner majorization investigated above to CV states with finite Wigner negativity.
This investigation is also motivated by the results of \cite{Koukoulekidis:2021ppu}, where the Wigner majorization between non Wigner-positive states has been studied for DV systems defined on finite-dimensional Hilbert spaces with odd dimensionality.
After discussing the main features of the our proposals, 
we extend  the results of Sec.\,\ref{sec:Wignermajo_Channels} to the case where the input state is a generic CV state with finite Wigner negativity.

\subsection{Three new notions of preorder among Wigner functions}
\label{subsec:3proposals}

In Sec.\,\ref{subsec:results_genericWigner} we introduce three preorders in the set of CV states with finite, not necessarily vanishing, Wigner negativity.
As discussed in Sec.\,\ref{subsec:results_genericWigner}, we exclude from our analysis states with non-absolutely integrable Wigner functions. 
An example in this class of states is the  single-mode pure state with wavefunction \footnote{We thank Ludovico Lami and an anonymous referee for pointing out this example.} (see also \cite{LiebWigner})
\begin{equation}
\psi(x)=\begin{cases}
1 & {\rm if }\, x\in[-1/2,1/2]
\\
0 & {\rm otherwise}
\end{cases}\,.
\end{equation}
Computing the corresponding Wigner function, we find 
\begin{equation}
W(x,p)=\begin{cases}
\frac{1}{\pi}\frac{\sin\left[p(1-2\vert x\vert)\right]}{p} & {\rm if }\, x\in[-1/2,1/2]
\\
0 & {\rm otherwise}
\end{cases}\,,
\end{equation}
which is not absolutely integrable.
Generally, these states are difficult to understand in the resource theory of Wigner negativity due to their ill-defined resourcefulness. For this reason, our analysis focuses on states with finite Wigner negativity, i.e. with absolutely integrable Wigner functions.
In this subsection, we discuss and compare the three generalized preorder proposals, illustrating also insightful examples.

\begin{figure*}[t!]
\vspace{.2cm}
\hspace{-0.927cm}
\includegraphics[width=0.5\textwidth]{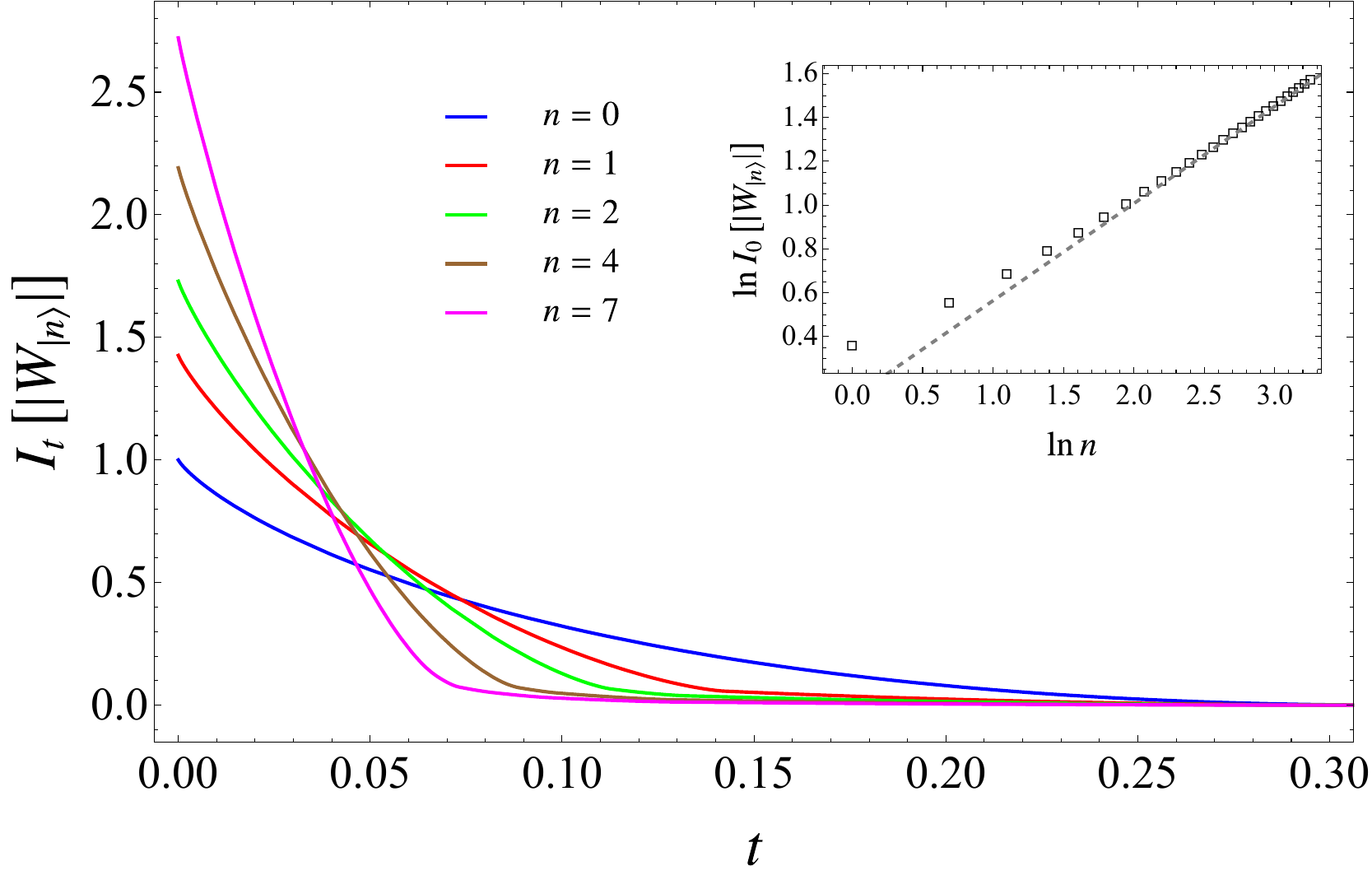}
\hspace{0.5cm}
\includegraphics[width=0.5\textwidth]{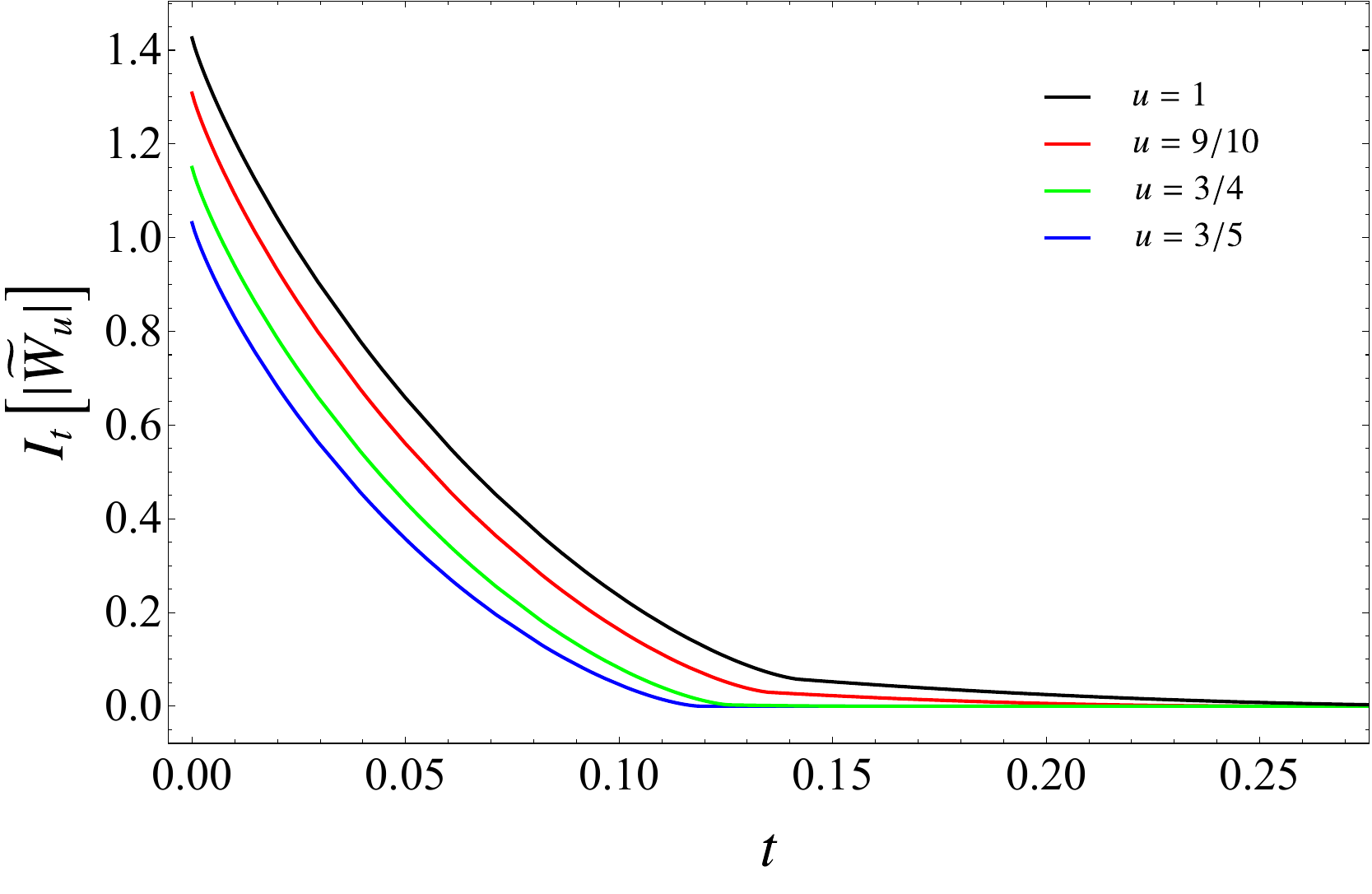}
\vspace{-0.3cm}
\caption{
Left: The functional in (\ref{eq:firstcond}) evaluated on the absolute value of the Wigner functions (\ref{eq:wignerHarmonic}) of the harmonic oscillator eigenstates and plotted as a function of $t$. Five different eigenstates are considered. The fact that each curve intersects all the others indicates the absence of Wigner majorization among these eigenstates.
The inset shows $\ln I_{0}$ evaluated on the same class of Wigner functions plotted as a function of $\ln n$. The dashed line corresponds to $0.44 \ln n 
+ 0.12 $ and is obtained through a fit procedure.
Right: The functional in (\ref{eq:firstcond}) is evaluated on the mixture (\ref{eq:Wigner_HCmix_ex}) for four choices of $u$. The non-intersection of the curves signals the relation $|\widetilde{W}_{1}|\ggcurly|\widetilde{W}_{9/10}|\ggcurly|\widetilde{W}_{3/4}|\ggcurly |\widetilde{W}_{3/5}|$  and the consequent Wigner majorization among the considered states.
}
\label{fig:proposal1}
\end{figure*}

\subsubsection{Proposal 1}
\label{subsec:proposal1}

The absolute value of any Wigner function is a positive function. Thus, we can employ the weak majorization relation $\ggcurly$ defined by one of the equivalent criteria 1-4 in Sec.\,\ref{subsec:reviewmajo} and induce an order relation among the absolute values of Wigner functions. Notice that we cannot impose the majorization relation $\succ$ between the absolute values of Wigner functions
because  their integral over the entire phase space is not equal to one and depends on the considered state. 
The weak majorization relation $\ggcurly$ between the absolute values of the Wigner functions of two generic CV states immediately induces 
\\

\noindent\fbox{%
    \parbox{\linewidth}{{
 {\bf Result 8 (Proposal 1):} Consider  two {\it generic} CV states with density matrices $\hat{\rho}_1$ and $\hat{\rho}_2$ and Wigner functions $W_1$ and $W_2$ respectively  and assume that $W_1,\, W_2\in L^1(\mathbb{R}^{2N})$.
 We propose to extend the majorization relation $\succ_{\textrm{\tiny W}}$ given in Sec.\,\ref{subsec:reviewmajo} to this more general set of states. We define that
$\hat{\rho}_1\succ_{\textrm{\tiny W}} \hat{\rho}_2$ if and only if $|W_1|\ggcurly |W_2|$, where the relation $\ggcurly$ is defined by \eqref{eq:firstcond}-\eqref{eq:def_decrrearrang_ext}.
}
 }}
 \\
 
\noindent
Since the Wigner function of a Wigner-positive state is equal to its absolute value, when evaluated between two Wigner-positive states, Proposal 1 reduces to the usual Wigner majorization employed in Secs.\,\ref{sec:MajorizationPositive} and \ref{sec:Wignermajo_Channels}.
Moreover, if two generic CV states differ by a Gaussian unitary, they are majorization equivalent according to this proposal. 
This is true because the Wigner functions corresponding to these states differ by a symplectic transformation on the phase space coordinates, 
which preserves the values of integrals as the ones in (\ref{eq:firstcond}) and (\ref{eq:majocondition2}), which are required to check the Wigner majorization via Proposal 1.

A remarkable feature of Proposal 1 is its simplicity in establishing whether  a Wigner majorization relation occurs between two states. 
In particular, this can be achieved by testing the condition (\ref{eq:firstcond}) on the absolute value of the Wigner functions of the considered states. 
For convenience, we introduce the functional
\begin{equation}
\label{eq:def_functionalI}
I_t[|W_{\hat\rho}|]\equiv \int [|W_{\hat\rho}(\boldsymbol{r})|-t]_+ d\boldsymbol{r}\,,
\end{equation}
on a generic Wigner function $W_{\hat\rho}$ of the CV state $\hat\rho$. 
Comparing with (\ref{eq:firstcond}), the functional $I_t$ allows to reformulate the condition 1 for the weak majorization $|W_1|\ggcurly |W_2|$ as
\begin{equation}
\label{eq:weakMajo_functionalI}
I_t[|W_1|]\geq I_t[|W_2|]\,,
\qquad
\forall\,t\geq 0\,.
\end{equation} 
In the following, we discuss some examples involving single-mode states.
A well-known class of single-mode CV states is made by the eigenstates of a harmonic oscillator, i.e. the Fock states. 
Denoting  by $|n\rangle$ the $n$-th oscillator eigenstate, obtained by acting $n$ times on the vacuum state with the creation operator,
the Wigner functions of these states are \cite{QuantumOpticsSchleich}
\begin{eqnarray}
\label{eq:wignerHarmonic}
W_{|n\rangle}(x,p)&=&\frac{(-1)^n}{\pi}e^{-(x^2+p^2)} L_n(2 (x^2+p^2))
\\
&=&\frac{(-1)^n}{\pi}e^{-r^2} L_n(2r^2)\equiv W_{|n\rangle}(r)
\nonumber\,,
\end{eqnarray}
where $L_n$ is the Laguerre polynomial of order $n$.
Notice that $W_{|n\rangle}$ is a positive function on the phase space only when $n=0$ and, therefore, the vacuum state is the only Wigner-positive oscillator eigenstate.

In the left panel of Fig.\,\ref{fig:proposal1}, we show $I_t$ in (\ref{eq:def_functionalI}) as a function of $t$ on the absolute values of oscillator eigenstate Wigner functions (\ref{eq:wignerHarmonic}) for five distinct values of $n$.
All the curves displayed  mutually intersect at one point. 
 Exploiting (\ref{eq:weakMajo_functionalI}) and Proposal 1 in Result 8, we conclude that there is no Wigner majorization order between any pair of oscillator eigenstates. We check this fact for the first fifteen oscillator eigenstates, but only the curves for five of them are shown in the panel to avoid clutter.
 In the inset of this panel, we report the logarithm of $I_0[W_{|n\rangle}]$ as a function of $\ln n$, where $n$ labels the oscillator eigenstates we study. To understand its behaviour as a function of $n$, we have fitted the fifteen data points with largest value of $n$ to the linear function $a\ln n+ b$. The values of $a$ and $b$ that best agree with the data are $a\simeq 0.44$ and $b\simeq 0.12$.
The Wigner logarithmic negativity of the Fock states was numerically studied in \cite{Kenfack:2004ges}, while, in \cite{Fawzi:2024plw} it was analytically shown that $\mathcal{N}_{W_{|n\rangle}}=\ln I_0[W_{|n\rangle}]$ scales as $1/2\ln n$ as $n\to\infty$. The parameters obtained via our fit are close, even if they do not match the analytical result, since we consider values of $n$ that are not large enough to capture the asymptotic behavior.

In order to showcase examples of CV states that exhibit Wigner majorization order according to Proposal 1, we construct a mixture of the vacuum state and the first excited state of a harmonic oscillator. The Wigner function of this family of mixed states read
\begin{equation}
\label{eq:Wigner_HCmix_ex}
\widetilde{W}_u(r)=(1-u)W_{|0\rangle}(r)+uW_{|1\rangle}(r)\,,
\end{equation}
where $W_{|0\rangle}$ and $W_{|1\rangle}$ are given in (\ref{eq:wignerHarmonic}) and $u\in[0,1]$.
We find it worth noticing that, when $u\leq 1/2$, the mixture described by $\widetilde{W}_u$ is a passive state, i.e. it does not allow work extraction via cyclic unitary processes (see \cite{vanherstraeten2021continuous,VanHerstraeten:2021nce} for discussions on this class of states).
The functional $I_t$ in (\ref{eq:def_functionalI}) can be evaluated on the  absolute values of $\widetilde{W}_u$ in (\ref{eq:Wigner_HCmix_ex}). The curves obtained for four choices of the parameter $u$ are reported in the right panel of Fig.\,\ref{fig:proposal1}. Differently from the left panel, the four curves do not intersect for any finite values of $t$. In particular, we find that
\begin{equation}
I_t\left[|\widetilde{W}_{1}|\right]\geq I_t\left[|\widetilde{W}_{9/10}|\right]\geq I_t\left[|\widetilde{W}_{3/4}|\right]\geq I_t\left[|\widetilde{W}_{3/5}|\right]
\,,
\end{equation}
$\forall\,t\geq 0$.
Using (\ref{eq:weakMajo_functionalI}), this chain of inequalities implies 
\begin{equation}
|\widetilde{W}_{1}|\ggcurly|\widetilde{W}_{9/10}|\ggcurly|\widetilde{W}_{3/4}|\ggcurly |\widetilde{W}_{3/5}|\,.
\end{equation}
Thus, by employing Proposal 1 in Result 8, we find the following Wigner majorization order 
\begin{equation}
\label{eq:Proposal1Examples}
{\hat{\rho}}_{\widetilde{W}_{1}}\succ_{\textrm{\tiny W}}{\hat{\rho}}_{\widetilde{W}_{9/10}}\succ_{\textrm{\tiny W}}{\hat{\rho}}_{\widetilde{W}_{3/4}}\succ_{\textrm{\tiny W}}{\hat{\rho}}_{\widetilde{W}_{3/5}}\,.
\end{equation}
Looking at \eqref{eq:Proposal1Examples}, we may suspect a Wigner majorization order along the family of states $\widetilde{W}_u$. 
This is not the case, as we can check by comparing, for instance, $I_t\left[|\widetilde{W}_{1/10}|\right]$ with any of the functional shown in the right panel of Fig.\,\ref{fig:proposal1}. By trying for several instances, we may conjecture that there is a majorization order along the family $\widetilde{W}_u$ only as long as $u>1/2$.
Interestingly, $\mathcal{N}_{\widetilde{W}_u}$ is different from zero and monotonically increasing when $u>1/2$, while it is vanishing when $u\leq 1/2$.
It would be interesting to understand this observation better analytically. We leave this question for future works.

The efficiency of Proposal 1 in establishing Wigner majorization relations is exploited later,
in the analysis of the Gaussian channels acting on {\it generic} CV states under the lenses of Wigner majorization.

\subsubsection{Proposal 2}

\label{subsec:proposal2}

The motivation for Proposal 2 comes from the DV case investigated by Koukoulekidis and Jennings in \cite{Koukoulekidis:2021ppu}. These authors extended the majorization relation between probability vectors to quasiprobability vectors, essentially by first employing the weak majorization criterion \eqref{partialsums} which is then promoted to a majorization relation due to the unit normalization of quasiprobability vectors that satisfies \eqref{vectornorm}. In \cite{Koukoulekidis:2021ppu} this was phrased as a "heretic" Lorentz curve criterion. A continuous version of the heretic Lorentz curve criterion would be to find an extension of the majorization condition 4 (see (\ref{eq:majo_cond_decrrearr}) and \eqref{eq:def_decrrearrang_ext}) by finding a way to construct decreasing rearrangements for non-positive functions. However, this attempt leads to pathologies. To bypass this problem, we try to generalize conditions \eqref{eq:firstcond} and \eqref{eq:majocondition3} to non-positive Wigner functions.
\\ 

\noindent\fbox{%
    \parbox{\linewidth}{{
   {\bf Alternative proposal (Proposal 2):}
    Consider  two {\it generic} CV states with density matrices $\hat{\rho}_1$ and $\hat{\rho}_2$ and Wigner functions $W_1$ and $W_2$ respectively. We propose an alternative extension of the majorization relation $\succ_{\textrm{\tiny W}}$ given in Sec.\,\ref{subsec:reviewmajo} to this more general set of states. To distinguish it from Proposal 1, we use the symbol $\succ_{\textrm{\tiny W}_\textrm{\tiny 2}}$ for this second proposal. We define that
    $\hat{\rho}_1\succ_{\textrm{\tiny W}_\textrm{\tiny 2}} \hat{\rho}_2$ if and only if one of the two (equivalent) conditions is verified
\begin{equation}
\label{eq:final_negMajo1_mt}
\int [W_1-t]_+ d\boldsymbol{r} - \int [W_2-t]_+d\boldsymbol{r}\geq 0\,,
\quad
\forall t\in\mathbb{R}\,,
\end{equation}
where $[\cdot]_+$ is defined in (\ref{eq:squarebracket}),
or
\begin{equation}
\label{eq:final_negMajo2_mt}
\int^\infty_t m_{W_1}(s) ds - \int^\infty_t m_{W_2}(s)ds\geq0\,,
\quad
\forall t\in\mathbb{R}\,,
\end{equation}
with $m_W$ given by (\ref{eq:levelfunctiondef}).
     }
 }
 }
  \\
  
\noindent
The functional $I_t$ evaluated on a positive Wigner function $W$ (without absolute value) diverges when $t<0$. 
Indeed, in this case $W-t>0$, hence $[W-t]_+=W-t$ and 
\begin{equation}
\label{eq:universaldivergence}
I_t[W]=1-t\,\int d\boldsymbol{r}\,,
\end{equation}
where the normalization (\ref{eq:normalizationW}) has been used. 
Crucially, the divergence in (\ref{eq:universaldivergence}) is independent of the considered Wigner function. Thus, given two positive Wigner functions $W_1$ and $W_2$, we have that
\begin{equation}
I_t[W_1]-I_t[W_2]=0\,,
\quad\forall\,t<0\,.
\end{equation}
This means that the criterion (\ref{eq:firstcond}) can be extended to Wigner majorization of positive Wigner functions to any real value of the parameter $t$. In contrast, if we consider a generic Wigner function $W$ that can take negative values, the functional $I_t[W]$ still diverges for $t<0$, but this divergence depends on the underlying state.
Thus, if we aim at introducing a Wigner majorization proposal based on the sign of $I_t[W_1]-I_t[W_2]$, as Proposal 2 \eqref{eq:final_negMajo1_mt}, it is necessary to check that this difference is well-defined for negative values of $t$.
If this condition is verified, we define that
$\hat{\rho}_1\succ_{\textrm{\tiny W}_\textrm{\tiny 2}} \hat{\rho}_2$ if and only if
\begin{equation}
\label{eq:final_negMajo1_v2}
I_t[W_1]-I_t[W_2]\geq 0\,,
\quad
\forall t\in\mathbb{R}\,.
\end{equation}
Given the Wigner function $W$ of a CV state with density matrix $\hat\rho$, let $\mathcal{A}^{(W)}_t\subset \mathbb{R}^{2N}$ be the set of points where $W<t$. Then, the functional $I_t$ applied to the Wigner function $W$ can be rewritten as
\begin{equation}
\label{eq:majoCondition_general_v1}
I_t[W]=\int_{\bar{\mathcal{A}}^{(W)}_t} (W-t) d\boldsymbol{r}
=
\int_{\bar{\mathcal{A}}^{(W)}_t} W d\boldsymbol{r}
-
t\int_{\bar{\mathcal{A}}^{(W)}_t}  d\boldsymbol{r}\,,
\end{equation}
where
$\bar{\mathcal{A}}^{(W)}_t\equiv \mathbb{R}^{2N}\setminus \mathcal{A}^{(W)}_t$. Since $W$ and its absolute value are  integrable on $\mathbb{R}^{2N}$, it is so also on $\bar{\mathcal{A}}^{(W)}_t$ and, therefore, the first term in (\ref{eq:majoCondition_general_v1}) is finite. The second term can be written as
\begin{equation}
\label{eq:majoCondition_general_second term}
- \, t\! \int_{\bar{\mathcal{A}}^{(W)}_t}  \! d\boldsymbol{r}=
- \, t\,\textrm{Vol}\,\bar{\mathcal{A}}^{(W)}_t
=
-\,t\,\textrm{Vol}\,\mathbb{R}^{2N}+t\,\textrm{Vol}\,\mathcal{A}^{(W)}_t\,,
\end{equation}
where  ${\rm Vol}\,\mathcal{M}$ denotes the volume of a subset $\mathcal{M}\subset \mathbb{R}^{2N}$ of the phase space.
Due to the integrability of $W$, if $t>0$, then $\textrm{Vol}\,\mathbb{R}^{2N}-\textrm{Vol}\,\mathcal{A}^{(W)}_t$ and (\ref{eq:majoCondition_general_second term}) are finite, 
leading to a well-defined $I_t[W]$. On the other hand, if $t<0$, then $\textrm{Vol}\,\mathbb{R}^{2N}-\textrm{Vol}\,\mathcal{A}_t^{(W)}$ is infinite, and, in general, we are not guaranteed that $\textrm{Vol}\,\mathcal{A}^{(W)}_t$ is finite. In the following, we assume the validity of this requirement, thus restricting our analysis to those states whose Wigner function satisfies this property.
Under this assumption, $-t\, \textrm{Vol}\,\mathbb{R}^{2N}$is the only divergent term in (\ref{eq:majoCondition_general_second term}) and in (\ref{eq:majoCondition_general_v1}), 
which does not depend on $W$, differently from $t\textrm{Vol}\,\mathcal{A}^{(W)}_t$. 
This allows us to conclude that the divergence of  $I_t[W]$ for $t<0$ is universal for the class of these CV states.

The universality of the divergence of $I_t[W]$ for negative values of $t$ suggests to introduce the following regularization for the phase space.
Let $\mathbb{R}^{2N}_\Lambda$ be a hypercube of linear size $\Lambda$ or a hypersphere with radius $\Lambda$, both centered in the origin of the phase space. 
The volumes of these spaces as functions of $N$ and of the regulator $\Lambda$ read respectively 
\begin{equation}
 \label{eq:regularized Phasespace}
\textrm{Vol}\,\mathbb{R}^{2N}_\Lambda= \Lambda^{2N}\,,
\quad\;
\textrm{Vol}\,\mathbb{R}^{2N}_\Lambda=\frac{\pi^{N}}{\Gamma(N+1)} \Lambda^{2N}\,,
\end{equation}
We regularize $I_t[W]$ by replacing $\mathbb{R}^{2N}$ with $\mathbb{R}^{2N}_\Lambda$ in the functional. 
Now, for any finite value of $\Lambda$ the functional $I_t[W]$ is finite for any $t\in\mathbb{R}$. Notice that choosing to regularize $I_t[W]$ through a cut-off phase space with spherical symmetry is more convenient when $W$ is rotationally invariant. Given two CV states $\hat\rho_{W_1}$ and $\hat\rho_{W_2}$ with Wigner functions $W_1$ and $W_2$ respectively,
we can rephrase (\ref{eq:final_negMajo1_mt}) in a more precise way by saying that $\hat\rho_{W_1}\succ_{\textrm{\tiny W}_\textrm{\tiny 2}}\hat\rho_{W_2}$ if and only if
\begin{equation}
\infty >\lim_{\Lambda\to\infty} I_t[W_1]-I_t[W_2]\geq 0\,,
\quad
\forall\,t\in\mathbb{R}\,.
\end{equation}
By using (\ref{eq:majoCondition_general_v1}) and (\ref{eq:majoCondition_general_second term}), we find that, for any $t\in \mathbb{R}$
\begin{eqnarray}\label{eq:regularized_proposal2}
& & \lim_{\Lambda\to\infty}  I_t[W_1]-I_t[W_2] \; =
\\
& & =\;
t\left(\textrm{Vol}\,\mathcal{A}^{(W_1)}_t-\textrm{Vol}\,\mathcal{A}^{(W_2)}_t\right)
\nonumber
\\
& &
\;\;\;\; + \!\! \int_{\bar{\mathcal{A}}^{(W_1)}_t} W_1 d\boldsymbol{r}\,
- \int_{\bar{\mathcal{A}}^{(W_2)}_t} W_2 d\boldsymbol{r}<\infty \,, 
\nonumber
\end{eqnarray}
meaning that the condition (\ref{eq:final_negMajo1_mt}) of Proposal 2 is well-defined. In the light of the analysis above, we find it worth stressing that Proposal 2 for Wigner majorization, 
differently from Proposal 1, can be applied only to a subclass of CV states having Wigner functions $W$ such that ${\rm Vol}\,\mathcal{A}^{(W)}_t<\infty$ for any $t<0$.
Notably, all the Wigner functions relevant for this manuscript and, in particular, the ones considered in the following examples satisfy the condition on the finite volume of $\mathcal{A}^{(W)}_t$ and, therefore, can be compared through Proposal 2 for Wigner majorization.

At this point, we can discuss the main features of Proposal 2 for Wigner majorization between generic CV states.
First notice that, similarly to what happens for Proposal 1, Proposal 2 reduces to the usual Wigner majorization introduced in \cite{vanherstraeten2021continuous} when evaluated on two Wigner-positive states. Moreover, by the same argument discussed in Sec.\,\ref{subsec:proposal1} for Proposal 1, two CV states are majorization-equivalent according to Proposal 2 if they differ by a Gaussian unitary transformation, i.e. their Wigner functions differ by a symplectic transformation of the phase space coordinates.

The condition (\ref{eq:firstcond}) is equivalent to (\ref{eq:majocondition3}), as reviewed in Sec.\,\ref{subsec:reviewmajo}. 
This equivalence is formally preserved also when $t<0$, provided that the integrals in (\ref{eq:majocondition3}) are well-defined.
Using the definition (\ref{eq:levelfunctiondef}), we can rewrite the level function $m_W(t)$ of a given Wigner function $W$ as
\begin{equation}
\label{eq:levelfunct_diver}
m_W(t)=\textrm{Vol}\,\mathbb{R}^{2N}-\textrm{Vol}\,\mathcal{A}^{(W)}_t\,.
\end{equation}
As explained above, if $t<0$, (\ref{eq:levelfunct_diver}) is divergent due to the volume of the entire phase space. Indeed, we recall that we are considering CV states with Wigner functions such that $\textrm{Vol}\,\mathcal{A}^{(W)}_t<\infty$ for $t<0$. The divergence of (\ref{eq:levelfunct_diver}) when $t<0$ makes the integrals in (\ref{eq:majocondition2}) ill-defined in this regime.
Thus, to make sense of (\ref{eq:majocondition2}) evaluated on two generic CV states, we first regularize the phase space and, consequently, the integrals of the two level functions, then we consider the differences of the two integrals in (\ref{eq:majocondition2}) and, finally, we remove the regulator verifying that the result is finite for any $t\in \mathbb{R}$. If this is true, the condition (\ref{eq:final_negMajo2_mt}) is well-defined and gives an equivalent criterion for Proposal 2 for Wigner majorization. To verify this well-definiteness, given a CV state with Wigner function $W$, we introduce a regularized level function
\begin{equation}
\label{eq:regularized-level_function}
m^{(\Lambda)}_W(t)=\textrm{Vol}\,\mathbb{R}^{2N}_\Lambda-\textrm{Vol}\,\mathcal{A}^{(W)}_t\,,
\end{equation}
where, also in this case, $\mathbb{R}^{2N}_\Lambda$ can be either a hypercube or a sphere of linear size $\Lambda$, whose volumes are reported in (\ref{eq:regularized Phasespace}).
Integrating (\ref{eq:regularized-level_function}), we have
\begin{equation}
\label{eq:integral_regularized-level_function}
\int^\infty_t m^{(\Lambda)}_W(s)ds=\int^\infty_t\left[\textrm{Vol}\,\mathbb{R}^{2N}_\Lambda-\textrm{Vol}\,\mathcal{A}^{(W)}_s\right]ds\,.
\end{equation}
When $t>0$, (\ref{eq:integral_regularized-level_function}) is finite in the limit $\Lambda\to\infty$, as expected from Wigner majorization of Wigner-positive states \cite{vanherstraeten2021continuous}. On the other hand, when $t<0$, we can rewrite (\ref{eq:integral_regularized-level_function}) as
\begin{eqnarray}
\label{eq:integral_levelfunc_negativet_v1}
\int^\infty_t m^{(\Lambda)}_W(s)ds&=&\int^\infty_0 m^{(\Lambda)}_W(s)ds+\int^0_t m^{(\Lambda)}_W(s)ds
\\
&=&\int^0_t m^{(\Lambda)}_W(s)ds+\textrm{finite as $\Lambda\to\infty$}\,.
\nonumber
\end{eqnarray}
Using (\ref{eq:regularized-level_function}), (\ref{eq:integral_levelfunc_negativet_v1}) becomes
\begin{equation}
\label{eq:integral_levelfunc_negativet_v2}
\int^\infty_t m^{(\Lambda)}_W(s)\, ds=- \, t\,\textrm{Vol}\,\mathbb{R}^{2N}_\Lambda+\int^0_t \textrm{Vol}\,\mathcal{A}^{(W)}_s ds+\textrm{finite}\,.
\end{equation}
Since $\textrm{Vol}\,\mathcal{A}^{(W)}_s$ is finite for the class of states we are considering, 
the integral in the right-hand side of (\ref{eq:integral_levelfunc_negativet_v2}) is always finite, also for $\Lambda\to\infty$. 
Notice that all the Wigner functions are bounded from below and therefore a finite value $\bar{t}$ exists 
such that $\mathcal{A}^{(W)}_{\bar{t}}=\emptyset$ for $t<\bar{t}$. In this range for the parameter $t$, the integral in the right-hand side of (\ref{eq:integral_levelfunc_negativet_v2}) is identically zero, and therefore we do not have to worry about the limit $t\to-\infty$ of this term. The analysis above implies that, in the limit $\Lambda\to\infty$, the divergence of the integral (\ref{eq:integral_levelfunc_negativet_v2}) is independent of the Wigner function $W$. 
Considering again two CV states $\hat\rho_{W_1}$ and $\hat\rho_{W_2}$, we can compute the difference between the two corresponding integrals (\ref{eq:integral_levelfunc_negativet_v2}),
obtaining
\begin{eqnarray}
\label{eq:difference_WignermajoProp2}
&&\int^\infty_t \left(m^{(\Lambda)}_{W_1}(s)-m^{(\Lambda)}_{W_2}(s)\right)ds
\\
&&
=
\int^0_t \left(\textrm{Vol}\,\mathcal{A}^{(W_1)}_s-\textrm{Vol}\,\mathcal{A}^{(W_2)}_s \right)ds
\nonumber
+
\textrm{finite terms}\,,
\end{eqnarray}
which is finite as $\Lambda\to\infty$.
This result implies that the condition (\ref{eq:final_negMajo2_mt}) in Proposal 2 is well-defined and can be used to check the Wigner majorization between CV states.

\begin{figure*}[t!]
\vspace{.2cm}
\hspace{-0.527cm}
\includegraphics[width=0.48\textwidth]{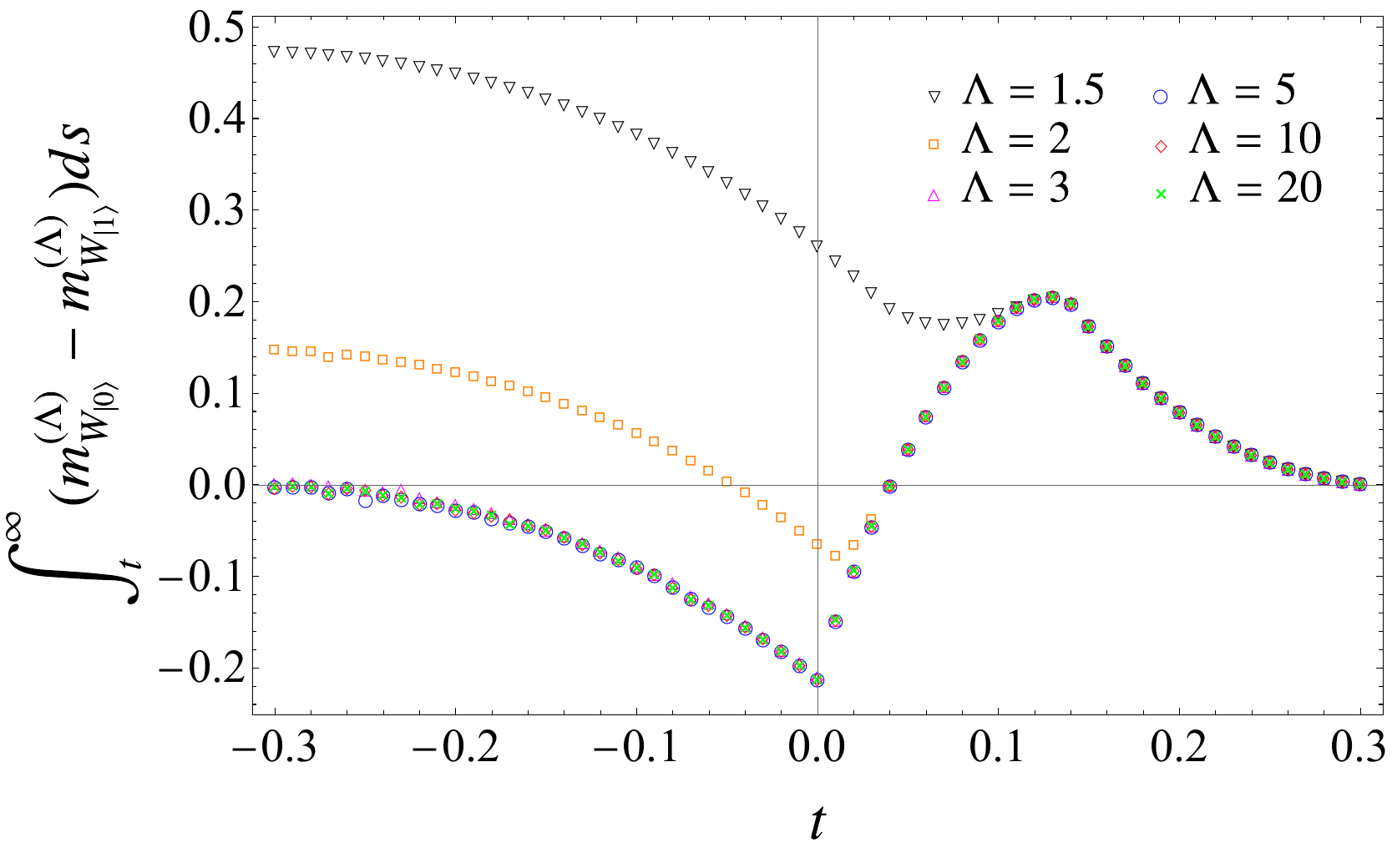}
\hspace{0.5cm}
\includegraphics[width=0.48\textwidth]{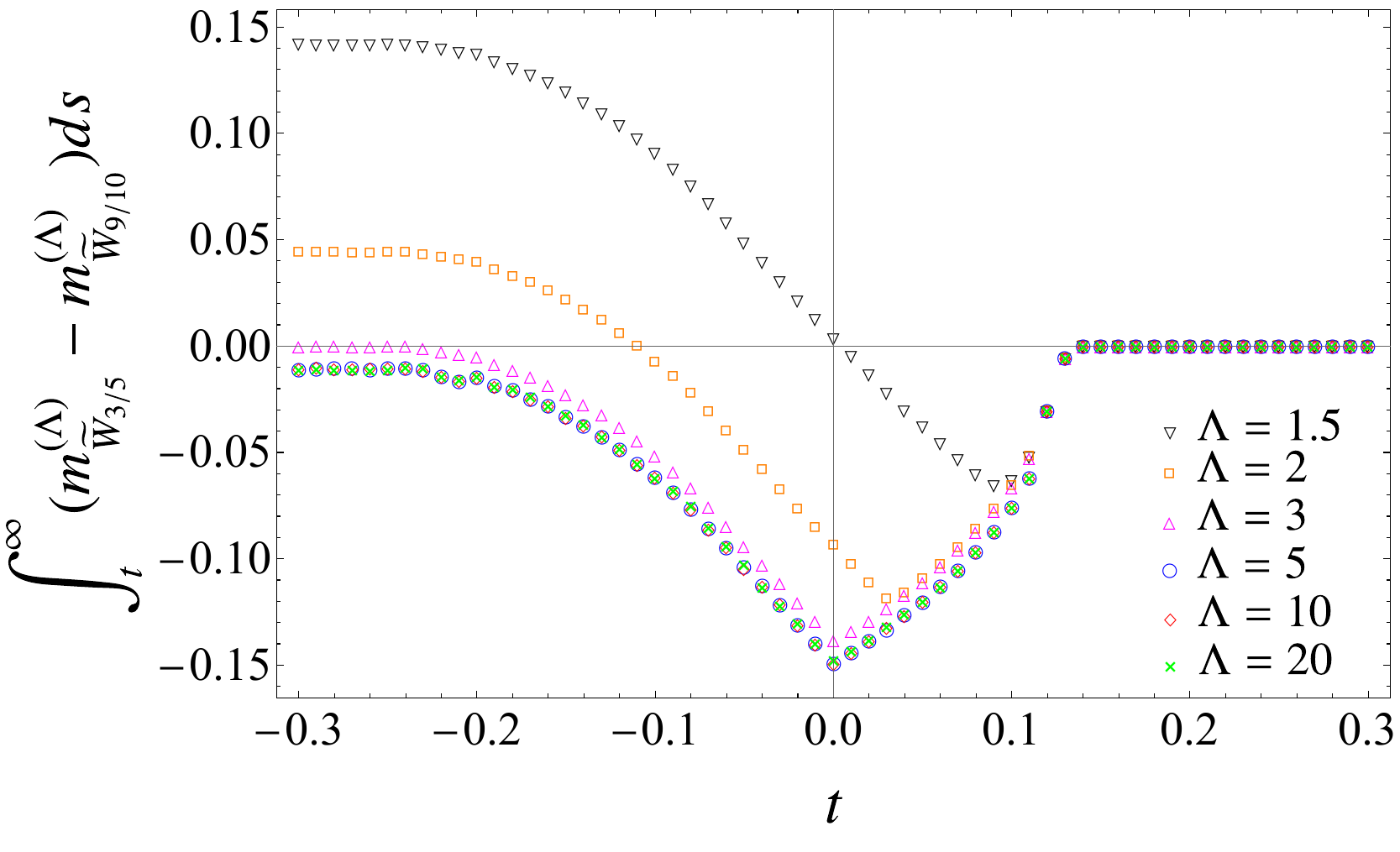}
\vspace{-0.3cm}
\caption{
Test of Proposal 2 for Wigner majorization
through the difference (\ref{eq:difference_WignermajoProp2}) evaluated for a pair of Wigner functions $W_1$ and $W_2$. 
Left: $W_1=W_{|0\rangle}$ and $W_2=W_{|1\rangle}$, where $W_{|n\rangle}$ is defined in (\ref{eq:wignerHarmonic}). 
Right: $W_1=\widetilde{W}_{3/5}$ and $W_2=\widetilde{W}_{9/10}$, where both the Wigner functions are defined in (\ref{eq:Wigner_HCmix_ex}).
 In both panels, the curves are shown as functions of $t$ for various values of $\Lambda$, exhibiting collapse for large $\Lambda$. 
 In the left panel the asymptotic curve has no definite sign, implying the absence of majorization order between $W_{|0\rangle}$ and $W_{|1\rangle}$,
while in the right panel the asymptotic curve takes only negative values, meaning that $W_2=\widetilde{W}_{9/10}\succ W_1=\widetilde{W}_{3/5}$.
}
\label{fig:proposal2}
\end{figure*}

In Fig.\,\ref{fig:proposal2}, the condition (\ref{eq:final_negMajo2_mt}) is exploited to test Wigner majorization according to Proposal 2. 
In particular, the difference in (\ref{eq:difference_WignermajoProp2}) is shown as function of $t$ for  $W_1=W_{|0\rangle}$ and $W_2=W_{|1\rangle}$ in the left panel, 
where $W_{|n\rangle}$ is given in (\ref{eq:wignerHarmonic}), 
and for $W_1=\widetilde{W}_{3/5}$ and $W_2=\widetilde{W}_{9/10}$ in the right panel, with $\widetilde{W}_u$ defined in (\ref{eq:Wigner_HCmix_ex}). 
The curves are obtained for six distinct values of $\Lambda$. In both panels, we observe that when the cut-off $\Lambda$ is large enough, the curves collapse on an asymptotic curve. 
From the resulting asymptotic curve in the right panel, we observe that the plotted quantity has a negative sign in the limit $\Lambda\to\infty$. Thus, from the condition (\ref{eq:final_negMajo2_mt}) we conclude that   
\begin{equation}
{\hat{\rho}}_{\widetilde{W}_{9/10}}\succ_{\textrm{\tiny W}_\textrm{\tiny 2}}{\hat{\rho}}_{\widetilde{W}_{3/5}}\,.
\end{equation}
On the other hand, the asymptotic curve in the left panel does not have a definite sign, 
meaning that the states ${\hat{\rho}}_{W_{|0\rangle}}$ and ${\hat{\rho}}_{W_{|1\rangle}}$ do not display a Wigner majorization relation according to Proposal 2.
Interestingly, both the presence and the absence of a Wigner majorization relation in the cases in Fig.\,\ref{fig:proposal2} are also found using Proposal 1 (see (\ref{eq:Proposal1Examples}) and Fig.\,\ref{fig:proposal1}). 
It would be interesting to delve further into the differences between Proposal 1 and Proposal 2, for instance, by finding whether they are equivalent or there are examples where these yield different majorization relations for the same pair of CV states. We leave this task for future investigations.

\subsubsection{Tautological preorder}

\label{subsec:proposal3}

In Sec.\,\ref{subsec:Wignermajo_thermonoiseChannels}, we show that CV output states obtained after applying a Gaussian channel and their corresponding input states do not have a fixed Wigner majorization relation.
In Sec.\,\ref{subsec:results_genericWigner},
we introduce an alternative preorder relation that, by construction, ensures that every output state obtained by applying a Gaussian channel is Wigner-majorized by the corresponding input state. 
\\

\noindent\fbox{
    \parbox{\linewidth}{{
   {\bf Alternative proposal (Tautological preorder):}
     Consider  two {\it generic} CV states with density matrices $\hat{\rho}_1$ and $\hat{\rho}_2$ and Wigner functions $W_1$ and $W_2$ respectively. We provide a new notion of preorder among these states. This relation differs from Propositions 1 and 2 and is denoted by the symbol $\succ_{\textrm{\tiny W}_\textrm{\tiny t}}$. We define that
    $\hat{\rho}_1\succ_{\textrm{\tiny W}_\textrm{\tiny t}} \hat{\rho}_2$ if and only if $W_1 \succ_{\textrm{t}} W_2$, namely
if there exists a kernel $k$ of the form (\ref{eq:k_kerneldef_mt}) such that 
\begin{equation}
\label{eq:tautologicalcondition_mt}
W_{2}(\boldsymbol{r})=\int d\boldsymbol{z} k(\boldsymbol{r},\boldsymbol{z})W_{1}(\boldsymbol{z})
\, .
\end{equation}
     }
 }}
  \\

\noindent
Because of its defining property, we refer to this proposal as tautological preorder.
This proposal applies to any pair of CV states because it requires checking the existence of a Gaussian channel connecting the two considered states. Given that it does not rely on any of the conditions 1-4 in Sec.\,\ref{subsec:reviewmajo}, we are not guaranteed that this relation identifies a preorder and, therefore, a full-fledged majorization relation. 
In Appendix \ref{app:preorder} we prove that the tautological preorder defines indeed a preorder, justifying the notation introduced.

This preorder relation with respect to the action of Gaussian channels has substantial differences to Proposal 1 and Proposal 2.
The first one concerns the case of Wigner-positive states. 
While Proposals 1 and 2 reduce to the Wigner majorization discussed in \cite{vanherstraeten2021continuous}, the tautological preorder does not. 
Consider the following example involving single-mode Gaussian states. In Sec.\,\ref{subsec:Wignermajo_thermonoiseChannels}, 
we showed that the output state obtained by applying a thermal noise channel is not necessarily majorized by the corresponding input state. 
However, using the tautological preorder, by definition, the input state  is always more ordered than the output state, in contrast to the other proposals.

 Another difference from the other proposals emerges when we try to characterize states that are equivalent through the tautological preorder. 
According to this proposal, $W_1$ and $W_2$ are majorization-equivalent if we can find a reversible Gaussian channel connecting them. As discussed in \cite{heinosaari2009semigroup}, a Gaussian channel is reversible if and only if $Y=0$. Thus, two Wigner functions are equivalent through tautological preorder if they are related by a Gaussian unitary channel.
This condition is more restrictive than in Proposal 1 and Proposal 2. The integrals in (\ref{eq:firstcond}) and (\ref{eq:majocondition3}) are left invariant not only by symplectic transformations. Hence, in Proposals 1 and 2, majorization-equivalent CV states can be connected by a larger class of transformations than Gaussian unitaries.
More differences arise when we restrict to the set of Gaussian states.  Two Gaussian states can always be connected by a Gaussian channel \cite{Serafini17book}. For instance, two thermal states can be related by a loss or an amplifier channel. Thus, \textit{all Gaussian states are equivalent according to the tautological preorder}. This is not the case for the Wigner majorization proposals, as follows from Result 1 (see \eqref{eq:MajorizationGaussianStates_mt}).

Another comparison between the three proposals in Sec.\,\ref{subsec:3proposals} concerns the validity of \eqref{eq:propweakmajo}. Exploiting the convexity of the functional  \eqref{eq:squarebracket}, we prove that \eqref{eq:propweakmajo} holds for both Proposal 1 and Proposal 2. The tautological preorder requires a more careful analysis. Consider three states with Wigner functions $W$, $W_1$ and $W_2$ such that $W\succ_{\textrm{t}} W_1$ and $W\succ_{\textrm{t}} W_2$. We now ask whether $ W\succ_{\textrm{t}} t W_1+(1-t)W_2$, with $t\in [0,1]$. From the majorization relations, we can write
\begin{equation}
\label{eq:probabilisticChannel}
t W_1+(1-t)W_2=
\int d\boldsymbol{z} [t k_1(\boldsymbol{r},\boldsymbol{z})+(1-t)k_2(\boldsymbol{r},\boldsymbol{z})]W(\boldsymbol{z})\,,
\end{equation}
where $k_1$ and $k_2$ are the kernels relating $W$ with $W_1$ and $W_2$ respectively. If $k_1$ and $k_2$ have the form \eqref{eq:k_kerneldef_mt}, $t k_1+(1-t)k_2$ does not, and, therefore, the property \eqref{eq:propweakmajo} does not hold for the tautological preorder. 

The differences between Proposals 1 and 2 and the tautological preorder should not be seen as a problem but rather as a hint of a fundamentally different role for the preorder $\succ_{\textrm{t}}$. 
Observe that we can map a generic input state $\hat{\rho}_{\textrm{\tiny in}}$ to a Gaussian state with a covariance matrix  $\gamma_{\textrm{\tiny out}}$ independently of the input state. This is achieved by applying the (degenerate
\footnote{
This channel is perfectly well-defined, as we can observe from the definition \eqref{eq:gaussian channel def} of a Gaussian channel acting on a Gaussian state.
We refer to it as degenerate given that $\det X=0$. Consequently, the second integral in \eqref{eq:propertykernel1} is divergent. However, this does not affect the action of the channel on a generic input Wigner function. Indeed, as a consequence of $X=0$, the kernel \eqref{eq:k_kerneldef} loses its dependence on $\mathbf{z}$, and the second condition in \eqref{eq:propertykernel1} is not necessary.
}) Gaussian channel defined by $Y=\gamma_{\textrm{\tiny out}}$ and $X$ given by the matrix with all vanishing entries (to verify this fact, we can straightforwardly use \eqref{eq:inputoutputNewW}).
This implies that \textit{the Gaussian states are less ordered than all the other CV states according to the tautological preorder}. In other words, Gaussian states are at the bottom of the tautological preorder. Thus, the relation $\succ_{\textrm{t}}$ can be legitimately seen as a preorder for non-Gaussianity, differently from other proposals characterizing disorder or uncertainty. We hope that these results will be useful in the context of the resource theory of non-Gaussianity.

\subsection{Relations with logarithmic Wigner negativity}
\label{subsec:relationWLN}

\begin{figure*}[t!]
\vspace{.2cm}
\hspace{-0.927cm}
\includegraphics[width=0.47\textwidth]{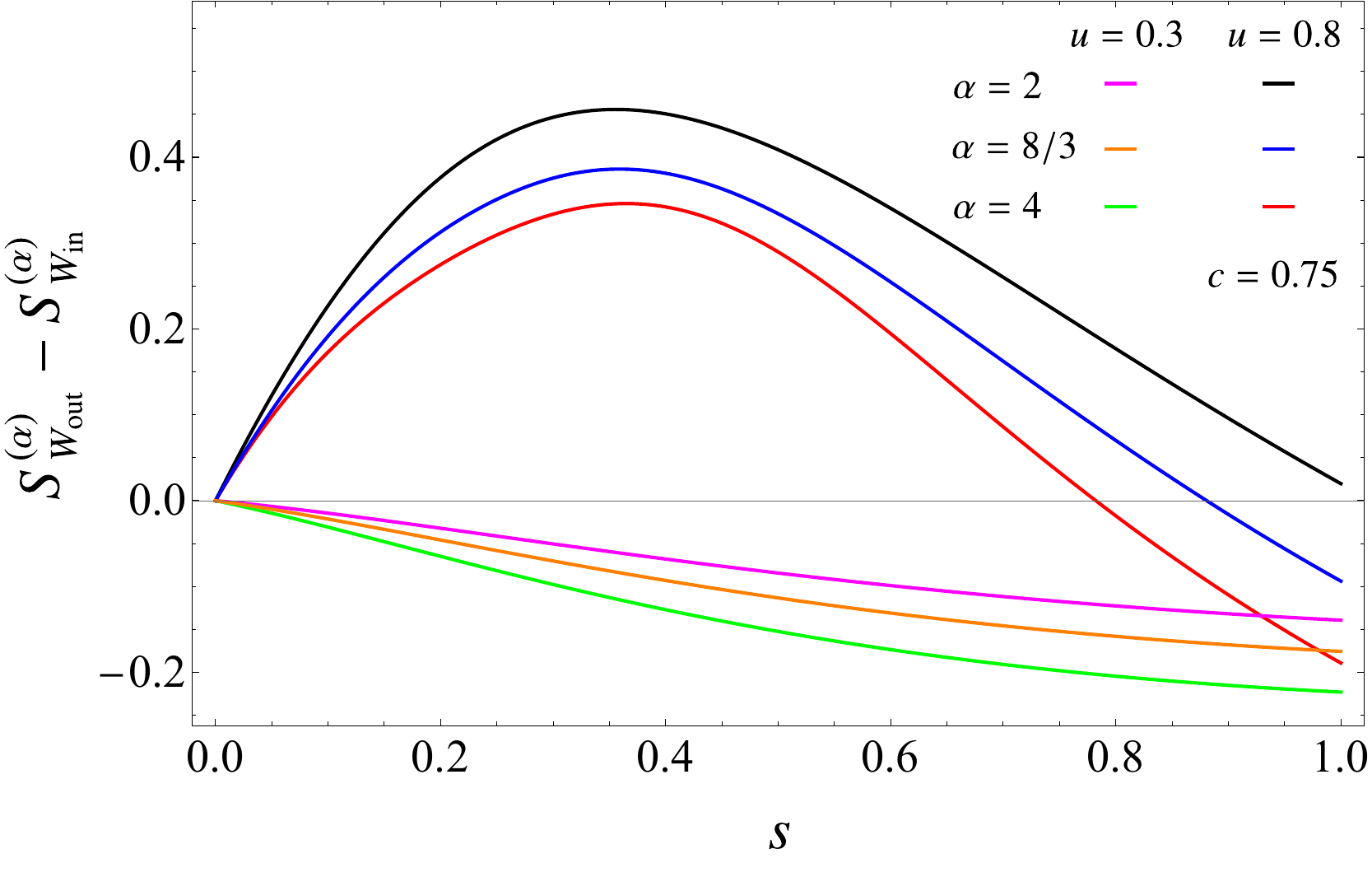}
\hspace{0.5cm}
\includegraphics[width=0.45\textwidth]{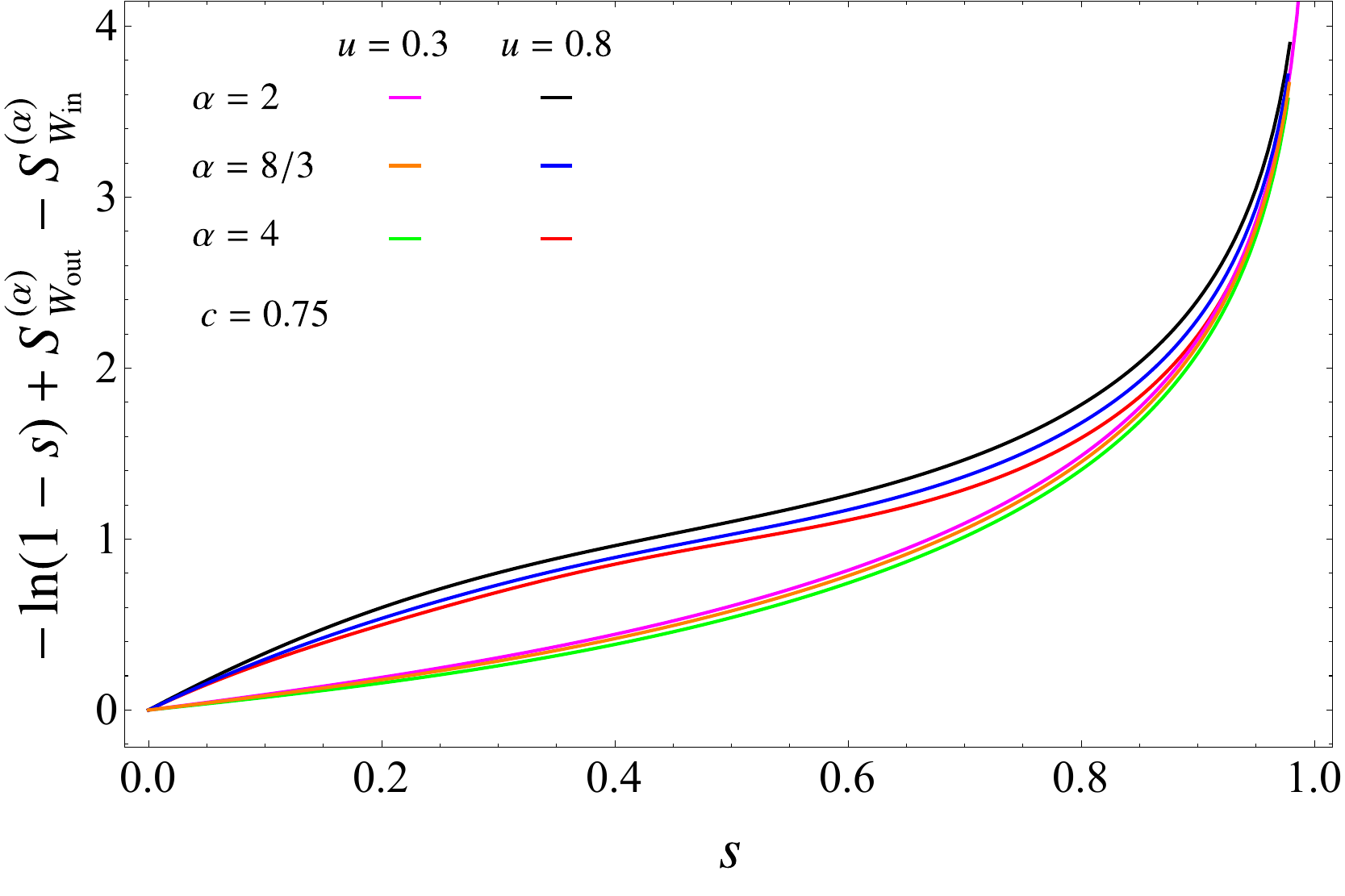}
\vspace{-0.3cm}
\caption{
Left: The difference between $S^{(\alpha)}_{W}$ evaluated on the output state obtained by applying a thermal noise channel (\ref{eq:noisechannel}) and the corresponding input state given by the mixture (\ref{eq:Wigner_HCmix_ex}). The curves are plotted as functions of the parameter $s$ of the thermal noise channel, with  $c=0.75$. 
Two different input states corresponding to two different values of $u$ 
and three distinct values of the Rényi index $\alpha=2p/(2q-1)$, with $p$ and $q$ integers, are considered. 
The fact that the curves have no definite sign shows that $S^{(\alpha)}_{W}$ (and the Wigner Renyi entropy with even indices) are not in general monotonic under Gaussian channels.
Right: The logarithm of the determinant of the matrix $X$ characterizing the thermal noise channel is subtracted from the quantity plotted in the left panel. The curves confirm the inequality (\ref{eq:ineq_Renyinew}).
}
\label{fig:WignerRenyi}
\end{figure*}

In this subsection, we show that whenever one of the preorder relations proposed in Sec.\,\ref{subsec:3proposals} occurs between two states, 
the corresponding Wigner logarithmic negativities are related in the same way. 
\\

\noindent\fbox{%
    \parbox{\linewidth}{{
{\bf Result 9}: Given two CV states $\hat{\rho}_A$ and $\hat{\rho}_B$ with Wigner functions $W_A$ and $W_B$ respectively, using any of the three preorder proposals reported above, we have that, for $*\in \{\textrm{W},\textrm{ W}_\textrm{2},\textrm{ W}_\textrm{t}\}$,
\begin{equation}
\hat{\rho}_A\succ_{*} \hat{\rho}_B\;\quad \Rightarrow \;\quad
\mathcal{N}_{W_{A}}\geq \mathcal{N}_{W_{B}}\,.
\end{equation}
}
}
}
\\

\noindent
 This means that the implication holds for the preorder induced by any of the three proposals discussed in Sec.\,\ref{subsec:3proposals}.

Given a state $\hat\rho$ with Wigner function $W_{\hat\rho}$, let us evaluate the functional (\ref{eq:def_functionalI}) for $t=0$ on $W_{\hat\rho}$ and $|W_{\hat\rho}|$. 
From (\ref{eq:Wlogneg_def}), we can rewrite both functionals in terms of the Wigner logarithmic negativity of $\hat\rho$ as follows
\begin{equation}
\label{eq:functionalabs_t0}
I_0[|W_{\hat\rho}|]= e^{\mathcal{N}_{W_{\hat\rho}}}\,,
\quad
I_0[W_{\hat\rho}]=\frac{e^{\mathcal{N}_{W_{\hat\rho}}}+1}{2}\,.
\end{equation}
Crucially, both the functions of the Wigner logarithmic negativity in (\ref{eq:functionalabs_t0}) are monotonically increasing. It is worth mentioning that the relations in (\ref{eq:functionalabs_t0}) could be exploited to gain insights into the Wigner logarithmic negativity. For instance, combining the first relation with the behavior observed in the inset of the left panel of Fig.\,\ref{fig:proposal1}, we can argue that, as already found in \cite{Fawzi:2024plw}, $\mathcal{N}_{W_{|n\rangle}}\sim  \ln n$ for $n\to\infty$, where $W_{|n\rangle}$ are the Wigner functions (\ref{eq:wignerHarmonic}) of the harmonic oscillator eigenstates.

Now, consider two CV states $\hat\rho_A$ and $\hat\rho_B$ with Wigner functions $W_A$ and $W_B$ respectively. 
From Proposal 1,
$\hat\rho_A\succ_{\textrm{\tiny W}} \hat\rho_B$ if 
\begin{equation}
\label{eq:proposal1_funcI}
I_t[|W_{A}|]- I_t[|W_{B}|]\geq 0\,,
\quad
\forall\,t\geq 0
\,,
\end{equation}
while, from Proposal 2, $\hat\rho_A\succ_{\textrm{\tiny W}_{\tiny 2}} \hat\rho_B$ if
\begin{equation}
\label{eq:proposal2_funcI}
I_t[W_{A}]- I_t[W_{B}]\geq 0\,,
\quad
\forall\,t\in \mathbb{R}
\,.
\end{equation}
Since (\ref{eq:proposal1_funcI}) and  (\ref{eq:proposal2_funcI}) must hold also in the special case given by $t=0$, 
by exploiting the monotonicity of the functions in (\ref{eq:functionalabs_t0}), 
we have that $\mathcal{N}_{W_A}\geq \mathcal{N}_{W_B}$
when either $\hat\rho_A\succ_{\textrm{\tiny W}} \hat\rho_B$ or $\hat\rho_A\succ_{\textrm{\tiny W}_{\tiny 2}} \hat\rho_B$.
This proves Result 9 for Proposals 1 and 2 of generalized Wigner majorization.

In order to prove also Result 9 for the tautological preorder, it is enough to recall that, when $\hat\rho_A\succ_{\textrm{\tiny W}_{\tiny t}} \hat\rho_B$, 
there exists a Gaussian channel $\mathcal{E}$ such that $\mathcal{E}(\hat\rho_A)=\hat\rho_B$. In \cite{parisferraro18, Takagi:2018rqp}, the authors prove that, in CV systems, the Wigner logarithmic negativity is monotonic under a class of operations called Gaussian protocols, which include the Gaussian channels. This leads to the conclusion that $\hat\rho_A\succ_{\textrm{\tiny W}_{\tiny t}} \hat\rho_B $ implies $\mathcal{N}_{W_A}\geq \mathcal{N}_{W_B}$ also for the tautological preorder.
This concludes the proof of Result 9. 
We find it worth mentioning that the Wigner logarithmic negativity 
\eqref{eq:Wlogneg_def} is a Schur-convex functional both of $\vert W\vert$ and $W$.
To show this fact, we first observe that $I_0[\cdot]$ is Schur-convex given that it is obtained by integrating the convex function $[\cdot]_+$ \cite{MaOlAr,vanherstraeten2021continuous}. Then, inverting both the equations in \eqref{eq:functionalabs_t0}, we find that the Wigner logarithmic negativity is written by composing a Schur-convex functional with a strictly increasing function (the logarithm), which still leads to a Schur-convex functional \cite{MaOlAr}. This concludes the proof. This property is consistent with Result 9, namely with the fact that majorization orders $\succ_{\textrm{\tiny W}} $ and $\succ_{\textrm{\tiny W$_2$}} $ imply the monotonicity of the Wigner logarithmic negativity.

In the rest of this subsection, we focus on another family of quantities that can be constructed from the Wigner function of a given state, namely the Wigner Renyi entropies defined in \eqref{eq:RenyiWigner}.
In \cite{Koukoulekidis:2021ppu}, the Wigner Renyi entropies have been studied in DV systems defined on Hilbert spaces with odd dimensionality. For these quantities to be well-defined for states that are not Wigner-positive, one can choose $\alpha=2p/(2q-1)$, with $p$ and $q$ integer numbers. 
With this choice, in \cite{Koukoulekidis:2021ppu} it has been proved that, given a DV state majorizing another DV state, the discrete Wigner Renyi entropies of the latter is larger than the one of the former. 

Motivated by this finding, in the following we try to establish an inequality between the Wigner Rényi entropies of CV states connected by Gaussian channels.
To the best of our knowledge, this analysis in the context of CV states has never been performed.
To include all CV states in our analysis, hereafter we restrict the discussion to Wigner Rényi with $\alpha=2p/(2q-1)$.
We stress that this constraint on the exponent in \eqref{eq:RenyiWigner} requires the choice of a branch for the multi-valued function. In the following, we will always choose the real branch of the integrand
\footnote{
To avoid the choice of the branch, we could have defined the Wigner Renyi entropies \eqref{eq:RenyiWigner} with an absolute value on the integrand. However, in this way, the Wigner function would have lost its normalization, and, therefore, the interpretation of the resulting quantity as a Renyi entropy would have been more difficult. We postpone the discussion of this potentially interesting new definition to future investigations.
}.
In the literature, we find other attempts to define a Wigner entropy for states whose Wigner function may admit negative values. For this purpose, in \cite{Cerf:2023jws}, a complex Wigner entropy has been introduced via an analytic continuation to a complex phase space.

Let us consider a CV state $\hat\rho_{\textrm{\tiny in}}$ with Wigner function $W_{\textrm{\tiny in}}$ and, given a Gaussian channel $\mathcal{E}$ identified by the matrices $X$ and $Y$, let us define $\hat\rho_{\textrm{\tiny out}}=\mathcal{E}(\hat\rho_{\textrm{\tiny in}})$ and $W_{\textrm{\tiny out}}$ its Wigner function. The Wigner functions $W_{\textrm{\tiny in}}$ and $W_{\textrm{\tiny out}}$ are related by (\ref{eq:inputoutputNewW}). By using the triangular inequality, we can write the inequality 
\begin{equation}
\label{eq:triangularinequality}
|W_{\textrm{\tiny out}}(\boldsymbol{r})|\leq
\int d\boldsymbol{z} k (\boldsymbol{r},\boldsymbol{z})|W_{\textrm{\tiny in}}(\boldsymbol{z})|\,,
\end{equation}
where we have also exploited that $k$ is a positive function.
Now let $\Phi$ be a non-negative increasing convex function
$\Phi:\mathbb{R}_+\to \mathbb{R}_+$ such that $\Phi(0)=0$. The following inequalities hold
\begin{eqnarray}
\label{eq:inequalitiesWignerrenyi}
 &&
 \nonumber\int d\boldsymbol{r}~\Phi \left(\det X |W_{\textrm{\tiny out}}(\boldsymbol{r})|\right)
 \\&\leq &   \int d\boldsymbol{r}~\Phi \left(\int d\boldsymbol{z}~\det X k(\boldsymbol{r},\boldsymbol{z}) |W_{\textrm{\tiny in}}(\boldsymbol{z})|\right) \nonumber \\
&\leq &\int d\boldsymbol{r} \int d\boldsymbol{z}~\det X k (\boldsymbol{r},\boldsymbol{z}) \Phi (|W_{\textrm{\tiny in}}(\boldsymbol{z})|) \nonumber \\
&=& \int  d\boldsymbol{z}~\det X  \int d\boldsymbol{r}~k(\boldsymbol{r},\boldsymbol{z})  \Phi (|W_{\textrm{\tiny in}}(\boldsymbol{z})|) \nonumber \\
&=& \int d\boldsymbol{z}~ \det X \Phi (|W_{\textrm{\tiny in}}(\boldsymbol{z})|) \,.
\end{eqnarray}
The first inequality follows from (\ref{eq:triangularinequality}) and $\Phi$ being non-negative and increasing. 
The second inequality is implied by Jensen’s inequality because, from (\ref{eq:propertykernel1_mt}), 
we have that $\det X k(\boldsymbol{r},\boldsymbol{z}) $ is a probability distribution over $\boldsymbol{z}$. Finally, the last equality comes from (\ref{eq:propertykernel1_mt}). The function $\Phi(x)=x^\alpha$ with $\alpha\geq 1$ and $x\in[0,\infty)$ satisfies all the assumptions and, therefore, by applying it to (\ref{eq:inequalitiesWignerrenyi}), yields 
\\

\noindent\fbox{%
    \parbox{\linewidth}{{
{\bf Result 10:}
Consider a CV state $\hat\rho_{\textrm{\tiny in}}$ with Wigner function $W_{\textrm{\tiny in}}$ and act on it with a Gaussian channels $\mathcal{E}$ determined by the matrices $X$ and $Y$. If we denote the output state as
$ \hat\rho_{\textrm{\tiny out}}=\mathcal{E}(\hat\rho_{\textrm{\tiny in}})$ and its Wigner function as $W_{\textrm{\tiny out}}$, the following inequality holds $\forall~\alpha\geq 1 $:
 \begin{equation}
 \label{eq:ineq_quasi_RenyiNegativity_mt}
\  (\det X)^{\alpha-1} \int d\boldsymbol{r}~|W_{\textrm{\tiny out}}(\boldsymbol{r})|^{\alpha} \leq \int d\boldsymbol{r}~|W_{\textrm{\tiny in}}(\boldsymbol{r})|^\alpha
 \,.
 \end{equation}
 }}}
 \\
 
 \noindent
Taking the logarithm of the two sides of (\ref{eq:ineq_quasi_RenyiNegativity_mt}) and dividing by the negative quantity $1-\alpha$ (we assume $\alpha=2p/(2q-1)\geq 1$), we find
\begin{equation}
\label{eq:ineq_Renyinew}
-\ln\det X+ S_{W_{\textrm{\tiny out}}}^{(\alpha)}\geq S_{W_{\textrm{\tiny in}}}^{(\alpha)}\,,
\end{equation}
where $S_W^{(\alpha)}$ is defined in (\ref{eq:RenyiWigner}).

For any value of $\alpha=2p/(2q-1)$, (\ref{eq:ineq_quasi_RenyiNegativity_mt}) exhibits a dependence on the matrix $X$ characterizing the Gaussian channel applied to the input state. 
This dependence does not allow to prove the monotonicity of $S_W^{(\alpha)}$ under Gaussian channels, suggesting also that this monotonicity might not be valid in the CV case. 
This hand-wavy argument is verified by explicit counterexamples reported in Fig.\,\ref{fig:WignerRenyi}.
In the left panel of Fig.\,\ref{fig:WignerRenyi}, 
we show the difference between $ S^{(\alpha)}_{W_{\textrm{\tiny in}}}$ with input state (\ref{eq:Wigner_HCmix_ex}) and $ S^{(\alpha)}_{W_{\textrm{\tiny out}}}$, 
where the output state is obtained by applying the thermal noise channel (\ref{eq:noisechannel}) to the input state. 
The results are plotted as functions of $s$ parameter in (\ref{eq:noisechannel}) with a fixed value of the other parameter $c=0.75$. We report the curves for two values of $u$, namely for two different input states, and for three values of the Rényi index $\alpha=2p/(2q-1)$, with $p$ and $q$ integer. The lack of definite sign for the curves in this panel confirms that $S^{(\alpha)}_{W}$
and the Rényi entropies with even index are not monotonic under Gaussian channels.
To check the validity of the inequality (\ref{eq:ineq_Renyinew}), in the right panel we subtract $\ln\det X=\ln(1-s)$ (see (\ref{eq:noisechannel})) from the difference $S^{(\alpha)}_{W_{\textrm{\tiny out}}}-S^{(\alpha)}_{W_{\textrm{\tiny in}}}$. As expected, all the curves are positive in this panel.
Thus, the analysis above and the results showed in Fig.\,\ref{fig:WignerRenyi} allows us to conclude that there is no monotonicity for the Wigner Rényi entropies (or any of their generalizations) under Gaussian channels. 
Finally, we find it worth commenting that, setting $\alpha=1$ in (\ref{eq:ineq_quasi_RenyiNegativity_mt}), the dependence on $X$ drops out and we obtain an alternative proof of the monotonicity of the Wigner logarithmic negativity under Gaussian channels, already shown in \cite{parisferraro18, Takagi:2018rqp}.

\subsection{Quantum channels on generic input Wigner functions}
\label{subsec:ChannelsGenericWigner}

In this subsection, we exploit all the three preorder relations proposed in Sec.\,\ref{subsec:3proposals} to investigate the relation between generic CV states and the output obtained by applying quantum operations. Similarly to the analysis in Sec.\,\ref{sec:Wignermajo_Channels}, we mainly focus on Gaussian channels.

\subsubsection{Gaussian channels}

Let us begin by considering the action of Gaussian channels on generic CV states and studying possible preorder relations between input and output states.
Among the three proposals Sec.\,\ref{subsec:3proposals}, the tautological preorder is the most straightforward to discuss in this respect. Indeed, as mentioned in Sec.\,\ref{subsec:proposal3}, this relation is established, by construction, only between CV states related by a Gaussian channel. More precisely, every output state obtained by applying a Gaussian channel to a CV state is less ordered than the corresponding input state. As discussed in Sec.\,\ref{subsec:results_genericWigner}, the tautological preorder realizes a version of Nielsen's theorem for a new preorder relation and, therefore, could be of potential use in the context of the resource theory of non-Gaussianity and Wigner negativity. We postpone investigations along this line for future works.

The same analysis on Proposals 1 and 2 requires more effort. Since these two proposals reduce to the Wigner majorization introduced in \cite{vanherstraeten2021continuous} when evaluated for a pair of Wigner-positive states, we already know from Sec.\,\ref{subsec:Wignermajo_thermonoiseChannels} that there is not a fixed relation between the output state obtained from a Gaussian channel and the corresponding Wigner-positive input state. Exploiting the computations in Appendix \ref{app:exampleoutputWigner function}, in Appendix \ref{app:moreonWignermajo} we report an analysis to verify this statement also for input states whose Wigner function can take negative values.

After these comments on the Wigner majorization order between a generic CV state and the output state obtained by applying Gaussian channels, it is natural to ask whether the class of channels with $\det X\geq 1$ introduced in Sec.\,\ref{subsec:majorizing channel positive} are Wigner-majorizing channels also when acting on generic CV states. This is indeed true and the statement corresponds to the content of Result 11. For simplicity, we prove this result restricting to Proposal 1 in Result 8.
Let us consider a generic CV state $\hat\rho_\mathrm{\tiny in}$ and act on it with a Gaussian channel $\mathcal{E}$ characterized by the matrices $X$ and $Y$. We denote the output state as $\hat\rho_\mathrm{\tiny out}=\mathcal{E}(\hat\rho_\mathrm{\tiny in})$. Let us call $W_\mathrm{\tiny in}$ and $W_\mathrm{\tiny out}$ the Wigner functions of input and output states, respectively.
To prove Result 11, we want to show that if $W_\mathrm{\tiny in}$ and $W_\mathrm{\tiny out}$ are related by (\ref{eq:inputoutputNewW}) with $k$ semi-doubly stochastic kernel, then $W_\mathrm{\tiny in}$ and $W_\mathrm{\tiny out}$ satisfy
\begin{equation}
\label{eq:conditionrequired}
\int [|W_{\textrm{\tiny out}}(\boldsymbol{r})|-t]_+ d\boldsymbol{r} \leq \int [|W_{\textrm{\tiny in}}(\boldsymbol{r})|-t]_+ d\boldsymbol{r} \,,
\quad
\forall t\geq 0\,,
\end{equation} 
namely $\hat\rho_\mathrm{\tiny in}\succ_{\textrm{\tiny W}} \hat\rho_\mathrm{\tiny out}$ according to Proposal 1. This would be sufficient to prove Result 11, given that $k$ in (\ref{eq:k_kerneldef_mt}) is semi-doubly stochastic if and only if $\det X \geq 1$ (see (\ref{eq:propertykernel1_mt})).

 We begin by noticing that the functional $[\cdot-t]_+$ is convex and monotonically increasing \cite{Chong74}.
Given the condition
\begin{equation}
\det X\int d\boldsymbol{z}k(\boldsymbol{r},\boldsymbol{z})=1\,,
\end{equation}
because of (\ref{eq:propertykernel1_mt}), we can apply Jensen's inequality and write the following inequality
\begin{eqnarray}
\label{eq:applyconvexity}
&&\int d\boldsymbol{r} \left[\det X\int d\boldsymbol{z} k(\boldsymbol{r},\boldsymbol{z})|W_{\textrm{\tiny in}}(\boldsymbol{z})|-t\right]_+
\leq
\\
\nonumber
&&
\det X\int d\boldsymbol{r} d\boldsymbol{z}k(\boldsymbol{r},\boldsymbol{z}) [|W_{\textrm{\tiny in}}(\boldsymbol{z})|-t]_+ \leq 
\\
\nonumber
&&
\det X\int  d\boldsymbol{z} [|W_{\textrm{\tiny in}}(\boldsymbol{z})|-t]_+\,,
\end{eqnarray}
where, in the last step, we have used the other condition in (\ref{eq:propertykernel1_mt}).
From (\ref{eq:triangularinequality}), the fact that $[\cdot-t]_+$ is monotonically increasing and the inequality $a\left[f(x)-t\right]_+\leq\left[a f(x)-t\right]_+$ that holds for any positive function $f$ and for any $a\geq 1$ and $t\geq 0$ \footnote{
To show this inequality, we first observe that
$\left[a f(x)-t\right]_+=a\left[ f(x)-t/a\right]_+$. Given that the right-hand side is monotonically decreasing as a function of $t/a$, if $a\geq 1$, we have $\left[a f(x)-t\right]_+\geq a\left[ f(x)-t\right]_+$, which is what we wanted to prove.}, we have
\begin{eqnarray}
&&
\det X\int d\boldsymbol{r}[|W_{\textrm{\tiny out}}(\boldsymbol{r})|-t]_+
\leq
\\
&&
\nonumber
\int d\boldsymbol{r} \left[\det X\int d\boldsymbol{z} k(\boldsymbol{r},\boldsymbol{z})|W_{\textrm{\tiny in}}(\boldsymbol{z})|-t\right]_+ \,,
\end{eqnarray}
which, combined with (\ref{eq:applyconvexity}), leads to (\ref{eq:conditionrequired}). Thus, we can conclude that 
\\

\noindent\fbox{%
    \parbox{\linewidth}{
{\bf Result 11:}
The class of Gaussian channels $\mathcal{E}_{X,Y}$ characterized by the matrices $X$ and $Y$ such that $\det X\geq 1$ is a class of Wigner-majorizing channels when applied to generic CV states. Thus, for any CV state $\hat\rho$, we have that,
according to Proposal 1 in Result 8,
\begin{equation}
\hat\rho\succ_{\textrm{\tiny W}}\mathcal{E}_{X,Y}(\hat\rho)\,.
\end{equation}
}
}
\\

\noindent

\subsubsection{Beyond Gaussian channels}

So far, in studying the interplay between Wigner majorization and quantum operations we have considered only Gaussian channels, i.e. channels that map Gaussian states into Gaussian states. To extend our analysis beyond this class of quantum channels, in this subsection, we consider the random unitary Gaussian channels defined in (\ref{eq:random Gaussian unitary channel}). These channels act on the input state by randomly applying Gaussian unitaries $U_{S,\boldsymbol{\bar{r}}}$ (associated to a symplectic transformation $S$ and displacement $\bar{\boldsymbol{r}}$) according to a given probability distribution $p(S,{\bar{\boldsymbol{r}}})$.
As discussed in Sec.\,\ref{subsec:results_genericWigner}, if the random operations acting on the input state amounts only to displacements, and, consequently, the probability distribution becomes $p(S,{\bar{\boldsymbol{r}}})=p({\bar{\boldsymbol{r}}})$, the resulting channel is called random displacement channel. Interestingly, if $p({\bar{\boldsymbol{r}}})$ is a Gaussian distribution, we obtain a classical mixing channel, discussed in detail in Sec.\,\ref{subsec:majorizing channel positive}.

In this subsection, we aim to prove that
\\

\noindent\fbox{%
    \parbox{\linewidth}{
{\bf Result 12:} For any CV state and any choice of channel of the form (\ref{eq:random Gaussian unitary channel}) we have that, according to Proposal 1 in Result 8,
\begin{equation}
\label{eq:inputmajorization_RGUC_mt}
\hat{\rho}\succ_{\textrm{\tiny W}}\mathcal{E}_p(\hat{\rho})\,,
\end{equation}
}}
\\

\noindent
where, for convenience, we report here the expression of the random unitary Gaussian channel
\begin{equation}
\label{eq:random Gaussian unitary channel bis}
\mathcal{E}_p(\hat{\rho})=\int dS\int d\boldsymbol{\bar{r}}\, p(S,\boldsymbol{\bar{r}}) U_{S,\boldsymbol{\bar{r}}} \hat{\rho} U_{S,\boldsymbol{\bar{r}}}^\dagger\,.
\end{equation}

Employing the usual notation, we consider an input state $\hat\rho_{\textrm{\tiny in}}$ and its Wigner function $W_{\textrm{\tiny in}}$. We denote by $\hat\rho_{\textrm{\tiny out}}=\mathcal{E}_p(\hat\rho_{\textrm{\tiny in}})$ the output state obtained by applying a random unitary Gaussian channel (\ref{eq:random Gaussian unitary channel bis}) with probability distribution $p$, and $W_{\textrm{\tiny out}}$ its Wigner function.
Since the channel (\ref{eq:random Gaussian unitary channel bis}) randomly implements Gaussian unitaries and displacements, its action on the input Wigner function can be written as
\begin{equation}
\label{eq:WinWoutRGUchannel}
W_{\textrm{\tiny out}}(\boldsymbol{r})
=
\int dS d\boldsymbol{\bar{r}} \; p\left(S,\boldsymbol{\bar{r}}\right)
  W_{\textrm{\tiny in}}(S\boldsymbol{r}+\boldsymbol{\bar{r}})
\,.
\end{equation}
Notice that the integral over $S$ is performed over all the phase space symplectic transformations associated with Gaussian unitaries.
 Now consider a generic convex non-increasing function $\Phi:\mathbb{R}_+\to \mathbb{R}_+$ such that $\Phi(0)=0$. Using (\ref{eq:WinWoutRGUchannel}), the triangular inequality and the non-increasing property of $\Phi$,
 we obtain
\begin{eqnarray}
\label{eq:ineqRGUchannel_1}
  &&\int d\boldsymbol{r} \, \Phi (|W_{\textrm{\tiny out}}(\boldsymbol{r})|) 
  \\&=& \int d\boldsymbol{r} \, \Phi \left(\bigg|\int dS d\boldsymbol{\bar{r}} \; p\left(S,\boldsymbol{\bar{r}}\right)
  W_{\textrm{\tiny in}}(S\boldsymbol{r}+\boldsymbol{\bar{r}})\bigg|\right)\nonumber
  \\
  &\leq&  \int d\boldsymbol{r} \, \Phi \left(\int dS d\boldsymbol{\bar{r}} \; p\left(S,\boldsymbol{\bar{r}}\right)
 |W_{\textrm{\tiny in}}(S\boldsymbol{r}+\boldsymbol{\bar{r}})|\right)\nonumber\,, \nonumber
  \end{eqnarray}
where we have further exploited that the probability distribution $p$ is positive.
Since $\Phi$ is also convex, 
we can apply Jensen's inequality to (\ref{eq:ineqRGUchannel_1}), finding
\begin{eqnarray}
&&\int d\boldsymbol{r} \, \Phi (|W_{\textrm{\tiny out}}(\boldsymbol{r})|) 
\\
&\leq & \int d\boldsymbol{r} \int dS \,d\boldsymbol{\bar{r}} \; p\left(S,\boldsymbol{\bar{r}}\right) \Phi (|W_{\textrm{\tiny in}}(S\boldsymbol{r}+\boldsymbol{\bar{r}})|) \nonumber \\
                                           &=& 
                                           \int dS \,d\boldsymbol{\bar{r}} \; p\left(S,\boldsymbol{\bar{r}}\right) \int d\boldsymbol{r} \,\Phi (|W_{\textrm{\tiny in}}(\boldsymbol{r})|)\,, \nonumber
\end{eqnarray}
where in the last step we have performed the change of variables  $S\boldsymbol{r}+\bar{\boldsymbol{r}}\to \boldsymbol{r}$, which does not alter the integration measure over $\boldsymbol{r}$.
Finally, using the normalization of $p\left(S,\boldsymbol{\bar{r}}\right)$, we obtain
\begin{equation}
\label{eq:ineqRGUchannel_fin}
\int d\boldsymbol{r} \, \Phi (|W_{\textrm{\tiny out}}(\boldsymbol{r})|) 
\leq
\int d\boldsymbol{r} \, \Phi (|W_{\textrm{\tiny in}}(\boldsymbol{r})|)
\,.
\end{equation}
Given that (\ref{eq:ineqRGUchannel_fin}) holds for any non-increasing convex positive function $\Phi$ which is vanishing in zero, using the condition 2 in Sec.\,\ref{subsec:reviewmajo} (see (\ref{eq:majocondition2})), we have that $|W_{\textrm{\tiny in}}(\boldsymbol{r})|\ggcurly|W_{\textrm{\tiny out}}(\boldsymbol{r})|$.
By definition of Proposal 1 in Result 8, this implies $\hat{\rho}_{\textrm{\tiny in}}\succ_{\textrm{\tiny W}}\hat{\rho}_{\textrm{\tiny out}}$, which proves Result 12.

Result 12 has insightful connections with Uhlmann's theorem. 
Recalling that an unital channel is a quantum channel that maps the identity into itself, Uhlmann's theorem states that the output state of any unital channel is density matrix-majorized (see Sec.\,\ref{subsec:reviewmajo}) by the corresponding input state. According to the definition, the channel (\ref{eq:random Gaussian unitary channel bis}) is unital. This fact can be shown by exploiting the normalization of the probability distribution $p$. Thus, Result 12 could be seen as a realization of an analog of Uhlmann's theorem for Wigner majorization between CV states. Extending this result to understand the Wigner majorization relation between CV states and the output state after applying a generic unital channel is a task that deserves future investigations.

\section{Discussion and outlook}
\label{sec:Discussion}

In this work we have developed the theory of continuous majorization of Wigner functions in quantum phase space, or Wigner majorization in short.  
We have extended previous work \cite{vanherstraeten2021continuous} to general $N$-mode Wigner-positive states. Moreover, we have proposed possible notions of preorder involving states with finite non-vanishing Wigner negativity.
We have also investigated majorization between
input and output states of Gaussian channels, and some more general channels. 

For Gaussian states, we found a simple criterion for majorization, which is equivalent to comparing the purity of the states. Consequently, there is always a Wigner majorization
relation between a pair of Gaussian states, but it does not add any qualitatively new information beyond purity. Moreover, for Gaussian channels, the Wigner majorization relation
between input and output states can work both ways, as we demonstrated with the simple thermal channel example. We also gave a partial proof of the conjecture made in \cite{vanherstraeten2021continuous}: we showed (Result 2) that in the convex hull of $N$-mode Gaussian states
the equivalence class of pure Gaussian states Wigner majorizes all other states. This is expected, given the relation between majorization and purity. One potentially useful result is the rewriting of the Gaussian channel map from input Wigner function of a  to output Wigner function \cite{Walschaers:2021zvx} as a convolution with a kernel  associated to the channel (Result 6). This kernel is well-defined, as opposed to one computed using the Choi-Jamiolkowski dual representation (\ref{eq:CJ-GaussianWinger}).

When we try to extend the definition of Wigner majorization to include Wigner-negative states with finite Wigner negativity, in addition to Wigner-positive states, more features begin to appear. First of all, in Sec.\,\ref{subsec:results_genericWigner}, we proposed three different notions of preorder for generic CV states, with two of them (Proposals 1 and 2) realizing majorization relations generalizing the existing one to states with finite non-vanishing Wigner negativity.
Proposal 2 was modeled as the direct counterpart of 
Wigner majorization in the DV case, but requires careful regularization procedures. This is why we focused on a more detailed study of Proposal 1. Both proposals
have the desired feature of implying monotonicity of Wigner logarithmic negativity (Result 9). However, from the point of view of resource theories of Wigner negativity or non-Gaussianity, 
both proposals fail to work as a natural preorder (since under Gaussian channels, we do not have a one-way relation of input always majorizing the output).
For this reason we also included the tautological preorder, where we just declare that establishing this preorder is equivalent to finding a Gaussian channel between the two states: this is the reason why we called it as "tautological" preorder. Since, as discussed at the end of Sec.\,\ref{subsec:proposal3}, the tautological preorder does not satisfy \eqref{eq:propweakmajo}, it would be interesting to improve this proposal in such a way it fulfills this property. A promising way is suggested by \eqref{eq:probabilisticChannel}: this extension requires including probabilistic combinations of Gaussian channels, characterized by a kernel that is a convex combination of kernels \eqref{eq:k_kerneldef}, as Wigner-majorizing channels.

Wigner majorization by Proposals 1 or 2  are natural concepts, and characterize the "randomness" of Wigner functions. A question for 
further study is to find interesting applications in some CV settings, 
and also investigate 
more in detail the differences between the two proposals (in particular, finding if there are examples where these give different majorization relations between the same pair of states).
Proposal 1 is based on applying the definition of weak majorization for absolute value of Wigner functions. Therefore all equivalent conditions \ref{item:firstcond}-
\ref{item:decrrearr} on page 3 are applicable, in particular condition \ref{item:second_cond}. It would be insightful to further clarify the connection of Proposals 1 and 2 to Schur-convex (or Schur-concave) functionals.
We also identified a subclass of Gaussian channels where the input always majorizes the output (Result 11),
and proved that the same is true for a class of non-Gaussian channels: the random Gaussian unitary channels (Result 12). Perhaps there is some interesting resource theory where such
operations are considered free, and Wigner majorization is the natural preorder.
It is intriguing to draw a parallelism between the random unitary Gaussian channels in \eqref{eq:random Gaussian unitary channel} and convex mixtures of Clifford unitaries mentioned in \cite{Koukoulekidis:2021ppu} as a promising class of operations to be exploited in magic resource theories. 
Following this idea, in the future, it would be insightful to make more contact between Result 12 and the resource theory of stabilizerness for CV systems.

Our initial motivation for this work came from the question what natural notions of majorization could exist in quantum field theories. 
Continuous majorization of Wigner functions is one such natural concept, defined from first principles, that can be applied to any perturbative quantum field theory where the 
notion of a Fock space is a starting point. What other natural notions could exist, and what applications could be found? 
For example, in \cite{vanLuijk:2024cop}, the notion of density matrix majorization has been studied in the framework of von Neumann operator algebras, which are at the core of the algebraic formulation of quantum field theories. 
Moreover, a recent work has investigated Wigner functions in Krylov space as a measure of the growth of complexity in chaotic dynamics \cite{Basu:2024tgg}.
One possible direction could be to explore whether Wigner majorization is well suited to characterize the growth of chaos and complexity.

\section*{Acknowledgments}

We are grateful 
to Nicolas Cerf and Ludovico Lami for helpful comments on this manuscript and Otto Veltheim for useful discussions.
We acknowledge the two anonymous referees for their insightful suggestions during the review process and for pointing out the notion of semi-doubly stochastic kernels, which led to the improvement of Results 7 and 11.
JdB is supported by the European Research Council under the European Unions Seventh Framework Programme (FP7/2007-2013), ERC Grant agreement ADG 834878.
GDG is supported by the ERC Consolidator grant (number: 101125449/acronym: QComplexity). Views and opinions expressed are however those of the authors only and do not necessarily reflect those of the European Union or the European Research Council. Neither the European Union nor the granting authority can be held responsible for them.
EKV acknowledges the financial support of the Research Council of Finland through the Finnish Quantum Flagship project (358878, UH).

\appendix

\section{Transformations of Wigner functions through Gaussian channels}
\label{app:proofinputoutputformula}

For completeness, in this appendix, we derive the formula (\ref{eq:inputoutputNewW}) in the main text, which was originally found in \cite{DEMOEN1977,DEMOEN197927}.

The first ingredient we exploit is that the characteristic function $\chi_{\textrm{\tiny out}}$ of an output state of a Gaussian channel $\mathcal{E}$  is related to the one of the input state $\chi_{\textrm{\tiny in}}$ as \cite{Serafini17book}
\begin{equation}
\label{eq:transformationRuleCharacteristic}
\chi_{\textrm{\tiny out}}(\boldsymbol{r})=e^{-\frac{1}{2}\boldsymbol{r}^\textrm{t}J^\textrm{t}Y J\boldsymbol{r}}\chi_{\textrm{\tiny in}}(J^\textrm{t}X^\textrm{t}J\boldsymbol{r})\,,
\end{equation} 
where $X$ and $Y$ are the matrices characterizing the Gaussian channel according to (\ref{eq:gaussian channel def}).
Notice the different factor $1/2$ instead of $1/4$ in the exponent of (\ref{eq:transformationRuleCharacteristic}), compared with the corresponding formula in \cite{Serafini17book}. This is due to the different definition of the covariance matrix, which (in Williamson's basis) is $\frac{1}{2}\boldsymbol{1}$ in our conventions and $\boldsymbol{1}$ in the convention of \cite{Serafini17book}.
The Wigner function of a state is related to its characteristic function by (\ref{eq:fromCharacteristictoWigner}).
To avoid carrying the parameter $N$ throughout the computation, we restrict the analysis to the single-mode case, i.e. $N=1$, stressing that for $N>1$, the calculation is identical. Combining (\ref{eq:transformationRuleCharacteristic}) and (\ref{eq:fromCharacteristictoWigner}),
we find the relation between the Wigner function of the output state and the characteristic function of the input state.
It reads
\begin{equation}
\label{eq:inputoutputchannelW_proof}
W_{\textrm{\tiny out}}(\boldsymbol{r})=\frac{1}{4\pi^2}\int d\boldsymbol{r'} e^{\mathrm{i}\boldsymbol{r'}J\boldsymbol{r}} e^{-\frac{1}{2}\boldsymbol{r'}^\textrm{t}J^\textrm{t}Y J\boldsymbol{r'}}\chi_{\textrm{\tiny in}}(J^\textrm{t}X^\textrm{t}J\boldsymbol{r'})\,.
\end{equation}
We want to prove (\ref{eq:inputoutputNewW}) with $N=1$ by showing that it is the same as (\ref{eq:inputoutputchannelW_proof}).
We begin by plugging (\ref{eq:fromCharacteristictoWigner}) connecting $W_{\textrm{\tiny in}}$ and $\chi_{\textrm{\tiny in}}$ into (\ref{eq:inputoutputNewW}) and we obtain
\begin{equation}
\label{eq:inputoutputchannelW_step3}
W_{\textrm{\tiny out}}(\boldsymbol{r})=\int d\boldsymbol{w}d\boldsymbol{v} \chi_{\textrm{\tiny in}}(\boldsymbol{v})
e^{\mathrm{i}\boldsymbol{v^{\textrm{t}}}J\boldsymbol{w}}
\frac{e^{-\frac{1}{2}(\boldsymbol{r}-X\boldsymbol{w})^{\textrm{t}}Y^{-1}(\boldsymbol{r}-X\boldsymbol{w})}}{8\pi^3\sqrt{\det Y}}\,.
\end{equation}
The integral over $\boldsymbol{w}$ is Gaussian and can be performed, yielding
\begin{eqnarray}
\label{eq:GaussianIntegral}
&&\int d\boldsymbol{w}e^{\left(\mathrm{i}\boldsymbol{v}^{\textrm{t}}J+\boldsymbol{r}^{\textrm{t}}Y^{-1}X\right)\boldsymbol{w}}e^{-\frac{1}{2}\boldsymbol{w}^{\textrm{t}}X^{\textrm{t}}Y^{-1}X\boldsymbol{w}}=
\\
&=&2\pi\frac{\sqrt{\det Y}}{\det X}e^{\frac{1}{2}\boldsymbol{r}^{\textrm{t}}Y^{-1}\boldsymbol{r}}e^{-\frac{1}{2}\boldsymbol{v}^{\textrm{t}}JX^{-1}YX^{-\textrm{t}}J^\textrm{t}\boldsymbol{v}}e^{\mathrm{i}\boldsymbol{v}^{\textrm{t}}J X^{-1} \boldsymbol{r}}\,,
\nonumber
\end{eqnarray}
where we have also used the fact that the matrix $Y$ is symmetric and so is its inverse.
Plugging (\ref{eq:GaussianIntegral}) into (\ref{eq:inputoutputchannelW_step3}), we find
\begin{equation}
W_{\textrm{\tiny out}}(\boldsymbol{r})=\int \frac{d\boldsymbol{v}}{4\pi^2} \frac{\chi_{\textrm{\tiny in}}(\boldsymbol{v})}{\det X}e^{-\frac{1}{2}\boldsymbol{v}^{\textrm{t}}JX^{-1}YX^{-\textrm{t}}J^\textrm{t}\boldsymbol{v}}e^{\mathrm{i}\boldsymbol{v}^{\textrm{t}}J X^{-1} \boldsymbol{r}}\,,
\end{equation}
which leads to (\ref{eq:inputoutputchannelW_proof})
once we perform the change of variables $\boldsymbol{r'}=JX^{-\textrm{t}}J^{\textrm{t}}\boldsymbol{v}$, with $d\boldsymbol{r'}=\frac{d\boldsymbol{v}}{\det X}$. Thus, we have proven (\ref{eq:inputoutputNewW}) for $N=1$. Notice that the derivation for generic $N$ follows the same steps.

\section{Choi–Jamiolkowski isomorphism for Gaussian operations}
\label{app:CJapplication to Majorization}

 In this appendix, we review the basics of the C-J isomorphism and how it is implemented for CV systems. 
The C-J isomorphism is a bijective mapping between quantum completely positive maps describing quantum operations and quantum states. 
Let us first review its realization in the context of quantum systems defined on finite-dimensional Hilbert space.
The basic idea is to double the Hilbert spaces where a quantum operation $\mathcal{E}$ acts and apply the resulting operation to a maximally entangled state defined on the doubled Hilbert space. Thus, if $\mathcal{E}$ acts on $\mathcal{H}$, the density matrix $\hat\rho_C$ of the state resulting from the C-J isomorphism, also called Choi state, belongs to $\mathcal{H}^{\otimes 2}$. Formally, the Choi state reads \cite{JAMIOLKOWSKI1972275,CHOI1975285,GeomQuantumStates_book}
\be
\label{eq:CJ_finitedimensional}
\hat\rho_C= (\mathcal{E}\otimes\boldsymbol{1}_{\mathcal{H}})(|\psi\rangle\langle\psi|)\,,
\ee
where $|\psi\rangle\langle\psi|$ is the state which maximally entangles the two copies of $\mathcal{H}$.
Crucially, if $\mathcal{E}$ is a completely positive operation, the density matrix $\hat\rho_C$ is positive. In addition, if $\mathcal{E}$ is also trace-preserving, $\hat\rho_C$ has trace equal to one.

In infinite-dimensional Hilbert spaces, such as the ones where CV states are defined, the maximally entangled state is not normalizable, and the mapping (\ref{eq:CJ_finitedimensional}) has to be adapted.
We first notice that in $N$-mode CV systems the maximally entangled state can be approximated by a limiting sequence of states $|\psi_\nu\rangle^{\otimes N}$, as $\nu\to\infty$. The states $|\psi_\nu\rangle$ are defined as \cite{Serafini17book}
\begin{equation}
|\psi_\nu\rangle=\frac{1}{\cosh \nu}\sum_{j}(\tanh \nu)^j |j\rangle\otimes |j\rangle\,,
\end{equation}
where $|j\rangle$ is the Fock state obtained by applying $j$ times the bosonic creation operator on the vacuum. In this context, the parameter $\nu$ can be seen as a regulator for the maximally entangled state.
At this point, the Choi reduced density matrix associated with a certain CV quantum operation $\mathcal{E}$ can be seen as the limit for $\nu\to\infty$ of the sequence of density matrices obtained as
\begin{equation}
\hat\rho_{C,\nu}=(\mathcal{E}\otimes\boldsymbol{1}_{\mathcal{H}})\left(|\psi_\nu\rangle^{\otimes N}\langle\psi_\nu|^{\otimes N}\right)\,.
\end{equation}
Given the expression of $\hat\rho_{C,\nu}$, one can compute the corresponding Wigner function for any value of $\nu$ using (\ref{eq:characteristicfuntion}) and (\ref{eq:fromCharacteristictoWigner}) and finally evaluate it in the regime $\nu\to\infty$. The result can be interpreted as the Wigner function associated with the quantum operation $\mathcal{E}$.

When the operation $\mathcal{E}$ is a Gaussian channel, the resulting Choi state is a Gaussian state, and the corresponding Wigner function is Gaussian \cite{Serafini17book}. This Gaussian distribution is supported on a quantum phase space parameterized by $4N$ coordinates, accounting for the doubling of the Hilbert space where the Choi state is defined. The expression of the Gaussian Wigner function is reported in (\ref{eq:CJ-GaussianWinger}), and it shows the dependence of the covariance matrix in the doubled phase space as a function of the matrices $X$ and $Y$ characterizing the Gaussian channel $\mathcal{E}$.

\section{Tautological preorder is a preorder}
\label{app:preorder}

In this appendix, we prove that the tautological preorder discussed in Sec.\,\ref{sec:MajorizationNegative} (see \eqref{eq:tautologicalcondition}) deserves the name of preorder. More precisely, we show that the relation $\succ_t$ identifies a preorder.
 To verify this statement, we show that the following conditions hold. 
\begin{enumerate}
\item For any Wigner function $W$
\begin{equation}
\label{eq:condition1preorder}
W \succ_\textrm{t} W,\,\quad \,\,\forall W\,;
\end{equation}
\item
Given three generic Wigner functions $W_1$, $W_2$ and $W_3$
\begin{equation}
\label{eq:condition2preorder}
{\rm if}\;\,\, W_1 \succ_\textrm{t} W_2,\,\, W_2 \succ_\textrm{t} W_3,\quad\Rightarrow\quad  W_1 \succ_\textrm{t} W_3\,.
\end{equation}
\end{enumerate}
To check the validity of (\ref{eq:condition1preorder}) we observe that, taking the limit $Y\to 0$, $X\to \boldsymbol{1}$ of the first expression in (\ref{eq:tautologicalcondition}) with $k$ defined in (\ref{eq:k_kerneldef}), we obtain
\begin{equation}
W_2(\boldsymbol{r})= \int d\boldsymbol{y} \delta(\boldsymbol{r}-\boldsymbol{y}) W_1(\boldsymbol{y})=W_1(\boldsymbol{r}) \,.
\end{equation}
In other words, the kernel $k$ reduces to the identity kernel ($\delta$-function), and, therefore, we can always find the tautological preorder relation between a Wigner function and itself. This fact is physically understood since the identity channel is a Gaussian channel. 
The second condition (\ref{eq:condition2preorder}) can be verified as follows.
Given three Wigner functions $W_1$, $W_2$ and $W_3$, the left side of (\ref{eq:condition2preorder}) can be rewritten as
\begin{equation}
W_1 \succ_\textrm{t} W_2 \quad\Rightarrow\quad 
W_{2}(\boldsymbol{r})
=
\int d\boldsymbol{z} k_{21}(\boldsymbol{r},\boldsymbol{z})W_{1}(\boldsymbol{z}) \,,
\end{equation}
and
\begin{equation}
W_2 \succ_\textrm{t} W_3 \quad\Rightarrow\quad 
W_{3}(\boldsymbol{r})
=
\int d\boldsymbol{z} k_{32}(\boldsymbol{r},\boldsymbol{z})W_{2}(\boldsymbol{z}) \,,
\end{equation}
where the kernels $k_{ij}$ are given by (\ref{eq:k_kerneldef_extended}) with matrices $X_{ij}$  and $Y_{ij}$ characterizing them.
Thus, we have
\begin{eqnarray}
\nonumber
W_{3}(\boldsymbol{r})
&=&
\int d\boldsymbol{z} d\boldsymbol{y} k_{32}(\boldsymbol{r},\boldsymbol{z})k_{21}(\boldsymbol{z},\boldsymbol{y})W_{1}(\boldsymbol{y})
\\
&=&
\int d \boldsymbol{y}K(\boldsymbol{r},\boldsymbol{y}) W_{1}(\boldsymbol{y})
\,,
\end{eqnarray}
where the kernel $K$ is given by (\ref{eq:k_kerneldef_extended}) with the following $X$ and $Y$
\begin{equation}
X=X_{32}X_{21}\,,
\quad\quad
Y=Y_{32}+X_{32}Y_{21}X_{32}^{\rm t}\,.
\end{equation}
This implies that $W_1 \succ_t W_3$. From the physical point of view, the validity of the condition (\ref{eq:condition2preorder}) comes from the fact that, by combining two Gaussian channels, we still obtain a Gaussian channel (the Gaussian channels form a semigroup \cite{heinosaari2009semigroup}).
We conclude that $\succ_t$ is a preorder, and we can refer to it as tautological preorder.

One might wonder whether $\succ_t$ is also a partial order, meaning that, given any pair of Wigner functions $W_1$ and $W_2$, the additional condition
\begin{equation}
\label{eq:condition2partialorder}
{\rm if}\;\,\, W_1 \succ_t W_2,\,\, W_2 \succ_t W_1,\quad\Rightarrow\quad  W_1 = W_2\,,
\end{equation}
has to be fulfilled.
This property is not true for $\succ_t$. Indeed, in the case of $Y\to 0$, namely when the Gaussian channel connecting $W_1$ and $W_2$ is a Gaussian unitary channel, we have that
\begin{equation}
W_2(\boldsymbol{y})=W_1(X^{-1}\boldsymbol{y})\,,
\end{equation}
where $X$ is a symplectic matrix.
The inverse transformation is, in this instance, well-defined and reads
\begin{equation}
W_2(X\boldsymbol{y})=W_1(\boldsymbol{y})\,,
\end{equation}
which implies that $ W_1 \succ_t W_2$ and $ W_2 \succ_t W_1$, even if $ W_1 \neq  W_2$. Thus, $ \succ_t$ is not a partial order.

\section{Examples of Wigner functions after Gaussian channels}
\label{app:exampleoutputWigner function}

In this appendix, we report the analytical expressions of the Wigner functions of output states obtained by acting with thermal noise channels and classical mixing channels on the classes of single-mode CV states relevant to this manuscript.

\subsection{Thermal noise channels and oscillator eigenstates}

We begin by considering the single-mode input state $\hat\rho_{\textrm{\tiny in}}=(1-u)|0\rangle\langle0|+u |1\rangle\langle 1|$, $u\in[0,1]$, where $|0\rangle$ and $|1\rangle$ are the vacuum and the first excited state of a harmonic oscillator respectively.
The Wigner function $\widetilde{W}_u$ of this state is reported in (\ref{eq:Wigner_HCmix_ex}). We want to compute the Wigner function $W_{\textrm{\tiny out}}$ of the output state obtained by applying to $\hat\rho_{\textrm{\tiny in}}$ the thermal noise channel defined by the matrices $X$ and $Y$ in (\ref{eq:noisechannel}). The strategy is to exploit (\ref{eq:inputoutputNewW}) evaluated on $\widetilde{W}_u$ with the kernel $k$ written in terms of the matrices (\ref{eq:noisechannel}).

Writing $\widetilde{W}_u$ in (\ref{eq:Wigner_HCmix_ex}) explicitly in terms of the Laguerre polynomial of order zero and one, it turns out that
the following integrals are useful for writing down the final result $W_{\textrm{\tiny out}}$
\begin{eqnarray}
\label{eq:useful_int1}
&&\int dx' dp'e^{-\frac{\left(x-x'\right)^2}{1-s}-\frac{\left(p-p'\right)^2}{1-s}}e^{-\frac{1}{2 c s}({x'}^2+{p'}^2)}
\nonumber
\\
&=&
\frac{2\pi c s(1-s)}{1+s(2c-1)}e^{-\frac{r^2}{1+s(2c-1)}}\,,
\end{eqnarray}
\begin{eqnarray}
\label{eq:useful_int2}
&&\int dx'dp'
\frac{\left(x-x'\right)^2}{1-s}
e^{-\frac{\left(x-x'\right)^2}{1-s}-\frac{\left(p-p'\right)^2}{1-s}}e^{-\frac{{x'}^2+{p'}^2}{2cs}}
\\
\nonumber
&=&
2\pi c s(1-s)\frac{x^2(1-s)+cs[1+s(2c-1)]}{[1+s(2c-1)]^3}e^{-\frac{r^2}{1+s(2c-1)}}
\,,
\end{eqnarray}
and 
\begin{eqnarray}
\label{eq:useful_int3}
&&\int dx'dp'
\frac{\left(p-p'\right)^2}{1-s}
e^{-\frac{\left(x-x'\right)^2}{1-s}-\frac{\left(p-p'\right)^2}{1-s}}e^{-\frac{{x'}^2+{p'}^2}{2cs}}
\\
\nonumber
&&=
2\pi c s(1-s)\frac{p^2(1-s)+cs[1+s(2c-1)]}{[1+s(2c-1)]^3}e^{-\frac{r^2}{1+s(2c-1)}}
\,,
\end{eqnarray}
where we have expressed the outcome in terms of the radial phase-space coordinate $r=\sqrt{x^2+p^2}$
whenever possible.
Combining these three integrals, the Wigner function of the output state reads
\begin{eqnarray}
\label{eq:outputWigner_thermal channel}
&&W_{\textrm{\tiny out}}(x,p)=W_{\textrm{\tiny out}}(r)= e^{-\frac{r^2}{1+s(2c-1)}}\times
\\
\nonumber
&&
\times\frac{[1+s(2c-1)]^2+2u(1-s)[r^2-1-s(2c-1)]}{\pi[1+s(2c-1)]^3}
\,.
\end{eqnarray}
Let us discuss the result (\ref{eq:outputWigner_thermal channel}) in two significant regimes.
When $s=0$, we have
\begin{equation}
W_{\textrm{\tiny out}}(r)=
e^{-r^2}\frac{1+2u(r^2-1)}{\pi}
=\widetilde{W}_u(r)\,,
\end{equation}
which is consistent with the interpretation that, at $s=0$, the thermal noise channel leaves the input state invariant.
On the other hand, when $s=1$, the thermal channel should transform the input state into a thermal state with eigenvalue $c$ with probability equal to one. Indeed, the resulting Wigner function reads
\begin{equation}
\label{eq:Wign_singlemode_thermal}
W_{\textrm{\tiny out}}(r)=
\frac{e^{-\frac{r^2}{2c}}}{2\pi c}\,.
\end{equation}
Finally, if $u=0$, we are mixing a thermal state with eigenvalue $c$ and the vacuum state of the harmonic oscillator with probabilities $s$ and $1-s$, respectively. Since both the states involved in this mixture are Gaussian, from the rule (\ref{eq:gaussian channel def}), we expect the output to be a Gaussian state with covariance matrix given by
\begin{equation}
\gamma_{\textrm{\tiny out}}=\frac{1-s}{2}+sc=\frac{1+s(2c-1)}{2}\,.
\end{equation}
By imposing $u=0$ in (\ref{eq:outputWigner_thermal channel}), we consistently find the Wigner function of the output state to be
\begin{equation}
\label{eq:wign_GS_mixthermal}
W_{\textrm{\tiny out}}(r)=
\frac{e^{-\frac{r^2}{1+s(2c-1)}}}{\pi [1+s(2c-1)]}\,.
\end{equation}
The only parameter not defined on a finite domain is $c\geqslant 1/2$. It is worth noticing that when $c$ is large enough, the output Wigner function can be approximated by 
\begin{equation}
W_{\textrm{\tiny out}}(r)\sim \frac{e^{-\frac{r^2}{2sc}}}{2\pi s c }
\qquad\qquad
\textrm{when } c\gg 1\,,
\end{equation} 
namely, the output state becomes Wigner-positive for large values of $c$.

For completeness, we can repeat this analysis for an input state given by another mixture of oscillator eigenstates, namely $\hat\rho_{\textrm{\tiny in}}=(1-u)|1\rangle\langle1|+u |2\rangle\langle 2|$, with $u\in[0,1]$. The Wigner state of this input state reads
\begin{equation}
\label{eq:Wignermixed_1and2}
W_{\textrm{\tiny in}}(r)=(1-u)W_{|1\rangle}(r)+uW_{|2\rangle}(r)\,,
\end{equation}
where $W_{|1\rangle}$ and $W_{|2\rangle}$ are given in (\ref{eq:wignerHarmonic}).
Also in this case, we want to apply (\ref{eq:inputoutputNewW}) to the input Wigner function (\ref{eq:Wignermixed_1and2}) with the kernel $k$ determined by the matices (\ref{eq:noisechannel}).
To simplify the computation, we can exploit the fact that the effect of the thermal noise channel applied to $W_{|1\rangle}$ has been computed before in this section (see (\ref{eq:useful_int1})-(\ref{eq:useful_int3})). Thus, the Wigner function $W_{\textrm{\tiny out}}$ of the resulting output state can be written as
\begin{eqnarray}
\label{eq:output_superposition12_v1}
&&W_{\textrm{\tiny out}}(x,p)=
(1-u) \times
\\
\nonumber
&&
\times\frac{e^{-\frac{r^2}{1+s(2c-1)}}}{\pi}\frac{s(2+s(4c^2-1))-1+2r^2(1-s)}{[1+s(2c-1)]^3}
\\
\nonumber
&+&
u \int dx'dp' W_{|2\rangle}\left(\frac{x-x'}{\sqrt{1-s}},\frac{p-p'}{\sqrt{1-s}}\right) \frac{e^{-\frac{x'^2+p'^2}{2cs}}}{cs(1-s)2\pi}\,.
\end{eqnarray}
By employing (\ref{eq:wignerHarmonic}), we can write explicitly
\begin{eqnarray}
&&
W_{|2\rangle}\left(\frac{x-x'}{\sqrt{1-s}},\frac{p-p'}{\sqrt{1-s}}\right) = \\
& & 
=\, \frac{4 e^{-\frac{\left(q-q'\right)^2}{1-s}-\frac{\left(p-p'\right)^2}{1-s}}}{\pi} \,
\Bigg[ \, \frac{1}{4}-\frac{\left(x-x'\right)^2+\left(p-p'\right)^2}{1-s}
\nonumber
\\
& & \hspace{3.7cm}
+\,
\frac{\big( (x-x' )^2+ (p-p' )^2\big)^2}{2(1-s)^2}
\,\Bigg] .
\nonumber
\end{eqnarray}
The second term of (\ref{eq:output_superposition12_v1}) can be computed 
by using the integrals (\ref{eq:useful_int1})-(\ref{eq:useful_int3}) and the following relations
\begin{widetext}
\begin{eqnarray}
\nonumber
&&
\frac{1}{\pi^2 c s(1-s)}\int dx'dp'
\frac{\left(x-x'\right)^4}{(1-s)^2}
e^{-\frac{\left(x-x'\right)^2}{1-s}-\frac{\left(p-p'\right)^2}{1-s}}e^{-\frac{1}{2cs}({x'}^2+{p'}^2)}=
\\
&=&
\frac{2e^{-\frac{r^2}{1+s(2c-1)}}}{\pi}\frac{x^4(1-s)^2+6cs(1-s)(1+s(2c-1))x^2+3c^2s^2[1+s(2c-1)]^2}{[1+s(2c-1)]^5}
\,,
\end{eqnarray}
\begin{eqnarray}
\nonumber
&&\frac{1}{\pi^2 c s(1-s)}\int dx'dp'
\frac{\left(p-p'\right)^4}{(1-s)^2}
e^{-\frac{\left(x-x'\right)^2}{1-s}-\frac{\left(p-p'\right)^2}{1-s}}e^{-\frac{1}{2cs}({x'}^2+{p'}^2)}=
\\
&=&
\frac{2e^{-\frac{r^2}{1+s(2c-1)}}}{\pi}\frac{p^4(1-s)^2+6cs(1-s)(1+s(2c-1))p^2+3c^2s^2[1+s(2c-1)]^2}{[1+s(2c-1)]^5}\,,
\end{eqnarray}
\begin{eqnarray}
\nonumber
&&\frac{2}{\pi^2 c s(1-s)}\int dx'dp'
\frac{\left(x-x'\right)^2\left(p-p'\right)^2}{(1-s)^2}
e^{-\frac{\left(x-x'\right)^2}{1-s}-\frac{\left(p-p'\right)^2}{1-s}}e^{-\frac{1}{2cs}({x'}^2+{p'}^2)}=
\\
&=&
\frac{4e^{-\frac{r^2}{1+s(2c-1)}}}{\pi}\frac{\left(x^2(1-s)+cs[1+s(2c-1)]\right)\left(p^2(1-s)+cs[1+s(2c-1)]\right)}{[1+s(2c-1)]^5}
\,.
\end{eqnarray}
\end{widetext}

Plugging the result back into (\ref{eq:output_superposition12_v1}), through a bit of algebra, we obtain
\begin{eqnarray}
\label{eq:output_superposition12_v2}
&& \hspace{-.4cm}
W_{\textrm{\tiny out}}(r)
= \frac{e^{-\frac{r^2}{1+s(2c-1)}}}{\pi[1+s(2c-1)]^5}\times
\\
\nonumber
&&
\Big\lbrace
(s + 2  c  s - 1)  (1 + (2  c - 1)  s)^2\times
\\
\nonumber
&& \;\;\;\; \times (1 - 2  u + 
   s  (2  c + 2  u - 1))+
   \nonumber
   \\
   \nonumber
&&\;\;\;\;  + \,
2  r^2  (1-s)  (1 + (2  c - 1)  s) \times
\\
\nonumber
&& \times (1 - 3  u + 
   s  (3  u + 2  c  (1 + u) - 1))
+ 2 u r^4(1-s)^2
\Big\rbrace
\,.
\end{eqnarray}
As expected, the Wigner function  (\ref{eq:output_superposition12_v2}) of the output state is normalized to one and is such that, when $s=0$, the input Wigner function (\ref{eq:Wignermixed_1and2}) is retrieved, while, when $s=1$, we obtain (\ref{eq:Wign_singlemode_thermal}).

\subsection{Thermal noise channels and cat states}

Other prototypical examples of bosonic states with non-vanishing Wigner logarithmic negativity are the {\it cat states}. The pure states in this class are defined by the superposition of two coherent states as follows
\begin{equation}
\label{eq:catstate_def}
|\textrm{cat}_\pm(\boldsymbol{\alpha})\rangle=\frac{|\boldsymbol{\alpha}\rangle\pm|-\boldsymbol{\alpha}\rangle}{\sqrt{2(1\pm e^{-|\boldsymbol{\alpha}|^2})}}\,,
\end{equation}
where the coherent state $|\boldsymbol{\alpha}\rangle$ is a pure Gaussian state whose Wigner function is given by (\ref{eq:Wigner}) with $\gamma=\frac{\boldsymbol{1}}{2}$ and $\boldsymbol{\bar{r}}=\boldsymbol{\alpha}$ and the vector $\boldsymbol{\alpha}\in\mathbb{R}^{2N}$ is called {\it size of the cat}.
The Wigner functions of these two states are known and read \cite{QuantumOpticsSchleich}
\begin{equation}
\label{eq:Wignercatstate}
W^{(\boldsymbol{\alpha})}_{\pm}(\boldsymbol{r})=
\frac{e^{-|\boldsymbol{\alpha}+\boldsymbol{r}|^2}+e^{-|\boldsymbol{\alpha}-\boldsymbol{r}|^2}\pm 2\cos\left(2 \boldsymbol{\alpha}\cdot \boldsymbol{r}\right)e^{-|\boldsymbol{r}|^2}}{2\pi\left(1\pm e^{-|\boldsymbol{\alpha}|^2}\right)}
\,.
\end{equation}
Notice that, as expected, setting $\boldsymbol{\alpha}=0$ in (\ref{eq:Wignercatstate}), we retrieve the Wigner function of the ground state, i.e.
\begin{equation}
\label{eq:W ground state}
W_0(\boldsymbol{r})=
\frac{e^{-|\boldsymbol{r}|^2}}{\pi}
\,.
\end{equation}
The goal of this subsection is to compute the Wigner function of the output state obtained by applying a thermal noise channel (\ref{eq:noisechannel}) to the single-mode cat state given by (\ref{eq:Wignercatstate}) with $N=1$. 

For this purpose, the Wigner functions (\ref{eq:Wignercatstate}) can be plugged into (\ref{eq:inputoutputNewW}) with the kernel $k$ in (\ref{eq:k_kerneldef}) defined in terms of the matrices (\ref{eq:noisechannel}). Performing some Gaussian integrals and manipulating the resulting expression, the Wigner function of the output state reads
\begin{eqnarray}
\label{eq:Wignercatstate_afterchannel}
W^{(\boldsymbol{\alpha})}_{\pm,\textrm{\tiny out}}(\boldsymbol{r})&=&
\frac{e^{-\frac{|\sqrt{1-s}\boldsymbol{\alpha}+\boldsymbol{r}|^2}{1+(2c-1)s}}+e^{-\frac{|\sqrt{1-s}\boldsymbol{\alpha}-\boldsymbol{r}|^2}{1+(2c-1)s}}}{2\pi\left(1\pm e^{-|\boldsymbol{\alpha}|^2}\right)(1+(2c-1)s)}
\\
&\pm & \frac{2\cos\left(\frac{2\sqrt{1-s} \boldsymbol{\alpha}\cdot \boldsymbol{r}}{1+(2c-1)s}\right)e^{-\frac{|\boldsymbol{r}|^2+2cs| \boldsymbol{\alpha}|^2}{1+(2c-1)s}}}{2\pi\left(1\pm e^{-|\boldsymbol{\alpha}|^2}\right)(1+(2c-1)s)}
\,.
\nonumber
\end{eqnarray}
Let us comment on some consistency checks for the formula (\ref{eq:Wignercatstate_afterchannel}).
When $s=0$, $W^{(\boldsymbol{\alpha})}_{\pm,\textrm{\tiny out}}(\boldsymbol{r})=W^{(\boldsymbol{\alpha})}_{\pm}(\boldsymbol{r})$, namely we retrieve the input state Wigner function. On the other hand, when $s=1$, $W^{(\boldsymbol{\alpha})}_{\pm,\mathrm{o}}(\boldsymbol{r})$ becomes the Wigner function (\ref{eq:Wign_singlemode_thermal}) of a thermal state with eigenvalue $c$. Finally, when $\boldsymbol{\alpha}=0$, we obtain the result of the application of a thermal noise channel to the ground state whose Wigner function is reported in (\ref{eq:wign_GS_mixthermal}).

\subsection{Classical mixing channels}

To conclude this appendix, we report the analytical expressions of the Wigner functions of the output states obtained by applying an example of classical mixing channel to the mixed states $(1-u)|0\rangle\langle0|+u |1\rangle\langle 1|$ and $(1-u)|1\rangle\langle1|+u |2\rangle\langle 2|$, with $u\in[0,1]$, and to the cat states (\ref{eq:catstate_def}).

Given a generic input state with Wigner function $W_{\textrm{\tiny in}}$, we first plug $X=\boldsymbol{1}$ into (\ref{eq:inputoutputNewW}) and, after the change of variable $\boldsymbol{z}\to \boldsymbol{r}-\boldsymbol{y}$, we obtain
\begin{equation}
\label{eq:convolutionWignerGaussian}
W_{\textrm{\tiny out}}(\boldsymbol{r})=\int d\boldsymbol{y} W_{\textrm{\tiny in}}(\boldsymbol{r}-\boldsymbol{y})\frac{e^{-\frac{1}{2}\boldsymbol{y}^{\textrm{t}}Y^{-1}\boldsymbol{y}}}{(2\pi)^N\sqrt{\det Y}}\,,
\end{equation}
namely, the Wigner function of the output state is the convolution between the Wigner function of the input state and a Gaussian distribution with a covariance matrix given by $Y$.

Since we want to apply these channels to single-mode input states, we restrict our analysis to phase spaces with $N=1$. For simplicity, we also consider the subclass of classical mixing channels with $Y=c\boldsymbol{1}_2$, $c>0$ and $\boldsymbol{1}_2$ being the $2\times 2$ identity matrix.
To apply this classical mixing channel to the mixtures of oscillator eigenstates mentioned above, it is useful first to compute the following integrals
\begin{equation}
\label{eq:classical mixingH0}
\int d\boldsymbol{y} W_{|0\rangle}(\boldsymbol{r}-\boldsymbol{y})\frac{e^{-\frac{1}{2c}\boldsymbol{y}^{\textrm{t}}\boldsymbol{y}}}{2\pi c}=
\frac{e^{-\frac{r^2}{1+2c}}}{\pi (1+2c)}
\,,
\end{equation}
\begin{equation}
\label{eq:classical mixingH1}
\int d\boldsymbol{y} W_{|1\rangle}(\boldsymbol{r}-\boldsymbol{y})\frac{e^{-\frac{1}{2c}\boldsymbol{y}^{\textrm{t}}\boldsymbol{y}}}{2\pi c}=
\frac{e^{-\frac{r^2}{1+2c}}}{\pi (1+2c)^3}\left(2r^2+4c^2-1\right) ,
\end{equation}
\begin{eqnarray}
\label{eq:classical mixingH2}
&&
\int d\boldsymbol{y} W_{|2\rangle}(\boldsymbol{r}-\boldsymbol{y})\frac{e^{-\frac{1}{2c}\boldsymbol{y}^{\textrm{t}}\boldsymbol{y}}}{2\pi c}
\\
&=&
\frac{e^{-\frac{r^2}{1+2c}}}{\pi (1+2c)^5}\left[1-4r^2+2(8c^4+r^4+8c^2r^2-4c^2)\right]
,
\nonumber
\end{eqnarray}
where $r^2=x^2+p^2$ is the radial coordinate in the phase space. Notice, as a consistency check, that when $c\to 0$ (\ref{eq:classical mixingH0})-(\ref{eq:classical mixingH2}) reduce to the Wigner function of the oscillator eigenstates on the left-hand side.
Plugging in (\ref{eq:convolutionWignerGaussian}) the Wigner function (\ref{eq:Wigner_HCmix_ex}) of $(1-u)|0\rangle\langle0|+u |1\rangle\langle 1|$ as input state and exploiting (\ref{eq:classical mixingH0}) and (\ref{eq:classical mixingH1}), we obtain
\begin{equation}
\label{eq:classicalmixing_mix01}
W_{\textrm{\tiny out}}(r)=\frac{e^{-\frac{r^2}{1+2c}}}{\pi (1+2c)^3}\left[(1+2c)^2+2u\left(r^2-2c-1\right)\right]\,.
\end{equation} 
On the other hand, taking $(1-u)|1\rangle\langle 1|+u |2\rangle\langle 2|$ as input state and plugging its Wigner function (\ref{eq:Wignermixed_1and2}) in (\ref{eq:convolutionWignerGaussian}), we find
\begin{eqnarray}
\label{eq:classicalmixing_mix12}
W_{\textrm{\tiny out}}(r)&=&\frac{e^{-\frac{r^2}{1+2c}}}{\pi (1+2c)^5}\left[(1+2c)^2(1-u)(2r^2+4c^2-1)
\right.
\nonumber
\\
&+&
u
\left(1-4r^2+16 c^2+2r^4+16c^2r^2-8c^2\right)\left.\right
]\,,
\nonumber
\\
\,
\end{eqnarray}
where we have used both (\ref{eq:classical mixingH1}) and (\ref{eq:classical mixingH2}).
As expected, both the Wigner function (\ref{eq:classicalmixing_mix01}) and (\ref{eq:classicalmixing_mix12}) become equal to the ones of the corresponding input states when $c=0$. Indeed, in that case, the channel we consider reduces to the identity channel.

Finally, we apply the classical mixing channel to the cat states (\ref{eq:catstate_def}). Plugging (\ref{eq:Wignercatstate}) into (\ref{eq:convolutionWignerGaussian}), in the single-mode case, we obtain the following Wigner function of the output state
\begin{eqnarray}
W^{(\boldsymbol{\alpha})}_{\pm,\textrm{\tiny out}}(\boldsymbol{r})&=&\frac{e^{-\frac{|\boldsymbol{r}|^2}{1+2c}}}{\pi(1+2c)}\Bigg[\frac{e^{2c\frac{|\boldsymbol{\alpha}|^2}{1+2c}}\cosh\left(\frac{2\boldsymbol{\alpha}\cdot \boldsymbol{r}}{1+2c}\right)}{e^{|\boldsymbol{\alpha}|^2}\pm 1}
\\
& & \hspace{2.5cm}
\pm\,
\frac{e^{\frac{|\boldsymbol{\alpha}|^2}{1+2c}}\cos\left(\frac{2\boldsymbol{\alpha}\cdot \boldsymbol{r}}{1+2c}\right)}{e^{|\boldsymbol{\alpha}|^2}\pm 1}
\Bigg]\,.
\phantom{x}
\nonumber
\end{eqnarray}
This Wigner function is normalized to one, according to (\ref{eq:normalizationW}), and becomes the input Wigner function (\ref{eq:Wignercatstate}) in the limit $c\to 0$,
as it should.

\section{More on Wigner majorization among states with Wigner negativity}
\label{app:moreonWignermajo}

\begin{figure*}[t!]
\vspace{.2cm}
\hspace{-0.927cm}
\includegraphics[width=0.31\textwidth]{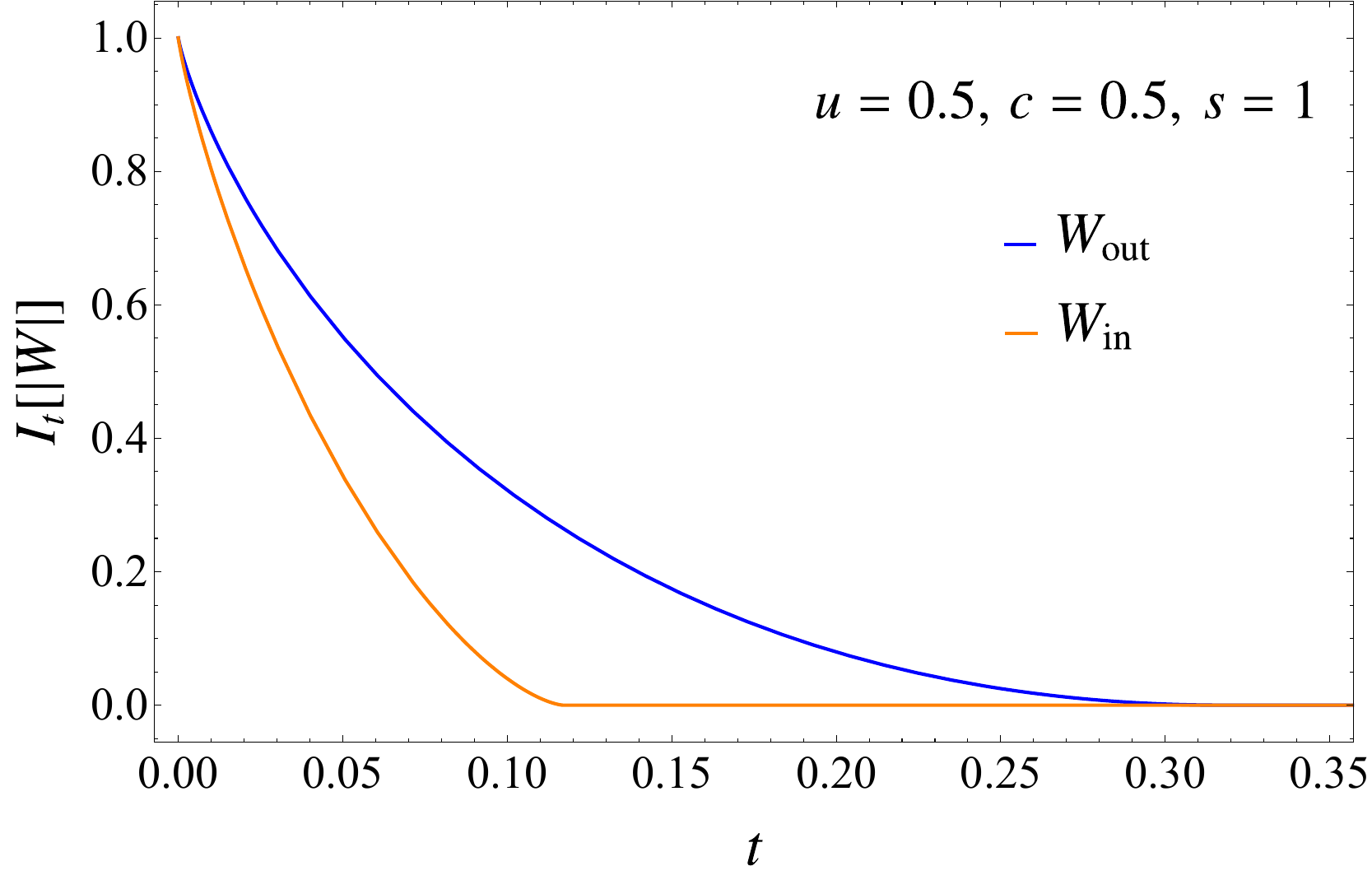}
\hspace{.2cm}
\includegraphics[width=0.31\textwidth]{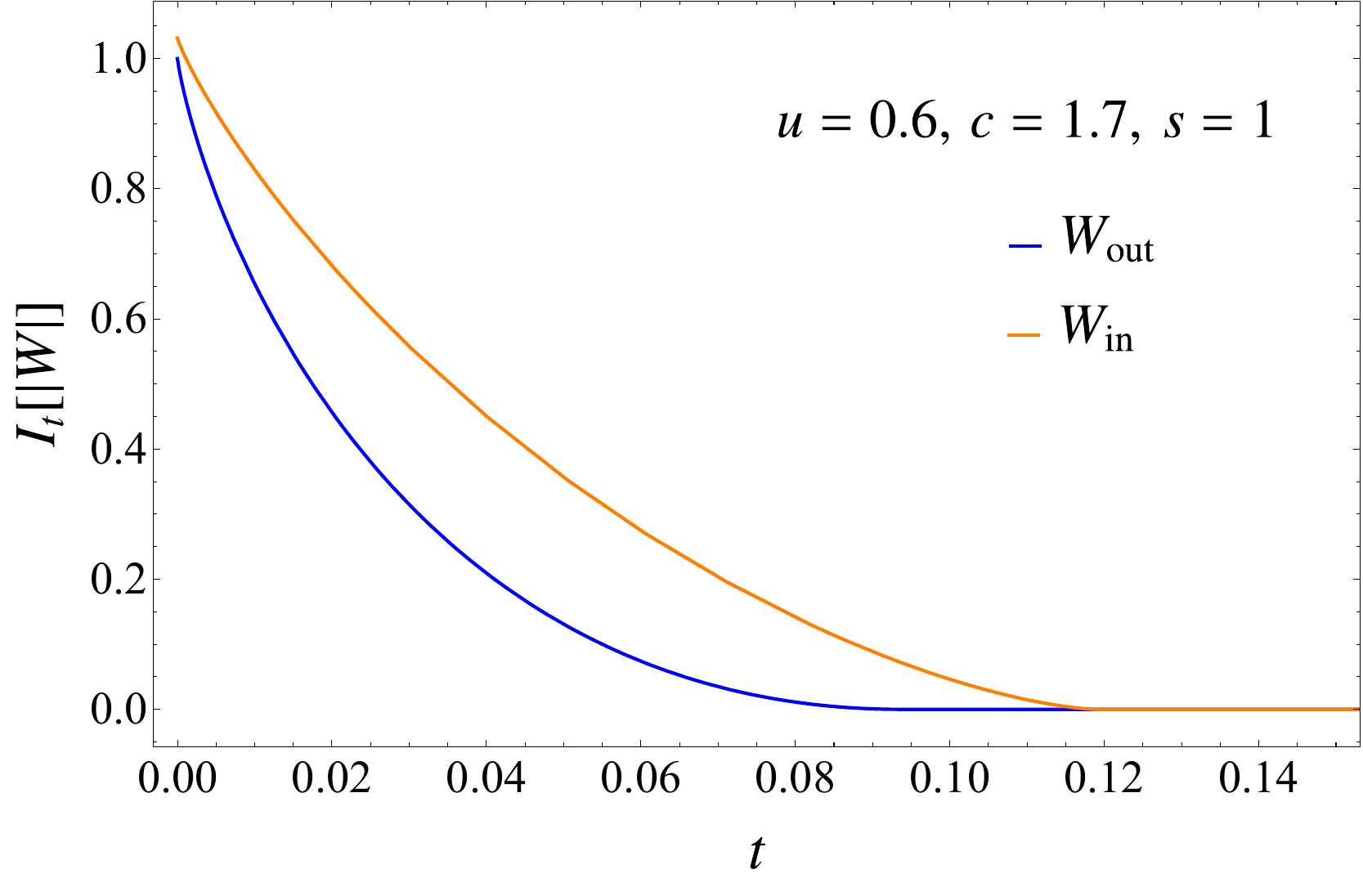}
\hspace{.2cm}
\includegraphics[width=0.31\textwidth]{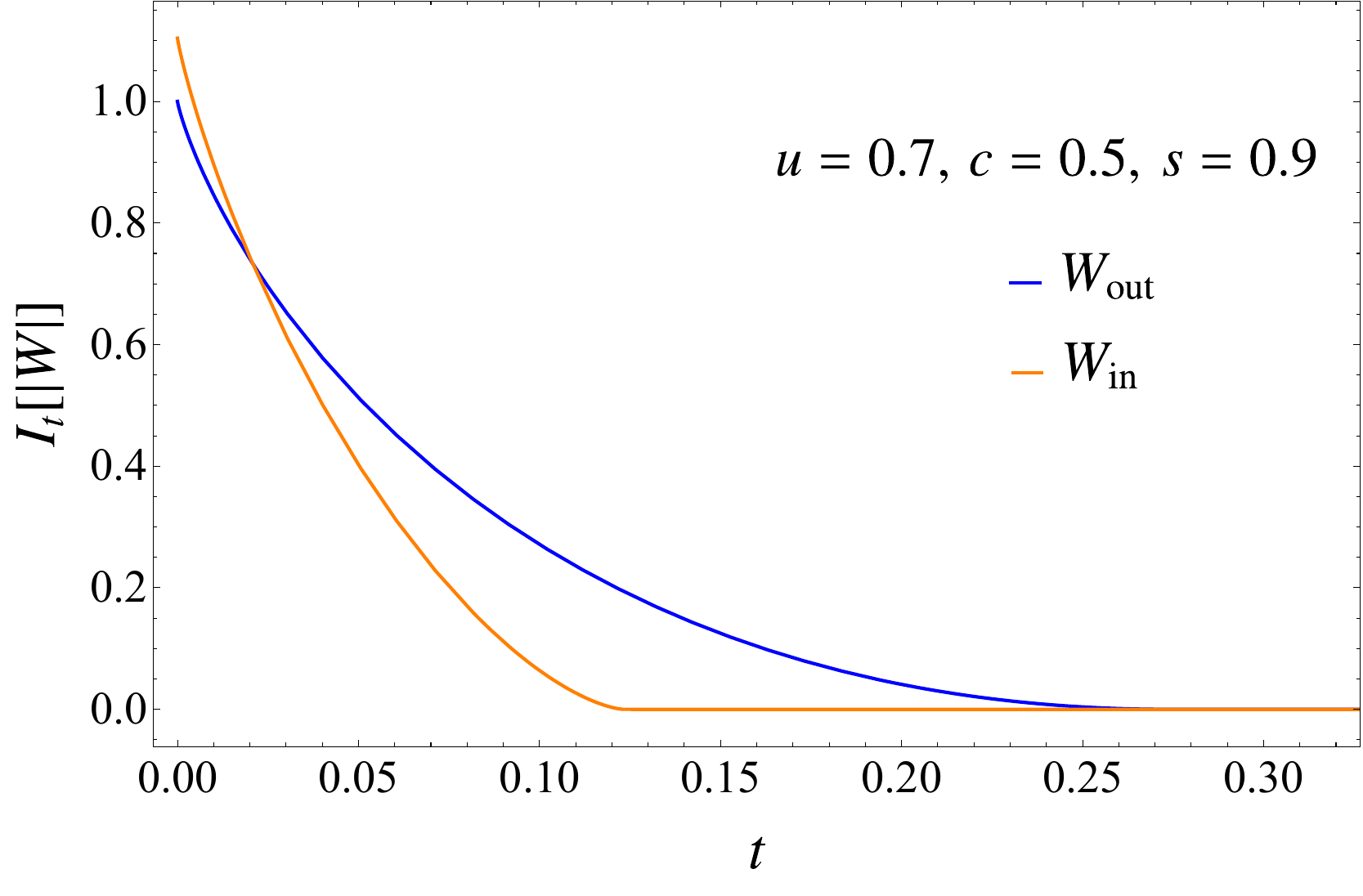}
\vspace{-0.3cm}
\caption{
Comparison of the functional defined in (\ref{eq:def_functionalI})  evaluated on the absolute values of input single-mode Wigner functions (\ref{eq:Wigner_HCmix_ex}) (orange) and of the corresponding output Wigner functions (blue) after the application of a thermal noise channel (\ref{eq:noisechannel}). 
The values of the parameters of the input state and the applied channel are reported in each panel. 
The panels show that input and output states do not have any fixed Wigner majorization relation according to Proposal 1 in Result 8. Wigner majorization can be observed depending on the choice of the parameters: $\hat\rho_{\textrm{\tiny out}}\succ_{\textrm{\tiny W}}\hat\rho_{\textrm{\tiny in}}$ in the left panel, $\hat\rho_{\textrm{\tiny in}}\succ_{\textrm{\tiny W}}\hat\rho_{\textrm{\tiny out}}$ in the middle panel, no Wigner majorization relation in the right panel. 
}
\label{fig:inputoutput}
\end{figure*}

In this appendix, we provide more evidence that Wigner majorization does not always occur between the output state from a Gaussian channel and the corresponding input state.
Since we want to include in our analysis input states that are not Wigner-positive, unless otherwise specified, we exploit Proposal 1 to test eventual Wigner majorization relation.

In Fig.\,\ref{fig:inputoutput}, we show the results obtained for three different choices of single-mode input states (\ref{eq:Wigner_HCmix_ex}) and thermal noise channels (\ref{eq:noisechannel}) applied to them.  
More precisely, we evaluate the functional $I_t$ in (\ref{eq:def_functionalI}) on the absolute value of the input Wigner functions $W_{\textrm{\tiny in}}$ (orange curves) and on the absolute value of the corresponding output Wigner functions $W_{\textrm{\tiny out}}$  (blue curves) and we plot the outcomes against $t$. The values of the parameters $u$ of the input state (\ref{eq:Wigner_HCmix_ex}) and $c$ and $s$ of the channel (\ref{eq:noisechannel}) applied to it are reported in the panels. The explicit expressions of the output Wigner functions obtained after applying (\ref{eq:noisechannel}) to the single-mode mixed state (\ref{eq:Wigner_HCmix_ex}) are reported in Appendix \ref{app:exampleoutputWigner function}.
In the left panel, we report an example where $|W_{\textrm{\tiny out}}|\ggcurly|W_{\textrm{\tiny in}}|$ and, therefore, $\hat\rho_{\textrm{\tiny out}}\succ_{\textrm{\tiny W}}\hat\rho_{\textrm{\tiny in}}$. On the other hand, in the middle panel, considering a different input state and a different thermal noise channel, we have $|W_{\textrm{\tiny in}}|\ggcurly|W_{\textrm{\tiny out}}|$ and $\hat\rho_{\textrm{\tiny in}}\succ_{\textrm{\tiny W}}\hat\rho_{\textrm{\tiny out}}$. Finally, in the right panel, the two curves intersect, signaling the absence of Wigner majorization order between $\hat\rho_{\textrm{\tiny in}}$ and $\hat\rho_{\textrm{\tiny out}}$
This figure explicitly shows the absence of a fixed Wigner majorization order through Proposal 1 between a generic input CV state and the output state obtained by applying a Gaussian channel. Analogous conclusions can be drawn for Proposal 2, although we do not report the results here for brevity.

To complement this finding, in the following we provide a sufficient criterion to rule out the existence of a Gaussian channel connecting two CV states or the presence of a Wigner majorization order.
\\

\noindent
{\bf Proposition 1}\,\, If there exists a Gaussian channel $\mathcal{E}$ such that $\mathcal{E}(\hat\rho_\mathrm{\tiny in})= \hat\rho_\mathrm{\tiny out}$ 
where $\hat\rho_\mathrm{\tiny in}$ and $\hat\rho_\mathrm{\tiny out}$ have  {\it different} Wigner logarithmic negativity, 
then $\hat\rho_\mathrm{\tiny out}$ cannot Wigner-majorize $\hat\rho_\mathrm{\tiny in}$ according to Proposals 1 and 2 in Result 8.
\vspace{0.05pt}

\noindent
{\it Proof:}\,\,
As shown in \cite{parisferraro18, Takagi:2018rqp} (see also Result 10 with $\alpha=1$), if $\mathcal{E}(\hat\rho_\mathrm{\tiny in})= \hat\rho_\mathrm{\tiny out}$, then $\mathcal{N}_{W_\mathrm{\tiny in}}>\mathcal{N}_{W_\mathrm{\tiny out}}$, where $W_\mathrm{\tiny in} $ and $W_\mathrm{\tiny out} $ are the Wigner functions of the input state and the output state respectively and $\mathcal{N}_{W}$ is defined in (\ref{eq:Wlogneg_def}). Notice that the previous inequality is strict by hypothesis.  If $\hat\rho_\mathrm{\tiny out}\succ_{\textrm{\tiny W}}\hat\rho_\mathrm{\tiny in}$, given the functional $I_t$ defined in (\ref{eq:def_functionalI}), we would have
either
\begin{equation}
I_t[|W_\mathrm{\tiny out}|] - I_t[|W_\mathrm{\tiny in}|]\geqslant0
\quad
\forall\,t\geqslant 0\,,
\end{equation}
or
\begin{equation}
I_t[W_\mathrm{\tiny out}] - I_t[W_\mathrm{\tiny in}]\geqslant 0
\quad
\forall\,t\in\mathbb{R}\,,
\end{equation}
for Proposal 1 and Proposal 2, respectively. In particular, we would have either $I_0[|W_\mathrm{\tiny out}|] \geqslant I_0[|W_\mathrm{\tiny in}|]$ or $I_0[W_\mathrm{\tiny out}] - I_0[W_\mathrm{\tiny in}] \geqslant 0$ for the two considered proposals. Using (\ref{eq:functionalabs_t0}), these inequalities would imply $\mathcal{N}_{\mathrm{\tiny in}}\leqslant\mathcal{N}_{\mathrm{\tiny out}}$, contradicting the hypotheses. Thus, the relations $\hat\rho_\mathrm{\tiny out}\succ_{\textrm{\tiny W}}\hat\rho_\mathrm{\tiny in}$ and $\hat\rho_\mathrm{\tiny out}\succ_{\textrm{\tiny W}_\textrm{\tiny 2}}\hat\rho_\mathrm{\tiny in}$  are ruled out,
according to Proposal 1 and Proposal 2 respectively.
\\

\noindent
{\bf Proposition 2}\,\, 
Given two CV states $\hat\rho_\mathrm{\tiny 1}$ and $\hat\rho_\mathrm{\tiny 2}$
with {\it different} Wigner logarithmic negativities and satisfying $\hat\rho_1 \succ_{\textrm{\tiny W}} \hat\rho_2$ or $\hat\rho_1 \succ_{\textrm{\tiny W}_\textrm{\tiny 2}} \hat\rho_2$, 
a Gaussian channel such that $\mathcal{E}(\hat\rho_2)= \hat\rho_1$ does not exist.

\vspace{0.2cm}

\noindent
{\it Proof:}\,\,
From the Proposals 1 or 2 and Result 9, 
 if $\hat\rho_1 \succ_{\textrm{\tiny W}} \hat\rho_2$ or $\hat\rho_1 \succ_{\textrm{\tiny W}_\textrm{\tiny 2}} \hat\rho_2$,
 we have  that $\mathcal{N}_{W_\mathrm{\tiny 1}}>\mathcal{N}_{W_\mathrm{\tiny 2}}$, 
 where $W_\mathrm{\tiny 1} $ and $W_\mathrm{\tiny 2} $ are the Wigner functions of $\hat\rho_1$ and $\hat\rho_2$ respectively 
 and $\mathcal{N}_{W}$ is defined in (\ref{eq:Wlogneg_def}). 
 Again, the inequality between the Wigner logarithmic negativities is strict by hypothesis.
As shown in \cite{parisferraro18}, the existence of a Gaussian channel $\mathcal{E}$ such that $\mathcal{E}(\hat\rho_2)= \hat\rho_1 $ would imply $\mathcal{N}_{W_\mathrm{\tiny 1}}\leq\mathcal{N}_{W_\mathrm{\tiny 2}}$. This would contradict the hypotheses, thus ruling out the existence of the channel $\mathcal{E}$.
\\

The two criteria reported above are not fully general due to the necessary assumptions on the Wigner negativities of the two states involved. We postpone refinement of these results towards a generalization with less stringent hypotheses to future works.

\section{Comparison between Proposal 2 and another proposal in the literature}
\label{app:Comparison}

After this work was accepted for publication, Ref. \cite{Upadhyaya:2025wdt} appeared, introducing a majorization relation valid for any quasi-probability distribution.
 More precisely, in \cite{Upadhyaya:2025wdt}, four conditions which generalize \ref{item:firstcond}-\ref{item:decrrearr} in Sec.\,\ref{subsec:reviewmajo} have been proven to be equivalent definitions of majorization between quasi-probability distributions, without constraints on their domain or negativity. In this appendix, we want to clarify the relation between the definition of majorization in \cite{Upadhyaya:2025wdt} and Proposal 2 in this work. We show that, when restricted to quasi-probability distributions with finitely supported negativity, the definition in \cite{Upadhyaya:2025wdt} is equivalent to Proposal 2 introduced in Sec.\,\ref{subsec:results_genericWigner} and discussed in Sec.\,\ref{subsec:proposal2}.

For this purpose, we start from condition 4 in Theorem 1 of \cite{Upadhyaya:2025wdt}, which states that, given two quasi-probability distributions $f$ and $g$ satisfying \eqref{eq:normalizationcondition},
we say $f\succ g$ if and only if, for any real $u\geq 0$, 
\begin{equation}
\label{eq:moregeneralWignerMajo}
   \int [f(\boldsymbol{r})-u]_+ d\boldsymbol{r} \geq \int [g(\boldsymbol{r})-u]_+d\boldsymbol{r} \,,
\end{equation}
and
\begin{equation}
\label{eq:moregeneralWignerMajo_2}
  \int [f(\boldsymbol{r})+u]_- d\boldsymbol{r} \leq \int [g(\boldsymbol{r})+u]_-d\boldsymbol{r}  \,,
\end{equation}
where $[y]_+ = \max (y,0)$ is defined in \eqref{eq:squarebracket}, while  $[y]_- \equiv \min (y,0)$.

Proposal 2 given by \eqref{eq:final_negMajo1} is considered in this paper only for those quasi-probability distributions whose negativity is finitely supported, i.e. when the phase subspace $\mathcal{A}^{(f)}_t$ introduced below \eqref{eq:final_negMajo1_v2} has finite volume for all $t$. We want to show that, when we restrict to this set of quasi-probability distributions, \eqref{eq:final_negMajo1}  and \eqref{eq:moregeneralWignerMajo}-\eqref{eq:moregeneralWignerMajo_2} are equivalent and, therefore, we can consider the results of the definition of \cite{Upadhyaya:2025wdt} as the generalization of Proposal 2 to the set of all quasi-probability distributions. For convenience, in this appendix, we will denote the quasi-probability distributions that we want to compare $f$ and $g$, instead of $W_1$ and $W_2$ as in \eqref{eq:final_negMajo1}.

The first part of the proof is straightforward given that \eqref{eq:moregeneralWignerMajo} is equivalent to \eqref{eq:final_negMajo1} when $t=u\geq 0$. Thus, to conclude the proof, we need to show that \eqref{eq:final_negMajo1} with $t< 0$ is equivalent to \eqref{eq:moregeneralWignerMajo_2} when $\textrm{Vol}\,\mathcal{A}^{(f)}_t$ and $\textrm{Vol}\,\mathcal{A}^{(g)}_t$ are finite.
Under these assumptions, in Sec.\,\ref{subsec:proposal2}, we have shown that we can rewrite
\eqref{eq:final_negMajo1} with $t< 0$ as \eqref{eq:final_negMajo1_v2} with a regularization that cancels out in \eqref{eq:regularized_proposal2}.
Thus for $t<0$ we can rewrite \eqref{eq:final_negMajo1} as 
\begin{eqnarray}\label{2ndtolastequation}
& &  I_t[f]-I_t[g] \; =
\\
& & =\;
t\left(\textrm{Vol}\,\mathcal{A}^{(f)}_t-\textrm{Vol}\,\mathcal{A}^{(g)}_t\right)
\nonumber
\\
& &
\;\;\;\; + \!\! \int_{\bar{\mathcal{A}}^{(f)}_t} f d\boldsymbol{r}\,
- \int_{\bar{\mathcal{A}}^{(g)}_t} g d\boldsymbol{r} \geq 0 \,.
\nonumber
\end{eqnarray}
Recalling that $f$ and $g$ are assumed to have the same normalization as in \eqref{eq:normalizationcondition} and that $\bar{\mathcal{A}}^{(f)}_t\equiv \mathbb{R}^{2N}\setminus \mathcal{A}^{(f)}_t$, \eqref{2ndtolastequation} implies 
\begin{equation}\label{lastequation}
\int_{\mathcal{A}^{(f)}_t} f\, d\boldsymbol{r} -t \textrm{Vol}\,\mathcal{A}^{(f)}_t \leq
\int_{\mathcal{A}^{(g)}_t} g\, d\boldsymbol{r} -t \textrm{Vol}\,\mathcal{A}^{(g)}_t
\,.
\end{equation}

Relabeling $t=-u<0$ and comparing with the definition of $[y]_- $, we find that \eqref{lastequation} is equivalent to \eqref{eq:moregeneralWignerMajo_2}. This concludes the proof that the definition of majorization, valid for any pair of quasi-probability distributions, proposed in \cite{Upadhyaya:2025wdt} is equivalent to Proposal 2 in Sec.\,\ref{subsec:results_genericWigner}, when we restrict to quasi-probability distributions with finitely supported negativity. Finally, we note that Theorem 1 of \cite{Upadhyaya:2025wdt} presents the definition in a mathematically better controlled form, and additionally proves the definition to be equivalent with three alternative definitions, thus extending the applicability of majorization for quasiprobability distributions.

\bibliographystyle{nb}
\bibliography{MajorizationLiterature}

\begin{thebibliography}{10}
\ifx\href\asklfhas\newcommand{\href}[2]{#2}\fi
\ifx\arxivref\asklfhas\newcommand{\arxivref}[2]{\href{http://arxiv.org/abs/#1}{#2}}\fi
\ifx\doiref\asklfhas\newcommand{\doiref}[2]{\href{http://dx.doi.org/#1}{#2}}\fi
\raggedright
\small
\parskip 0pt

\bibitem{Serafini17book}
A.~Serafini,
\textit{``{Quantum Continuous Variables: A Primer of Theoretical Methods}''},
CRC Press (2017).

\bibitem{Weedbrook12b}
C.~Weedbrook, S.~Pirandola, R.~Garc\'{\i}a-Patr\'on, N.~J.~Cerf, T.~C.~Ralph,
  J.~H.~Shapiro and S.~Lloyd,
\textit{``{Gaussian quantum information}''},
\textsf{\doiref{10.1103/RevModPhys.84.621}{Rev.~Mod.~Phys.~84,~621~(2012)}},
\texttt{\arxivref{1110.3234}{arxiv:1110.3234}}.

\bibitem{Adesso14}
G.~Adesso, S.~Ragy and A.~R.~Lee,
\textit{``{Continuous Variable Quantum Information: Gaussian States and
  Beyond}''},
\textsf{\doiref{10.1142/s1230161214400010}{Open~Syst.~Inf.~Dyn.~21,~1440001~(2014)}},
\texttt{\arxivref{1401.4679}{arxiv:1401.4679}}.

\bibitem{Brau-vanLoock}
S.~L.~Braunstein and P.~van~Loock,
\textit{``Quantum information with continuous variables''},
\textsf{\doiref{10.1103/RevModPhys.77.513}{Rev.~Mod.~Phys.~77,~513~(2005)}}.

\bibitem{Lloyd-Braunstein}
S.~Lloyd and S.~L.~Braunstein,
\textit{``Quantum Computation over Continuous Variables''},
\textsf{\doiref{10.1103/PhysRevLett.82.1784}{Phys.~Rev.~Lett.~82,~1784~(1999)}}.

\bibitem{Menicucci-etal}
N.~C.~Menicucci, P.~van~Loock, M.~Gu, C.~Weedbrook, T.~C.~Ralph and
  M.~A.~Nielsen,
\textit{``Universal Quantum Computation with Continuous-Variable Cluster
  States''},
\textsf{\doiref{10.1103/PhysRevLett.97.110501}{Phys.~Rev.~Lett.~97,~110501~(2006)}}.

\bibitem{Chitambar_2019}
E.~Chitambar and G.~Gour,
\textit{``Quantum resource theories''},
\textsf{\doiref{10.1103/RevModPhys.91.025001}{Rev.~Mod.~Phys.~91,~025001~(2019)}}.

\bibitem{Plenio:2007zz}
M.~B.~Plenio and S.~S.~Virmani,
\textit{``{An Introduction to Entanglement Theory}''},
\textsf{\doiref{10.1007/978-3-319-04063-9_8}{Quant.~Inf.~Comput.~7,~001~(2007)}},
\texttt{\arxivref{quant-ph/0504163}{quant-ph/0504163}}.

\bibitem{nielsen}
M.~A.~Nielsen,
\textit{``{Conditions for a Class of Entanglement Transformations}''},
\textsf{\doiref{10.1103/PhysRevLett.83.436}{Phys.~Rev.~Lett.~83,~436~(1999)}},
\texttt{\arxivref{quant-ph/9811053}{quant-ph/9811053}}.

\bibitem{Veitch_2014}
V.~Veitch, S.~A.~H.~Mousavian, D.~Gottesman and J.~Emerson,
\textit{``The resource theory of stabilizer quantum computation''},
\textsf{\doiref{10.1088/1367-2630/16/1/013009}{New~Journal~of~Physics~16,~013009~(2014)}}.

\bibitem{BSS}
S.~Bravyi, G.~Smith and J.~A.~Smolin,
\textit{``Trading Classical and Quantum Computational Resources''},
\textsf{\doiref{10.1103/PhysRevX.6.021043}{Phys.~Rev.~X~6,~021043~(2016)}}.

\bibitem{How-Campb}
M.~Howard and E.~Campbell,
\textit{``Application of a Resource Theory for Magic States to Fault-Tolerant
  Quantum Computing''},
\textsf{\doiref{10.1103/PhysRevLett.118.090501}{Phys.~Rev.~Lett.~118,~090501~(2017)}}.

\bibitem{Wang_2019}
X.~Wang, M.~M.~Wilde and Y.~Su,
\textit{``Quantifying the magic of quantum channels''},
\textsf{\doiref{10.1088/1367-2630/ab451d}{New~Journal~of~Physics~21,~103002~(2019)}}.

\bibitem{seddon2019}
J.~R.~Seddon and E.~T.~Campbell,
\textit{``Quantifying magic for multi-qubit operations''},
\textsf{Proceedings~of~the~Royal~Society~A~475,~20190251~(2019)}.

\bibitem{LOH22}
L.~Leone, S.~F.~Oliviero and A.~Hamma,
\textit{``{Stabilizer R{\'e}nyi entropy}''},
\textsf{Phys.~Rev.~Lett.~128,~050402~(2022)}.

\bibitem{garcia2024}
R.~J.~Garcia, G.~Bhole, K.~Bu, L.~Chen, H.~Arthanari and A.~Jaffe,
\textit{``On the Hardness of Measuring Magic''},
\texttt{\arxivref{2408.01663}{arxiv:2408.01663}}.

\bibitem{Koukoulekidis:2021ppu}
N.~Koukoulekidis and D.~Jennings,
\textit{``{Constraints on magic state protocols from the statistical mechanics
  of Wigner negativity}''},
\textsf{\doiref{10.1038/s41534-022-00551-1}{npj~Quantum~Inf.~8,~42~(2022)}},
\texttt{\arxivref{2106.15527}{arxiv:2106.15527}}.

\bibitem{parisferraro18}
F.~Albarelli, M.~Genoni, M.~Paris and A.~Ferraro,
\textit{``{Resource theory of quantum non-Gaussianity and Wigner
  negativity}''},
\textsf{\doiref{doi.org/10.1103/PhysRevA.98.052350}{Phys.~Rev.~A~98,~052350~(2018)}},
\texttt{\arxivref{1804.05763}{arxiv:1804.05763}}.

\bibitem{Takagi:2018rqp}
R.~Takagi and Q.~Zhuang,
\textit{``{Convex resource theory of non-Gaussianity}''},
\textsf{\doiref{10.1103/PhysRevA.97.062337}{Phys.~Rev.~A~97,~062337~(2018)}}.

\bibitem{Kenfack:2004ges}
A.~Kenfack and K.~\.Zyczkowski,
\textit{``{Negativity of the Wigner function as an indicator of
  non-classicality}''},
\textsf{\doiref{10.1088/1464-4266/6/10/003}{J.~Opt.~B~6,~396~(2004)}}.

\bibitem{HLP29}
G.~H.~Hardy, J.~E.~Littlewood and G.~Pólya,
\textit{``{Some simple inequalities satisfied by convex functions}''},
\textsf{Messenger~Math.~58,~145–152~(1929)}.

\bibitem{Chong74}
K.-M.~Chong,
\textit{``{Some extensions of a theorem of Hardy, Littlewood and Polya and
  their applications}''},
\textsf{Can.~J.~Math.~26,~1321~(1974)}.

\bibitem{Hickey84}
R.~Hickey,
\textit{``{Continuous majorisation and randomnes}''},
\textsf{J.~Appl.~Prob.~26,~924~(1984)}.

\bibitem{Joe87}
H.~Joe,
\textit{``{Majorization, Randomness and Dependence for Multivariate
  Distributions}''},
\textsf{The~Annals~of~Probability~26,~1217~(1987)}.

\bibitem{vanherstraeten2021continuous}
Z.~Van~Herstraeten, M.~G.~Jabbour and N.~J.~Cerf,
\textit{``{Continuous majorization in quantum phase space}''},
\textsf{\doiref{10.22331/q-2023-05-24-1021}{Quantum~7,~1021~(2023)}},
\texttt{\arxivref{2108.09167}{arxiv:2108.09167}}.

\bibitem{Upadhyaya:2025wdt}
T.~Upadhyaya, Z.~Van~Herstraeten, J.~Davis, O.~Hahn, N.~Koukoulekidis and
  U.~Chabaud,
\textit{``{Majorization theory for quasiprobabilities}''},
\texttt{\arxivref{2507.22986}{arxiv:2507.22986}}.

\bibitem{HLPbook}
G.~H.~Hardy, J.~E.~Littlewood and G.~Pólya,
\textit{``{Inequalities}''},
Cambridge University Press (1988).

\bibitem{MaOlAr}
A.~W.~Marshall, I.~Olkin and B.~C.~Arnold,
\textit{``{Inequalities: Theory of Majorization and Its Applications}''},
Springer Series in Statistics (2011).

\bibitem{Manjegani23}
S.~Manjegani and S.~Moein,
\textit{``{Majorization and semidoubly stochastic operators on $L^1(X)$}''},
\textsf{\doiref{https://doi.org/10.1186/s13660-023-02935-z}{J.~Inequal.~Appl.~2023,~27~(2023)}},
\texttt{\arxivref{2110.12031}{arxiv:2110.12031}}.

\bibitem{Cerf:2024sok}
N.~J.~Cerf and T.~Haas,
\textit{``{Information and majorization theory for fermionic phase-space
  distributions}''},
\texttt{\arxivref{2401.08523}{arxiv:2401.08523}}.

\bibitem{Broecker95}
T.~Bröcker and R.~F.~Werner,
\textit{``{Mixed states with positive Wigner functions }''},
\textsf{\doiref{https://doi.org/10.1063/1.531326}{J.~Math.~Phys.~36,~62–75~(1995)}}.

\bibitem{Williamson36}
J.~Williamson,
\textit{``{On the Algebraic Problem Concerning the Normal Forms of Linear
  Dynamical Systems}''},
\textsf{American~Journal~of~Mathematics~58,~141~(1936)}.

\bibitem{Bourassa:2021ezn}
J.~E.~Bourassa, N.~Quesada, I.~Tzitrin, A.~Sz\'ava, T.~Isacsson, J.~Izaac,
  K.~K.~Sabapathy, G.~Dauphinais and I.~Dhand,
\textit{``{Fast Simulation of Bosonic Qubits via Gaussian Functions in Phase
  Space}''},
\textsf{\doiref{10.1103/PRXQuantum.2.040315}{PRX~Quantum~2,~040315~(2021)}},
\texttt{\arxivref{2103.05530}{arxiv:2103.05530}}.

\bibitem{Jabbour15PRA}
M.~G.~Jabbour, R.~Garc\'{\i}a-Patr\'on and N.~J.~Cerf,
\textit{``{Interconversion of pure Gaussian states requiring non-Gaussian
  operations}''},
\textsf{\doiref{10.1103/PhysRevA.91.012316}{Phys.~Rev.~A~91,~012316~(2015)}},
\texttt{\arxivref{1409.8217}{arxiv:1409.8217}}.

\bibitem{Giedke:2002lqi}
G.~Giedke and J.~I.~Cirac,
\textit{``{Characterization of Gaussian operations and distillation of Gaussian
  states}''},
\textsf{\doiref{10.1103/PhysRevA.66.032316}{Phys.~Rev.~A~66,~032316~(2002)}}.

\bibitem{Walschaers:2021zvx}
M.~Walschaers,
\textit{``{Non-Gaussian Quantum States and Where to Find Them}''},
\textsf{\doiref{10.1103/PRXQuantum.2.030204}{PRX~Quantum~2,~030204~(2021)}},
\texttt{\arxivref{2104.12596}{arxiv:2104.12596}}.

\bibitem{Mari:2012ypq}
A.~Mari and J.~Eisert,
\textit{``{Positive Wigner Functions Render Classical Simulation of Quantum
  Computation Efficient}''},
\textsf{\doiref{10.1103/PhysRevLett.109.230503}{Phys.~Rev.~Lett.~109,~230503~(2012)}}.

\bibitem{Mancini20Capacity}
A.~Arqand, L.~Memarzadeh and S.~Mancini,
\textit{``{Quantum capacity of a bosonic dephasing channel}''},
\textsf{\doiref{10.1103/PhysRevA.102.042413}{Phys.~Rev.~A~102,~042413~(2020)}},
\texttt{\arxivref{12007.03897}{arxiv:12007.03897}}.

\bibitem{Lami:2022bjh}
L.~Lami and M.~M.~Wilde,
\textit{``{Exact solution for the quantum and private capacities of bosonic
  dephasing channels}''},
\textsf{\doiref{10.1038/s41566-023-01190-4}{Nature~Photon.~17,~525~(2023)}},
\texttt{\arxivref{2205.05736}{arxiv:2205.05736}}.

\bibitem{Uhlmann-thm}
A.~Uhlmann,
\textit{``{S\"atze \"uber Dichtematrizen}''},
\textsf{Wiss.~Z.~Karl-Marx-Univ.~Leipzig,~Math.-Nat.~R.~20,~633~(1971)}.

\bibitem{Hertz_2017}
A.~Hertz, M.~G.~Jabbour and N.~J.~Cerf,
\textit{``{Entropy-power uncertainty relations: towards a tight inequality for
  all Gaussian pure states}''},
\textsf{\doiref{10.1088/1751-8121/aa852f}{J.~Phys.~A~50,~385301~(2017)}},
\texttt{\arxivref{1702.07286}{arxiv:1702.07286}}.

\bibitem{Adesso:2012ni}
G.~Adesso, D.~Girolami and A.~Serafini,
\textit{``{Measuring Gaussian quantum information and correlations using the
  R\'enyi entropy of order 2}''},
\textsf{\doiref{10.1103/PhysRevLett.109.190502}{Phys.~Rev.~Lett.~109,~190502~(2012)}},
\texttt{\arxivref{1203.5116}{arxiv:1203.5116}}.

\bibitem{VanHerstraeten:2021nce}
Z.~Van~Herstraeten and N.~J.~Cerf,
\textit{``{Quantum Wigner entropy}''},
\textsf{\doiref{10.1103/PhysRevA.104.042211}{Phys.~Rev.~A~104,~042211~(2021)}},
\texttt{\arxivref{2105.12843}{arxiv:2105.12843}}.

\bibitem{Lami17Logdetinequalities}
L.~Lami, C.~Hirche, G.~Adesso and A.~Winter,
\textit{``{From Log-Determinant Inequalities to Gaussian Entanglement via
  Recoverability Theory}''},
\textsf{\doiref{10.1109/TIT.2017.2737546}{IEEE~Trans.~Inf.~Theory~63,~7553~(2017)}},
\texttt{\arxivref{1703.06149}{arxiv:1703.06149}}.

\bibitem{vanbever2021}
L.~Vanbever,
\textit{``{In Wigner phase space, convolution explains why the vacuum majorizes
  mixtures of Fock states}''},
\texttt{\arxivref{2104.14996}{arxiv:2104.14996}}.

\bibitem{Arias:2023kni}
R.~Arias, G.~Di~Giulio, E.~Keski-Vakkuri and E.~Tonni,
\textit{``{Probing RG flows, symmetry resolution and quench dynamics through
  the capacity of entanglement}''},
\textsf{\doiref{10.1007/JHEP03(2023)175}{JHEP~2303,~175~(2023)}},
\texttt{\arxivref{2301.02117}{arxiv:2301.02117}}.

\bibitem{Orus:2005jq}
R.~Orus,
\textit{``{Entanglement and majorization in (1+1)-dimensional quantum
  systems}''},
\textsf{\doiref{10.1103/PhysRevA.73.019904}{Phys.~Rev.~A~71,~052327~(2005)}},
\texttt{\arxivref{quant-ph/0501110}{quant-ph/0501110}},
[Erratum: Phys.Rev.A 73, 019904 (2006)].

\bibitem{Holevo:2001zsx}
A.~S.~Holevo and R.~F.~Werner,
\textit{``{Evaluating capacities of bosonic Gaussian channels}''},
\textsf{\doiref{10.1103/PhysRevA.63.032312}{Phys.~Rev.~A~63,~032312~(2001)}}.

\bibitem{Giovannetti2004}
V.~Giovannetti, S.~Lloyd, L.~Maccone, J.~H.~Shapiro and B.~J.~Yen,
\textit{``{Minimum R\'enyi and Wehrl entropies at the output of bosonic
  channels}''},
\textsf{\doiref{10.1103/PhysRevA.70.022328}{Phys.~Rev.~A~70,~022328~(2004)}}.

\bibitem{JAMIOLKOWSKI1972275}
A.~Jamiołkowski,
\textit{``{Linear transformations which preserve trace and positive
  semidefiniteness of operators}''},
\textsf{\doiref{https://doi.org/10.1016/0034-4877(72)90011-0}{Reports~on~Mathematical~Physics~3,~275~(1972)}}.

\bibitem{CHOI1975285}
M.-D.~Choi,
\textit{``{Completely positive linear maps on complex matrices}''},
\textsf{\doiref{https://doi.org/10.1016/0024-3795(75)90075-0}{Linear~Algebra~and~its~Applications~10,~285~(1975)}}.

\bibitem{EisertWolf05}
J.~Eisert and M.~M.~Wolf,
\textit{``{Gaussian quantum channels}''},
\textsf{\doiref{10.1103/PRXQuantum.2.030204}{Quantum~Information~with~Continous~Variables~of~Atoms~and~Light~10,~M.~M.~Wolf~(2007)}},
\texttt{\arxivref{quant-ph/0505151}{quant-ph/0505151}}.

\bibitem{HornJohnsonBook}
R.~Horn and C.~Johnson,
\textit{``{Matrix Analysis}''},
Cambridge (2013).

\bibitem{LiebWigner}
E.~Lieb and Y.~Ostrover,
\textit{``{Localization of multidimensional Wigner distributions}''},
\textsf{\doiref{10.1063/1.3486068}{J.~Math.~Phys.~51,~Y.~Ostrover~(2010)}},
\texttt{\arxivref{1007.1796}{arxiv:1007.1796}}.

\bibitem{QuantumOpticsSchleich}
W.~P.~Schleich,
\textit{``Quantum Optics in Phase Space''},
Wiley-VCH (2001).

\bibitem{Fawzi:2024plw}
O.~Fawzi, A.~Oufkir and R.~Salzmann,
\textit{``{Optimal Fidelity Estimation from Binary Measurements for Discrete
  and Continuous Variable Systems}''},
\texttt{\arxivref{2409.04189}{arxiv:2409.04189}}.

\bibitem{heinosaari2009semigroup}
T.~Heinosaari, A.~S.~Holevo and M.~M.~Wolf,
\textit{``{The semigroup structure of Gaussian channels}''},
\textsf{\doiref{doi.org/10.1103/PhysRevA.97.062337}{Quantum~Inf.~Comp.~10,~0619~(2010)}},
\texttt{\arxivref{0909.0408}{arxiv:0909.0408}}.

\bibitem{Cerf:2023jws}
N.~J.~Cerf, A.~Hertz and Z.~Van~Herstraeten,
\textit{``{Complex-valued Wigner entropy of a quantum state}''},
\textsf{\doiref{10.1007/s40509-024-00325-8}{Quant.~Stud.~Math.~Found.~11,~331~(2024)}},
\texttt{\arxivref{2310.19296}{arxiv:2310.19296}}.

\bibitem{vanLuijk:2024cop}
L.~van~Luijk, A.~Stottmeister, R.~F.~Werner and H.~Wilming,
\textit{``{Pure state entanglement and von Neumann algebras}''},
\texttt{\arxivref{2409.17739}{arxiv:2409.17739}}.

\bibitem{Basu:2024tgg}
R.~Basu, A.~Ganguly, S.~Nath and O.~Parrikar,
\textit{``{Complexity growth and the Krylov-Wigner function}''},
\textsf{\doiref{10.1007/JHEP05(2024)264}{JHEP~2405,~264~(2024)}},
\texttt{\arxivref{2402.13694}{arxiv:2402.13694}}.

\bibitem{DEMOEN1977}
B.~Demoen, P.~Vanheuverzwijn and A.~Verbeure,
\textit{``{Completely positive maps on the CCR-algebra}''},
\textsf{\doiref{https://doi.org/10.1007/BF00398582}{Lett.~Math.~Phys~2,~161–166~(1977)}}.

\bibitem{DEMOEN197927}
B.~Demoen, P.~Vanheuverzwijn and A.~Verbeure,
\textit{``{Completely positive quasi-free maps of the CCR-algebra}''},
\textsf{\doiref{https://doi.org/10.1016/0034-4877(79)90049-1}{Reports~on~Mathematical~Physics~15,~27~(1979)}}.

\bibitem{GeomQuantumStates_book}
I.~Bengtsson and K.~{\.Z}yczkowski,
\textit{``Geometry of quantum states: an introduction to quantum
  entanglement''},
Cambridge University Press (2017).

\bibitem{Gottesman:1998hu}
D.~Gottesman,
\textit{``{The Heisenberg representation of quantum computers}''},
\texttt{\arxivref{quant-ph/9807006}{quant-ph/9807006}},
in: \textit{``{22nd International Colloquium on Group Theoretical Methods in
  Physics}''},
32--43p.

\bibitem{Veitch_2012}
V.~Veitch, C.~Ferrie, D.~Gross and J.~Emerson,
\textit{``Negative quasi-probability as a resource for quantum computation''},
\textsf{\doiref{10.1088/1367-2630/14/11/113011}{New~J.~Phys.~14,~113011~(2012)}}.

\bibitem{Veitch_2013}
V.~Veitch, N.~Wiebe, C.~Ferrie and J.~Emerson,
\textit{``Efficient simulation scheme for a class of quantum optics experiments
  with non-negative {Wigner} representation''},
\textsf{\doiref{10.1088/1367-2630/15/1/013037}{New~J.~Phys.~15,~013037~(2013)}}.

\bibitem{Garttner:2022axh}
M.~G\"arttner, T.~Haas and J.~Noll,
\textit{``{Detecting continuous-variable entanglement in phase space with the Q
  distribution}''},
\textsf{\doiref{10.1103/PhysRevA.108.042410}{Phys.~Rev.~A~108,~042410~(2023)}},
\texttt{\arxivref{2211.17165}{arxiv:2211.17165}}.

\end{thebibliography}

\end{document}